\documentclass{article}
\usepackage{PRIMEarxiv}
\usepackage[utf8]{inputenc} 
\usepackage[T1]{fontenc}    
\usepackage{url}            
\usepackage{booktabs}       
\usepackage{amsfonts}       
\usepackage{nicefrac}       
\usepackage{fourier,amssymb,bbm,amsmath,gensymb}
\usepackage{microtype}      
\usepackage{lipsum}
\usepackage{fancyhdr}       
\usepackage{graphicx,subcaption}       
\graphicspath{{media/}}     

\usepackage{booktabs}
\usepackage{textcomp,gensymb}
\usepackage[flushleft]{threeparttable} 
\usepackage{lscape}
\usepackage{multirow} 
\usepackage{graphicx} 
\usepackage{hyperref}
\hypersetup{
    colorlinks=true,
    linkcolor=blue,
    filecolor=magenta,      
    urlcolor=cyan}
\usepackage{natbib} 
\usepackage[nottoc]{tocbibind} 
\usepackage{tablefootnote}

\usepackage{siunitx} 
\usepackage{chngcntr} 
\usepackage{copyrightbox} 
\usepackage{chngpage} 

\title{The effect of ambient air pollution on birth outcomes in Norway

}
\author{
  Xiaoguang Ling \\
  Department of Economics \\
  University of Oslo \\
  Oslo, Norway\\
  \texttt{lingxiaoguangnju@gmail.com} \\
}

\begin{document}
\maketitle

\begin{abstract}

Ambient air pollution is harmful to the fetus even in countries with relatively low levels of pollution. In this paper, I examine the effects of ambient air pollution on birth outcomes in Norway. I find that prenatal exposure to ambient nitric oxide in the last trimester causes significant birth weight and birth length loss under the same sub-postcode fixed effects and calendar-month fixed effects, whereas other ambient air pollutants such as nitrogen dioxide and sulfur dioxide appear to be at safe levels for the fetus in Norway. In addition, the marginal adverse effect of ambient nitric oxide is larger for newborns with disadvantaged parents. Both average concentrations of nitric oxide and occasional high concentration events can adversely affect birth outcomes. The contributions of my work include: first, my finding that prenatal exposure to environmental nitric oxide has an adverse effect on birth outcomes fills a long-standing knowledge gap. Second, with the large sample size and geographic division of sub-postal codes in Norway, I can control for a rich set of spatio-temporal fixed effects to overcome most of the endogeneity problems caused by the choice of residential area and date of delivery. In addition, I study ambient air pollution in a low-pollution setting, which provides new evidence on the health effects of low ambient air pollution.
\end{abstract}


\newpage
\section{Introduction}
\label{sec:intro}

Ambient air pollution has become one of the major threats to human health. According to the World Health Organization \citep{who2021global}, ambient air pollution causes millions of premature deaths each year. In addition to inducing cardiovascular and respiratory diseases such as heart attacks, strokes, and lung cancer, ambient air pollution has also been found to negatively affect the health of newborns through prenatal exposure. It is well documented that ambient air pollutants such as nitrogen dioxide ($NO_2$), suspended particulate matter ($PM$) and sulfur dioxide ($SO_2$) can harm unborn fetuses and thus negatively affect their long-term health status. However, less is known about the effects of ambient nitric oxide ($NO$).

This paper examines the effects of prenatal exposure to ambient air pollution in the first trimester on birth outcomes (e.g., birth weight and length) and attempts to fill gaps in knowledge about environmental $NO$ that have not been well studied in the literature. After controlling for a rich set of spatio-temporal fixed effects, my paper uses the variance in ambient air pollutant concentrations over narrow time intervals and in a small geographic area of Norway to determine how prenatal air pollution exposure affects birth outcomes. My data contain extensive information about parents as well as meteorological conditions that can be used to control for potential confounding factors. In addition, because parents can choose where to live and when to deliver, prenatal exposure to ambient air pollution is endogenous.
The abundance of data and the empirical strategy of using small geographic units and changes in pollutants over small time intervals limit the possibility of biasing the estimated effects due to such endogeneity problems.

My main finding is that a one standard deviation increase exposure to ambient $NO$ ($25.4 \mu g / m^3$) in the last trimester of pregnancy reduces birth weight by $1\%$ and birth length by $0.3\%$ in infants. In areas with high ambient $NO$ concentrations, the loss of birth weight and birth length can be as high as $1\%$ and $4\%$, respectively. The magnitude of adverse effects is greater for male than female newborns. I also find that the marginal adverse effect of ambient $NO$ is greater among newborns whose parents are disadvantaged in terms of immigrant background/nationality and income. Notably, both average ambient $NO$ pollution levels and occasional high $NO$ pollution events during the last trimester of pregnancy negatively affect birth outcomes. I find that ambient air pollution does not affect infants' APGAR1 and APGAR5 scores.
\footnote{The APGAR is a rapid test at 1 and 5 minutes after birth (APGAR1 and APGAR5) to determine if the newborn needs help breathing or has a heart problem (National Library of Medicine (NLM): \href{https://medlineplus.gov/ency/article/003402.htm}{https://medlineplus.gov/ency/article/003402.htm}). The test scores range from 1 to 10, with 10 being the highest. Norway's APGAR scores are concentrated at a higher level with little variation.}
In addition, prenatal exposure to other ambient air pollutants such as $PM$ does not appear to affect birth outcomes in Norway.

This paper contributes to the literature on the effects of prenatal exposure to ambient air pollution in the following ways: (i) I have identified the effects of ambient $NO$ rather than other pollutants that have been well studied in the literature, which fills a long-standing knowledge gap. In Norway, the relatively low concentrations of other ambient air pollutants, especially $NO_2$ and their temporal variability that differs from $NO$, make it possible to isolate the effects of ambient $NO$. (ii) Using weekly ambient air pollution data, which covers at least 46\% of all births in Norway between 2000 and 2016, and detailed registration data, I can identify how ambient air pollutants vary over a narrow time interval in a very small area. Together with detailed parental demographic information, my study largely overcomes the endogeneity problems caused by the choice of birth location and birth time, and the omission of variables that influence these choices. (iii) My study provides new evidence on the health effects of ambient air pollution in countries with relatively low pollution levels.

The remainder of the paper is organized as follows.Section \ref{sec:review} provides a review of the literature on the adverse effects of ambient air pollution on neonatal health. This section also discusses the particular patterns of ambient air pollution in Norway relative to the literature. After presenting the raw data that comprise my sample in Section \ref{sec:construct}, Section \ref{sec:inter_des} describes how I interpolated monitoring station-level data to the sub-postal district level and assessed the performance of the interpolation. Section \ref{sec:identify} presents my identification strategy and regression model specification for the effect of maternal prenatal air pollution exposure on birth outcomes. Sections \ref{sec:result} and \ref{sec:robust} present my main findings and robustness tests, respectively. In Section \ref{sec:hetero}, I investigate the heterogeneity of adverse health effects of air pollution on birth outcomes in terms of demographics and exposure levels. The final section concludes the paper.

\section{Literature review}
\label{sec:review}

\subsection{Ambient air pollution and birth outcomes}
\label{subsec:birth_air}

A large body of literature suggests that prenatal exposure to air pollutants, such as carbon monoxide ($CO$), nitrogen oxides (${NO_x}$), ozone (${O_3}$), sulfur oxides ($SO_X$), and particulate matter ($PM$), during the prenatal period, especially in the last trimester, is associated with poor birth outcomes.
\footnote{A study based on infants born in Oslo, Norway, between 1999 and 2002 found no significant association between full-term birth weight and exposure to traffic pollution ($NO_2$ and $PM$) during pregnancy \citep{RN315}. I obtained the same results when I restricted the birth outcome data to between 2000 and 2002 (results not shown). This may be due to the small number of observations (26,780 in the literature) and the lack of variation over such a short period of time.}
\cite{RN319} find that an inter-quartile increase in prenatal exposure to $NO_2$, $CO$ and $PM$ is associated with a $9 \sim 16 g$ reduction in birth weight.
\footnote{The study area (Connecticut and Massachusetts, USA) also has  very low levels of ambient air pollution. Daily concentrations of $NO_2$, $SO_2$ and $PM$ at the county level  are all below $40$ $\mu g /m^3$ during the study period, meeting most countries' air quality guidelines}
\cite{RN320} also find similar effects for $PM$: a $10$ $\mu g /m^3$ increase in daily $PM_{10}$ ($PM$ of 10 microns or less in diameter) concentration in the third trimester is correlated with an $11 g$ reduction in birth weight. Ambient pollutants such as $O_3$ \citep{salam2005birth} and $SO_2$ \citep{RN288,RN263} are also associated with low birth weight. Aside from birth weight, ambient air pollution also reduces birth length and fetal head circumference. Each $1 \mu g / m^3$ increase in $NO_2$ and $PM$ reduces  fetal head circumference by $0.12 \sim 0.18 mm$ \citep{RN289}. A $10 \mu g / m^3$ increase in $NO_2$ during pregnancy correlates with a $0.9 mm$ decrease in birth length \citep{estarlich2011residential}.

On the other hand, birth outcomes such as birth weight and birth length are strong indicators of fetal and neonatal mortality as well as a variety of other long-term health outcomes.
For example, below the ideal birth weight ($3,500 \sim 4,000g$), the lower the birth weight, the higher the fetal and neonatal mortality \citep{RN300,RN301}.
\footnote{Fetal macrosomia (oversized fetus) does not seem to be an issue in my study since it is very rare in the data and has a limited impact on mortality compared to low birth weight.}
In the long run, adults born small or disproportionate (too thin or short)  have a high risk for coronary heart disease, high blood pressure, high cholesterol concentrations, and abnormal glucose-insulin metabolism \citep{RN325}. Children with a birth weight of below $3,000g$ and a body mass index (BMI) of below $3.4$  at birth are associated with reduced visual acuity and impaired hearing \citep{RN326}. Children with birth weights below $2,500g$ (conventionally referred to as Low Birth Weight, LBW) have higher rates of subnormal growth and neuro-developmental problems that persist into adolescence \citep{RN302}. In addition to the health and developmental difficulties faced by individuals, low birth weight can also result in high economic costs for families and society. The expected costs of delivery and initial care for a baby weighing $1,000g$ at birth can exceed $\$100,000$ (in $2000$ dollars) \citep{RN329}.

The adverse effects of air pollution on birth outcomes may be heterogeneous. For example, \cite{jedrychowski2009gender} find that boys are more affected by ambient air pollution than girls. An  average increase of $30 \mu g /m^3$ in prenatal exposure to $PM$ during pregnancy is associated with a birth weight loss of $189g$ and a birth length loss of $1.1 cm$ for male newborns, compared to $17g$ and $0.4 cm$ for female newborns. Like other types of pollution, ambient air pollution may be more dangerous for disadvantaged people (in terms of race, income, or education) not only because these people may work and live in areas with high levels of pollution, but also because the adverse effects of air pollution may be nonlinear, and the marginal effects of higher levels of pollution may be greater. Meanwhile, the advantaged may also suffer higher exposure to ambient air pollution due to higher rates of participation in outdoor sports \citep{almond2018childhood}. 

The epidemiological studies mentioned above use linear or logistic regression models to compare odds ratios between groups with different levels of air pollution exposure, conditional on the demographics of pregnant women. However, family location and time of birth, i.e., the environment to which one is exposed, are chosen by individuals, and the reasons behind these choices are likely to influence birth outcomes as well. Limited by data, this literature often does not control for detailed parental information, making it difficult to determine the reasons for these self-selections. This leads to the possibility of endogeneity problems in these studies, regardless of the precision of the measurement of ambient air pollution concentrations. Thus, the correlations between ambient air pollution and health outcomes identified in the literature are not necessarily causal. In contrast to the literature, I do not exploit the spatial variation in air pollution. Instead, I employ a rich set of spatio-temporal fixed effects to capture all spatial variation in ambient air pollution and compare only infants born at the same location within a specific time interval (e.g., within the same calendar-month).

Furthermore, when it comes to nitrogen oxides ($NO_x$), the aforementioned papers only examine the effect of ambient $NO_2$, whereas little is known about the health effects of ambient $NO$ pollution. 
\footnote{Nitrogen oxides ($NO_x$) refer to a family of compounds composed of nitrogen and oxygen, such as nitrous oxide $N_2O$. Regarding ambient air pollution, the two main pollutants in the $NO_x$ family are nitrogen dioxide (${NO_2}$) and nitric oxide ($NO$) because of their high toxicity to human health. In gaseous form, $NO_2$ has a reddish-brown color and a strong odor, and is a major component of visible photo-chemical smog . In contrast, $NO$ is a colorless gas with a sweet odor.}
A systematic review of the literature shows that half of the $62$ studies on ambient air pollution, birth weight, and preterm birth specify the adverse effects of $NO_2$, but no papers examine $NO$ specifically \citep{stieb2012ambient} . On the website of the American Journal of Epidemiology, one of the top journals in the field of epidemiology, 46 journal articles examine the effects of $NO_2$, while only two are related to $NO$, suggesting that research on the adverse effects of ambient $NO$ on newborns is inadequate.
\footnote{These two studies find that the increase in ambient $NO$ exposure during pregnancy is associated with a higher risk of 
low birth weight \citep{ghosh2012assessing} and childhood acute lymphoblastic leukemia \citep{ghosh2013prenatal}. If we expand the topic from ``birth outcomes" to general health outcomes, there are studies on the relationship between $NO$ and diseases such as asthma \citep{kim2004traffic}.}
There is a strand of literature that examines $NO_x$ as a whole, rather than distinguishing $NO$ and $NO_2$, because the two pollutants are correlated (often categorized as ``traffic pollution") and appear to work together.
\footnote{Despite having the same origin and being interrelated, $NO$ and $NO_2$ have very different toxicity to human health and different temporal variations. The toxicity, chemical properties and temporal variation of these two pollutants are discussed in sections \ref{subsec:toxic} and \ref{subsec:concen}.}
A review of 41 studies on ambient air pollution and birth outcomes mentions three papers that examines $NO_x$, but since ambient $NO_x$ mainly includes $NO$ and $NO_2$ (as well as other nitrogen oxides), it is difficult to know whether $NO$ or $NO_2$ is important  \citep{shah2011air}. 
\footnote{It should be noted that the notation of $NO$ ambiguously represents $NO_x$ in this literature review, whereas the notation $NO_x$ is used in the other papers and in my study.}
To make matters worse, ambient $NO$ and $NO_2$ concentrations tend to be positively correlated. For example, if only $NO_2$ has negative health effects and $NO$ does not, then using $NO_x$ as the independent variable would underestimate the adverse effects of $NO_2$ because part of the variation in $NO_x$ is caused by the non-toxic confounder $NO$. In summary, as noted by the World Health Organization(WHO), ``Comparisons of $NO$ and $NO_2$ are scarce and still not conclusive with regard to their relative degree of toxicity'' \citep{who2000}. ``Although several studies have attempted to focus on the health risks of $NO_2$, the contributing effects of these other highly correlated co-pollutants are often difficult to rule out'' \citep{RN324}.

\subsection{Toxicity of $NO$}
\label{subsec:toxic}

Although the effect of ambient $NO$ has not been thoroughly examined in the literature, its toxicity makes it dangerous to ignore it as an ambient air pollutant. The toxicology of $NO$ is complex. At very low levels, $NO$ plays a key role in our cardiovascular, neurological, and immune systems \citep{lowenstein1994nitric}. Low doses of
inhaled $NO$ therapy is often used as an effective vasodilator in the treatment of certain respiratory diseases. 
\footnote{The safe dose of inhaled $NO$ (i$NO$) therapy  in neonates has not been fully established, but most studies start with a dose of 25 $\mu g/{{m}^{3}}$ and gradually decrease the dose. A does of 50 $\mu g/{{m}^{3}}$ has been used in adults.
See also: \cite{RN340} and \cite{RN341}. In my study, the ambient $NO$ concentration can be much higher than such a level.}
This is a reason why ambient $NO$ is rarely considered to be hazardous to human health.
\footnote{Another reason is that $NO$ is relatively unstable and can be oxidized to $NO_2$ by $O_2$ and $O_3$ in the ground atmosphere. This will be discussed in Subsection \ref{subsec:concen}.}

However, $NO$ is far from harmless. Similar to $NO_2$, $NO$ shows genotoxicity and can induce DNA structural alterations and DNA strand breaks \citep{weinberger2001toxicology}. 
\footnote{$NO_2$ is also known to induce DNA mutations and strand breaks in the respiratory tract \citep{koehler2010aspects}. Except for genotoxicity, the effects of $NO_2$ seem to be limited to the respiratory system. Inhalation of high concentration of $NO_2$ may result in acute bronchospasm, delayed pulmonary edema, and late bronchiolitis obliterans. Chronic exposure to low concentrations of $NO_2$ appears to induce pulmonary fibros and inhibit pulmonary defense mechanisms \citep{guidotti1978higher}.}
What makes $NO$ different from other pollutants, such as $NO_2$, is its very high affinity for hemoglobin. $NO$ has a much greater affinity for hemoglobin than oxygen.
\footnote{$NO$ has a 1500 times higher affinity for hemoglobin than $CO$, another air pollutant that is known to have a high hemoglobin affinity and impede the transport of oxygen \citep{RN337}.}
In blood, $NO$ binds to reduced hemoglobin (deoxyhemoglobin) $5 \sim 20$ times faster than it reacts with oxygen \citep{RN337}. Therefore,  inhaled $NO$ that diffuses into our blood through the alveoli and the capillaries will immdiately oxidize the Fe(II) of erythrocyte hemoglobin (Hb) to the Fe(III) state, forming methemoglobin (MetHb).
\footnote{The increase in maternal methemoglobin is also a biomarker for determining when a pregnant woman's health is threatened by toxic substances in the environment \citep{RN306}. In fact, methemoglobinemia is a well-known side effect of nitric oxide therapy mentioned above, and this therapy requires close monitoring for methemoglobin level \citep{RN309}.} 
The increase in methemoglobin impairs oxygen transport due to its lack of ability to bind oxygen reversibly \citep{RN308}.  To make matters worse, the fetus is more likely to be exposed to methemoglobin through the placental barrier \citep{RN305}. Based on the toxicology of $NO$, it is important to investigate whether prenatal exposure to ambient $NO$ may adversely affect the health status of the newborn. In a broader sense, my research also adds new evidence to the literature on ambient air pollution and human health.

In addition, recent studies are beginning to realize the role of genetic pleiotropy in birth outcomes. That is, birth weight loss and other long-term health problems may be the results of certain genetic defects. For example, children of diabetic fathers are, on average, lighter than children of non-diabetic fathers \citep{rn327,RN328}, and infants of mothers at risk for late onset diabetes are heavier \citep{tyrrell13}. The genotoxicity of $NO$ and $NO_2$ and the evidence of a positive association between air pollution and the risk of type II diabetes \citep{rn331,rn330,rn332} make it worthwhile to investigate whether genetic pleiotropy is a mechanism by which ambient air pollution reduces birth weight. The established literature on air pollution and neonatal health outcomes is understudied on this issue. At the same time, omitting heritable traits would lead to omitted-variable bias when ambient air pollution exposure is associated with certain parental genetic characteristics.  In this paper, with the rich registry data, I can observe the parents' history of diabetes, which helps me overcome this problem of omitted variable bias.

\subsection{Ambient $NO$ and $NO_2$ concentration}
\label{subsec:concen}

\begin{figure}[htbp]
    \centering
    \includegraphics[width=1\textwidth]{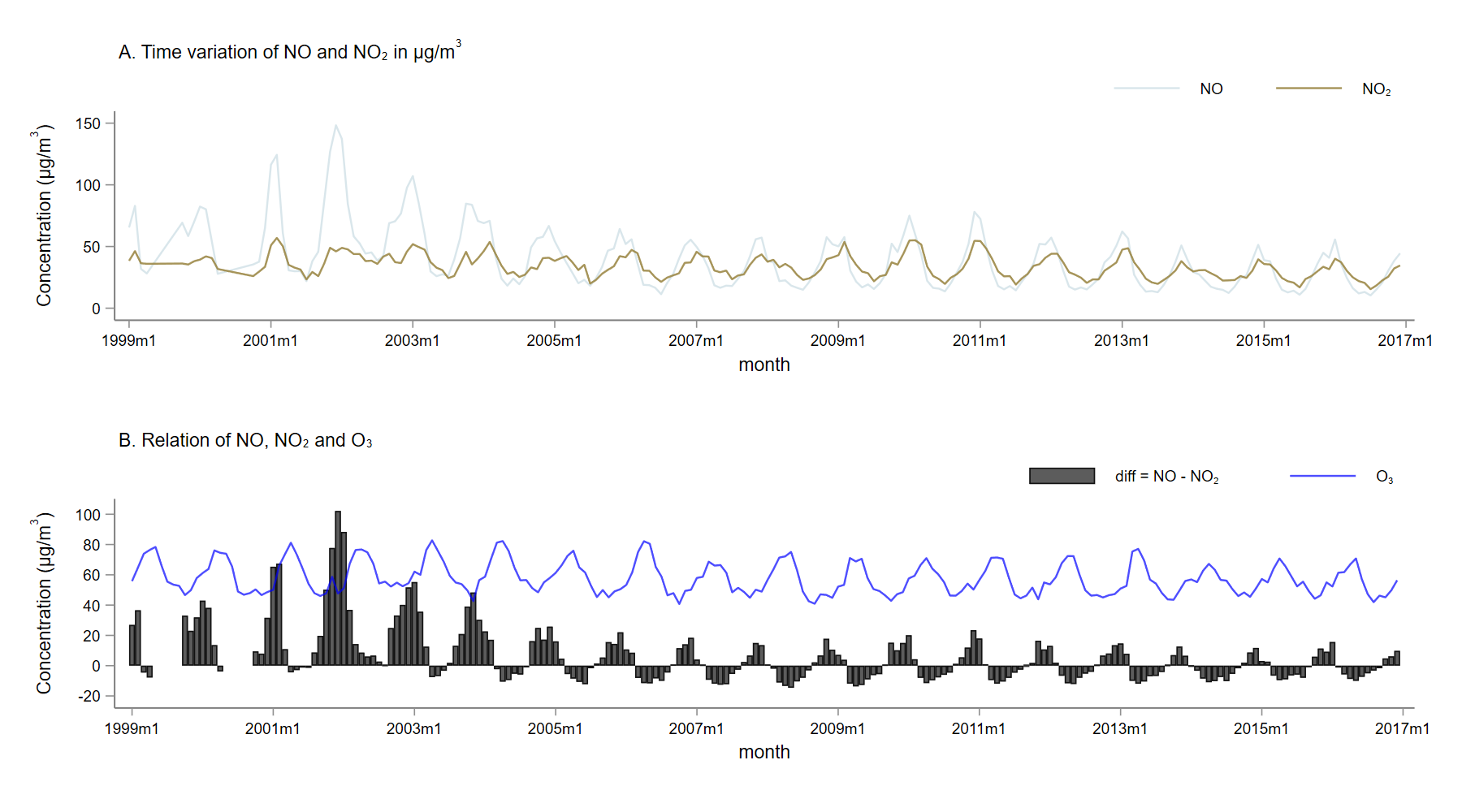}
    \caption{Monthly average $NO$, $NO_2$ and $O_3$ concentrations in Norway, 1999-2016}
    \label{fig:photo}
\end{figure}

There are two main obstacles to examining the effect of ambient $NO$ in the literature: (i) The concentrations of $NO$ and $NO_2$ are highly correlated. (ii) $NO$ is less toxic at very low concentrations compared with $NO_2$, and in many areas studied in the literature, ambient $NO$ concentrations are lower than $NO_2$. Interestingly, the characteristics of the Norwegian ambient air pollution depicted in Figure \ref{fig:photo} make it possible to overcome these two obstacles.
\footnote{It should be noted that the detection of $NO$ and $NO_2$ is not difficult. Sensors for accurate detection of $NO$  using chemiluminescent metho have been commercially available since the 1970s. 
Taking the example of air pollution in California, which has been studied extensively in the literature. From the database provided by \href{https://calepa.ca.gov/}{California Environmental Protection Agency}, we can easily obtain hourly average concentrations of ambient $NO$, $NO_2$ and other $NO_x$ for counties such as Los Angeles and San Diego as early as 1963.}  
 
The gray and brown curves in Figure \ref{fig:photo}. Part A Figure of \ref{fig:photo} shows the temporal variation of monthly average ambient $NO$ and $NO_2$ concentrations in Norway from 1999 to 2016. It is clear that $NO$ and $NO_2$ concentrations are strongly seasonal and often positively correlated.
\footnote{Since ambient $NO$ and $NO_2$ are products of the reaction of nitrogen and oxygen at high temperatures, such as combustion processes in motor vehicles, power plants, and manufacturing, these two ambient air pollutants are positively correlated with each other. In Norway, domestic shipping accounts for about one-third of $NO_x$ emissions. Oil/gas extraction and road traffic each account for a quarter of total $NO_x$ emissions. The rest of the $NO_x$ emissions are contributed by industrial (10\%) and agricultural (3\%) production, among others. For more information, please see: \href{https://www.ssb.no/statbank/table/08941/}{https://www.ssb.no/statbank/table/08941/}.}
However, the amplitude of the $NO$ concentration curve is larger than the variation of $NO_2$, i.e., the fluctuation of $NO$ concentration is larger than that of $NO_2$ during one year. Specifically, in winter, both $NO_2$ and $NO$ concentrations showed an increasing trend, but the increase in $NO$ is greater; in summer, both $NO_2$ and $NO$ concentrations showed a decreasing trend, but the decrease in $NO$ is again greater. The different seasonal variations in $NO$ and $NO_2$ levels imply that infants born in the same calendar quarter or even calendar-month may be exposed to ambient air with different $NO/NO_2$ ratios in the last trimester of pregnancy, a critical period for the fetus. This difference allows me to study the effects of these two pollutants separately.

The seasonal fluctuations of $NO$ and $NO_2$ are mainly caused by photochemical reactions and ambient ozone ($O_3$). In the terrestrial atmosphere, $NO$ is less stable than $NO_2$ and can be rapidly oxidized to $NO_2$ by $O_3$.
\footnote{Ground level ambient $O_3$ is another environmental pollutant that can damage the human respiratory system due to its strong oxidizing power. Although both oxygen ($O_2$) and $O_3$ can oxidize $NO$ to $NO_2$, $O_2$ and $NO$ react very slowly in air. In the laboratory, $O_2$ oxidizes slowly (in days) to $NO_2$ at room temperature when $NO$ is at a concentration of $100 \mu g / m^3$, while $O_3$ can complete the oxidation process in a few hours \citep{RN303}.} 
Therefore, $NO$ is known as a precursor of $NO_2$. Since $O_3$ at the ground level is formed mainly through photochemical reactions, when summer temperatures and solar irradiance are high, active photochemical reactions increase the concentration of $O_3$ in the environment, resulting in more $NO$ being oxidized to $NO_2$  \citep{spicer1977photochemical}.
\footnote{Precisely, in the ground atmosphere, with the participation of sunlight irradiation and other pollutants, relatively small amounts of $NO_2$ can in turn be decomposed into atomic oxygen (rapidly forming ${{O}_{3}}$) and $NO$, due to photolysis. The level of $NO_x$ concentration in the environment detected by the air quality sensor is actually in dynamic equilibrium (the so-called photostasis).} 
In winter, low temperatures and low solar irradiance prolong the life time of both $NO$ and $NO_2$ in the atmosphere.
\footnote{In tunnels with low solar irradiation and O $3$ levels, the ambient $NO$ concentration is often 5-10 times higher due to weak photochemical reactions. This has been documented by the  Norwegian Public Roads Administration: \href{https://vegvesen.brage.unit.no/vegvesen-xmlui/handle/11250/2656305}{https://vegvesen.brage.unit.no/vegvesen-xmlui/handle/11250/2656305}.} 
Less vegetation activity and higher use of heating energy in winter also contribute to high ambient $NO$ and $NO_2$ concentrations.

The above relationships between $NO$, $NO_2$ and $O_3$ are shown in Panel B of Figure \ref{fig:photo}. The black bars in Panel B indicate the difference between $NO$ and $NO_2$ (i.e. $NO-NO_2$ in $\mu g / m^3$), while the blue curve indicates the ambient $O_3$ concentration.
This may be more pronounced at high latitudes (e.g., Norway) because photochemical reactions are much weaker in cold and dark winters \citep{bowling1986climatology}, while long solar days promote photochemical reactions in summer \citep{schjoldager1979observations}.

It is noteworthy that the concentration of $NO$ has dropped sharply since 2005, as shown in Figure \ref{fig:photo}. This appears to be the result of new air pollution regulations in Norway, which came into effect in 2002, replacing regulations established in 1997. In addition to establishing stricter ambient air pollution limits, the new regulations control previously unregulated pollutants. Under the new regulations, certain air quality objectives must be met between 2005 and 2010. 
\footnote{More information about the new regulations (in Norwegian) can be found at: \href{https://lovdata.no/dokument/LTI/forskrift/2002-10-04-1088}{https://lovdata.no/dokument/LTI/forskrift/2002-10-04-1088}.}

Another feature that distinguishes Norway from other regions is that the annual average ambient $NO$ concentrations are higher than $NO_2$. In more than 30 European cities, ambient $NO$ concentrations are lower than $NO_2$ \citep{cyrys2012variation,henschel2015trends}.
\footnote{The two exceptions in the literature are Athens and Glasgow, which have a $NO/NO_2$ ratio of 1.22 $\sim$ 1.39, but both cities have very high air pollution levels ($NO_x > 150 \mu g/m^3$ per day), unlike Norway.} 
As mentioned above, in New Jersey and Las Vegas, the annual average concentration of ambient $NO$ is only about 1/4 of $NO_2$ \citep{roberts2012seasonal,kimbrough2017no}. Given that $NO_2$ is more toxic than $NO$ at low concentrations, ambient $NO$ pollution is less important in these areas. Nevertheless, in Figure \ref{fig:photo}, we can find that in most cases, the ambient $NO$ concentrations in Norway are similar to or even higher than $NO_2$. Until 2005, ambient $NO$ concentrations arehigher than $NO_2$ almost all year round. In 2002, $NO$ concentrations are even three times higher than $NO_2$. This suggests that the environmental $NO$ problem may be more severe in Norway than in other countries.

In addition to the weak photochemical reactions, the unusually high $NO/NO_2$ ratio in Norway can be explained by the chemical properties of $NO$ and $NO_2$ and the special climate of Norway. The freezing and condensation points of $NO_2$ are about $21 \degree{C}$ and $-10 \degree{C}$, respectively. In Norway, the average monthly temperature is below $10 \degree{C}$ for 8 months of the year. Even in summer, the maximum monthly average temperature stays below $21 \degree{C}$. Although $NO_2$ exists in gaseous form in air under normal ambient conditions due to its low partial pressure in the atmosphere ($908 mmHg$ at $25 \degree{C}$) \citep{world2010guidelines},  the gas will be compressed and much heavier than air if enough $NO_2$ molecules are present in the ambient air. Therefore, $NO_2$ is more common in low-lying areas. In contrast, $NO$ is lighter than $NO_2$, and its condensation point of $-152 \degree{C}$ means that its density is less affected by the ambient climate at the surface.

In conclusion, the above discussion implies that ambient $NO$ may be a serious ambient air pollutant in Norway (and possibly in other high-latitude areas). The high concentration of ambient $NO$ relative to $NO_2$ and the different seasonal patterns of variation of the two pollutants make it possible to determine their health effects separately.

\section{Data and sample selection}
\label{sec:construct}
To estimate the effect of ambient air pollution on birth outcomes, data on birth outcomes and prenatal ambient air pollution exposure, i.e., the level of air pollution in the maternal residence during pregnancy, are required. Since weather has an effect on both ambient air pollutant concentrations (Subsection \ref{subsec:concen}) and birth outcomes \citep{beltran2014associations}, meteorological conditions during pregnancy are necessary information. In addition, I need information on parental demographics, as these same parental characteristics may influence both contaminant exposure and birth outcomes. In this section, I describe how my data is constructed and how my baseline sample is selected.

\subsection{Birth outcome and parental demographic data}
\label{subsec:mfr_data}

The birth outcome data, such as birth weight, birth length and APGAR score,
is from \href{https://www.fhi.no/hn/helseregistre-og-registre/mfr/}{Medical Birth Registry (MFR)}, a national health registry that records all births in Norway. The mother's location in the year of delivery and the parents' demographics are provided by \href{https://www.ssb.no/en}{Statistics Norway (SSB)}. The mother's address is at the sub-postcode level, and its definition is presented in Subsection \ref{subsec:grunn}. The parental demographics include age, education level, nationality, immigration background, income, and wealth.
\footnote{Because financial status may be affected by family planning and therefore endogenous, I use information on income, wealth, and debt registered in the two years prior to the birth. The same endogeneity may be true for parents' education levels. However, since annual education registration data are not available, I use the highest education level registered in the dataset. Fortunately, when I restrict the sample to observations where the parents' education level is registered at least two years prior to the birth, the results remain the same (results not shown), meaning that there is unlikely to be an endogeneity problem.}

Since only infants born between 2000 and 2016 can be matched to their mother's location in my data, all newborns in my data (approximately 1 million in total) are born during this period. However, because I cannot observe ambient air pollution levels in all regions of Norway, my baseline sample contains only 46\% of these newborns. In Section \ref{sec:inter_des} I will show how the baseline sample is selected from the entire population. As a means of assessing the representativeness of my sample, Section \ref{sec:inter_des} also provides a statistical description of the population and the baseline sample.

\subsection{Sub-postcode unit (\textit{grunnkrets}) in Norway}
\label{subsec:grunn}

In Norway, there is a sub-postcode geographic unit, known in Norwegian as ``\textit{grunnkrets}'' , which means ``basic statistical unit''. These geographic units are delineated by Statistics Norway to facilitate statistical analysis. According to Statistic Norway, \textit{grunnkrets} are geographically cohesive and shall be as homogeneous as possible with respect to nature and economic base, communication conditions, and building structure.
These small, stable geographical units can serve as a flexible basis for the presentation of regional statistics.
\footnote{The definition of \textit{grunnkrets} is available on Statistic Norway's webpage: \href{https://www.ssb.no/a/metadata/conceptvariable/vardok/135/en}{\nolinkurl {https://www.ssb.no/a/metadata/conceptvariable/vardok/135/en}}
} 
On average, a ``\textit{grunnkrets}'' is around one-third the size of a postcode zone, and the entire country is divided into more than 14,000 \textit{grunnkrets}. 
\footnote{My baseline sample consists of 5,330 \textit{grunnkrets}, covering 38.6\% of Norway's area (118 municipalities and 1,455 postcode zones) and 46\% of the national population. As part of the robustness check, the regression in Appendix Table \ref{tab:rbst_dis} contains as many as 7853 \textit{grunnkrets} (56.8\% of the country's area).}
In the remainder of this paper, I refer to these basic statistical units as \textit{grunnkrets} directly.

\begin{figure}[htbp]
    \centering
    \includegraphics[width=0.8\textwidth]{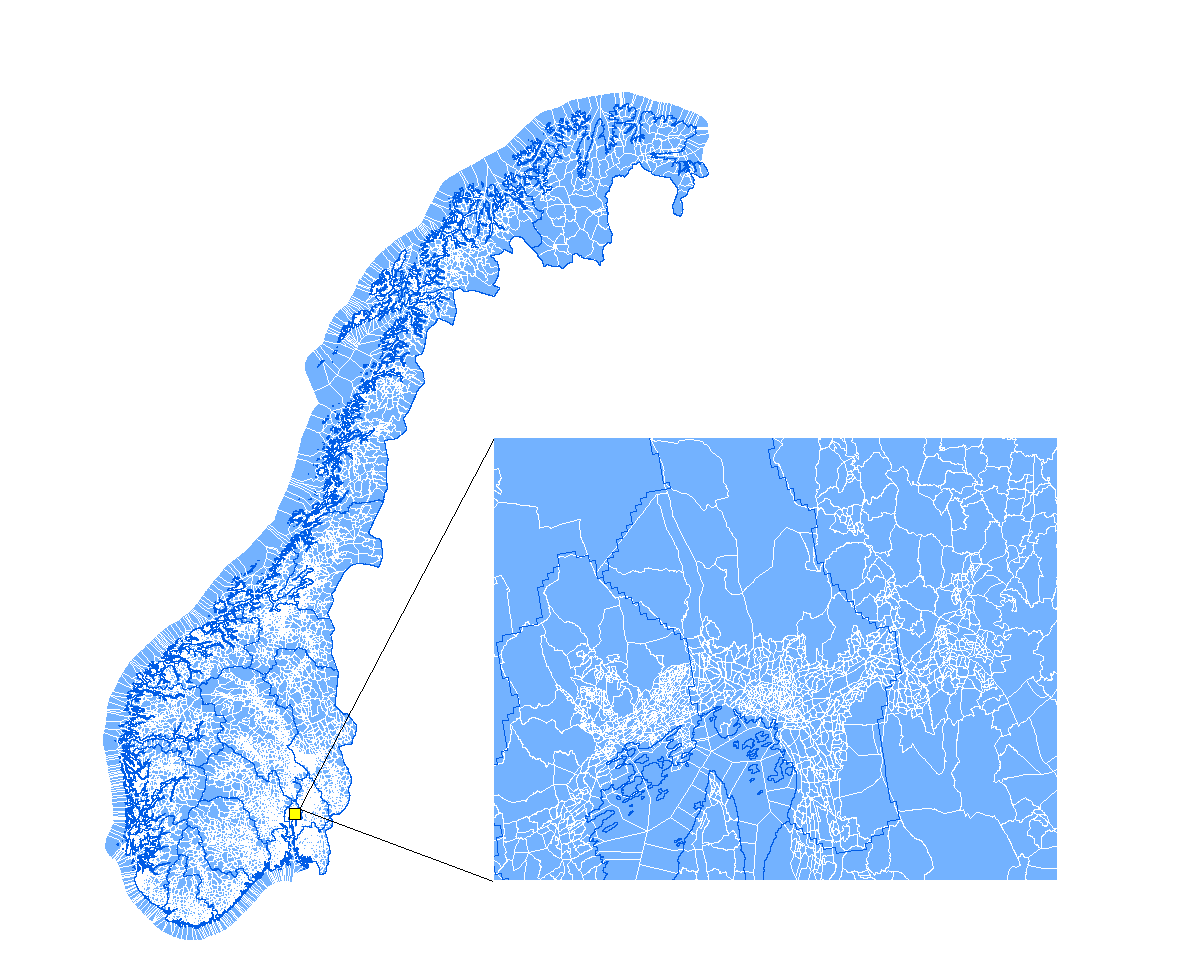}
    \caption{Sub-postcode unit \textit{Grunnkrets} in Norway}
    \label{fig:grunn}
\end{figure}

\begin{figure}[htbp]
    \centering
    \copyrightbox[b]{    \includegraphics[width=1\textwidth]{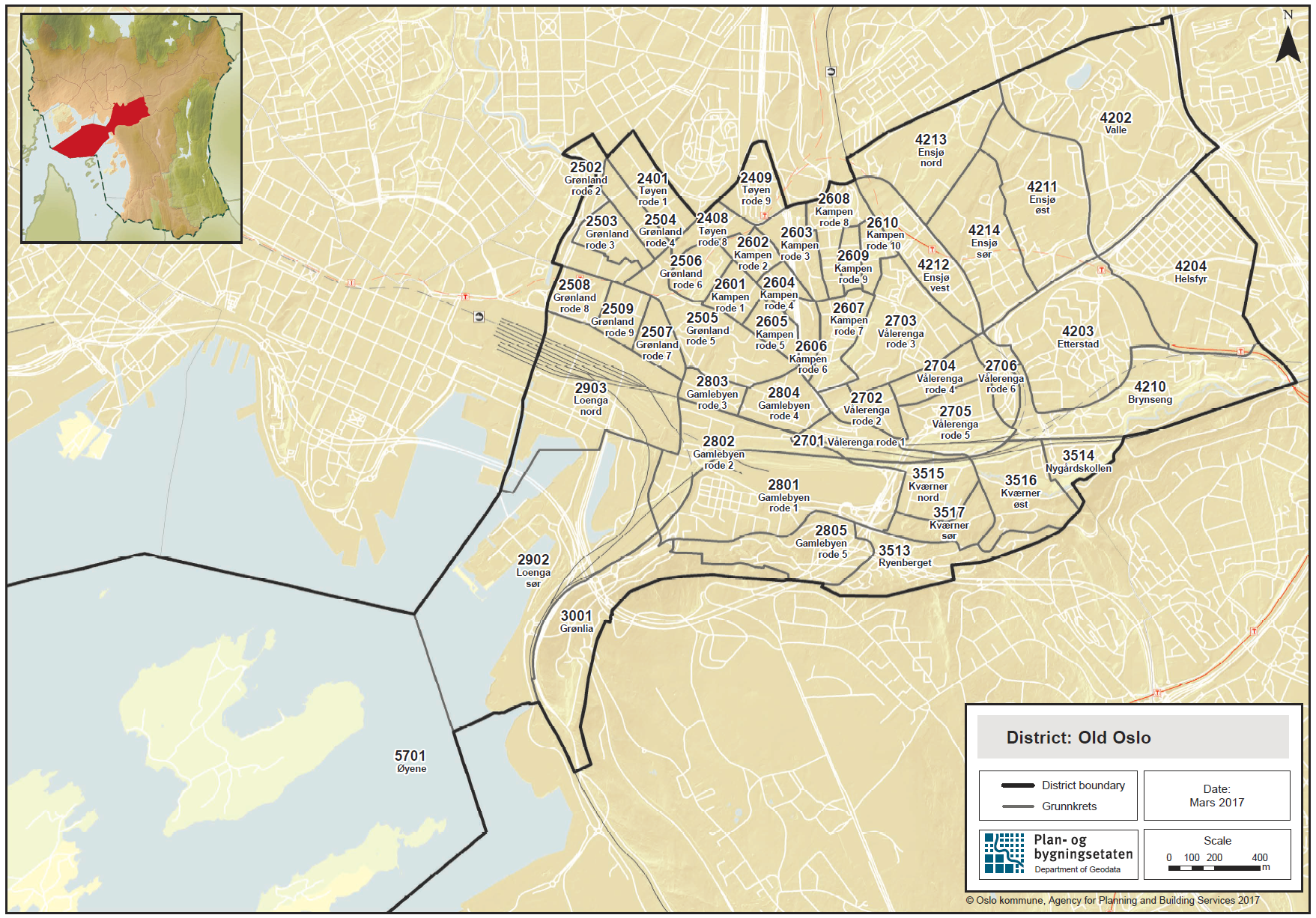}}{\scriptsize Figure obtained from Oslo municipality: \href{https://www.oslo.kommune.no/statistikk/geografiske-inndelinger/}{ https://www.oslo.kommune.no/statistikk/geografiske-inndelinger/}. }
    \caption{\textit{Grunnkrets} in Old Oslo district of Oslo city}
    \label{fig:grunn_2}
\end{figure}


Figure \ref{fig:grunn} displays a map of \textit{grunnkrets} (small blue polygons with white outlines) in Norway. As can be seen, a \textit{grunnkrets} is very small, and its size varies with population density.
\footnote{\textit{grunnkrets} are also very small in terms of population. By the end of 2016, there are on average 549 people per \textit{grunnkrets} in my baseline sample. The first three quartiles of the population per \textit{grunnkrets} are 196, 400 and 732. In fact, 99\% of \textit{grunnkrets} have fewer than 2,435 population, and these \textit{grunnkrets} cover a total of 97\% of country's population. This suggests that most people reside in small \textit{grunnkrets} (i.e., as shown in Figure \ref{fig:grunn}, the more densely populated the area, the smaller the \textit{grunnkrets}). 
Notably, the population of the most populous \textit{grunnkrets} in my sample increases from 3,455 in 1999 to 6,052 in 2016, while the average population per \textit{grunnkrets} only increases from 444 to 549 over the same period. This indicates that the population growth is uneven across \textit{grunnkrets}. The Norwegian population has gradually become more concentrated.} 
For example, in the less populated outskirts of Oslo, the capital city of Norway, \textit{grunnkrets} are larger than in the city center (see the zoomed-out part of Figure \ref{fig:grunn}). Figure \ref{fig:grunn_2} from the Oslo Municipality shows that there are about 50 \textit{grunnkrets} in the Old Oslo area (part of Oslo city center and seaside), ranging in size from about 0.04 $km^2$ to about 3 $km^2$. The largest \textit{grunnkrets} (number 5701) in Figure \ref{fig:grunn_2} contains several inhabited islands. 

As a geographic fixed effect, \textit{grunnkrets} are more effective in eliminating spatial endogeneity than zip-code zones, which are commonly employed in the literature \citep{salam2005birth,RN296}, because they are substantially smaller in size and are intentionally designed to be internally homogeneous by Statistic Norway. Individuals may be more inclined to select where to live within a postcode zone for unobservable reasons, but moving within a \textit{grunnkrets} is less meaningful. Compared with postcodes, it is more plausible to use infants born in the same area but at different times (and thus exposed to different levels of prenatal air pollution before birth) as counterfactuals to each other.  

\subsection{Ambient air pollution data}
\label{subsec:nilu}

\begin{figure}
    \centering
    \copyrightbox[b]
    {\begin{subfigure}[b]{0.48\linewidth}        
        \centering
        \includegraphics[width=\linewidth]{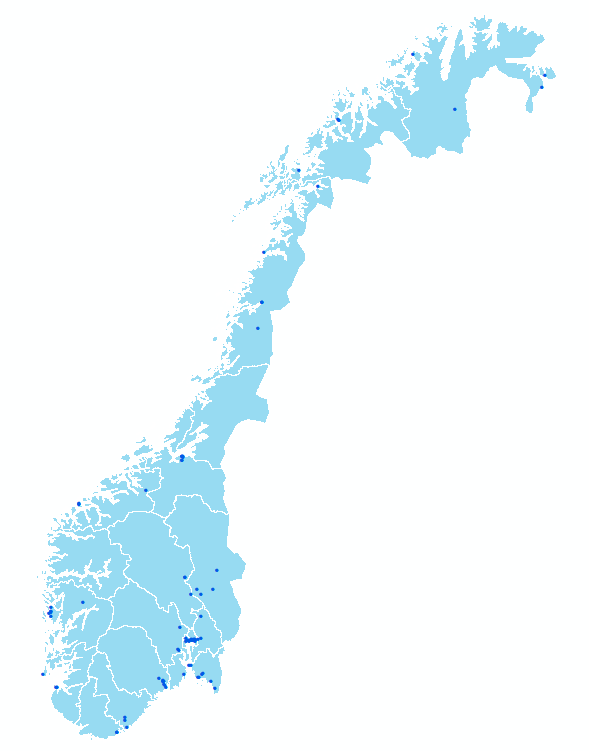}
        \caption{Air pollution monitoring stations}
        \label{fig:A}
    \end{subfigure}
    \begin{subfigure}[b]{0.495\linewidth}        
        \centering
        \includegraphics[width=\linewidth]{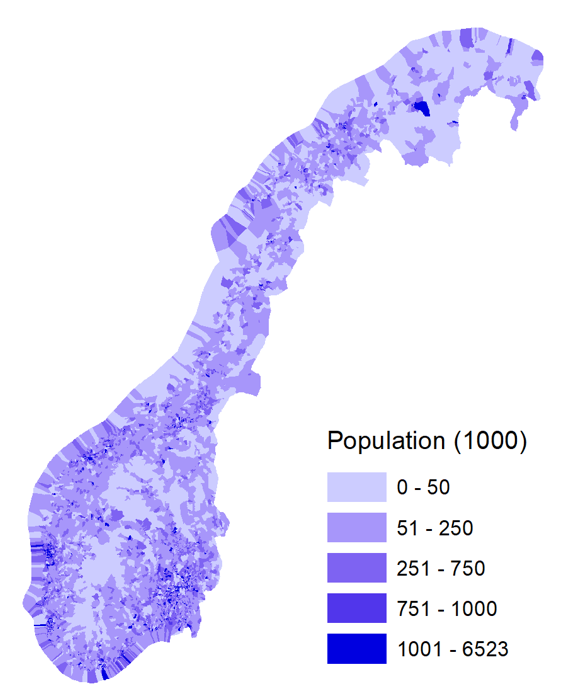}
        \caption{Population by \textit{Grunnkrets}}
        \label{fig:B}
    \end{subfigure}}{\scriptsize Notes: Population in 2022. Data from Geonorge, Norway's national website for map data: \url{https://register.geonorge.no}}
    \caption{Air pollution monitoring stations are located in areas with high population density}
    \label{fig:nilu}
\end{figure}

The ambient air pollution data is provided by \href{https://www.nilu.no/en/about-nilu/}{Norwegian Institute for Air Research (NILU)}, an independent, non-profit institution dedicated to the study of atmospheric composition, climate change, air quality, and environmental pollutants in Norway.
\footnote{The data collected by NILU is also an important source of data for \href{https://www.miljodirektoratet.no/}{Norwegian Environment Agency}.} 
During my study period, there are  in total 103 ambient air pollution monitoring stations in operation or previously in operation.
\footnote{Examples of the monitoring stations: \href{https://www.nilu.com/facility/nilus-observatories-and-monitoring-stations/}{https://www.nilu.com/facility/nilus-observatories-and-monitoring-stations/}.}
They are located in areas with high population density in Norway (excluding Svalbard). The location of the monitoring stations is depicted as dark blue dots in Part (a) of Figure \ref{fig:nilu}. Most of the monitoring stations are located along the Norwegian coastline, as the vast inland areas are mountainous and sparsely populated, as shown in Part (b) of Figure \ref{fig:nilu}. Due to the distribution of the monitoring stations, the values detected by the stations are mainly representative of pollution levels in urban areas. The pollution concentration data utilized in my study spans the years 1999 to 2016 to cover prenatal exposures of infants born between 2000 and 2016.
\footnote{In Section\ref{sec:inter_des}, I explain how I interpolate the air pollution concentrations at the \textit{grunnkrets} level based on the station-level panel data.}

The monitoring stations use a commercial Differential Optical Absorption Spectroscopy (DOAS) instrument (OPSIS AR500 analyzer) to measure the concentrations of ambient air pollutants such as $CO$, $NO$, $NO_2$, $O_3$, $SO_2$, $PM_{10}$, $PM_{2.5}$ and $PM_{1}$. 
The instrument performs well in detecting the aforementioned air pollutants \citep{xie2004intercomparison}.
\footnote{$NO$ ($O_3$) values detected by DOAS may be lower (higher) than those detected by various traditional point sampling techniques. This may be caused by the steep vertical gradient of the $NO-O_3-NO_2$ system (the $NO-O_3-NO_2$ reaction rapidly changes the components of the air after $NO$ is released). If this is the case, the value detected by DOAS should be closer to the actual pollutant concentration \citep{xie2004intercomparison}.} 
It should be noted that these monitoring stations are established (or closed) over time, and the types of pollutants that a station can detect can change over the study period. Therefore, the pollutant records generated by the monitoring stations are actually unbalanced panel data.
\footnote{The detection of $CO$ and $PM_1$ starts quite late and only covers a very small fraction of my sample. I therefore exclude the two pollutants from my study. Omitting the two ambient air pollutants should not affect my estimation, since the ambient $CO$ concentration is very low in Norway. As a type of $PM$ (particulate matters), $PM_1$ is highly correlated with other $PM$ such as $PM_{10}$, which is has been detected for many years.}

The ambient air pollutant data provided by NILU is daily averages. I further average the data into weekly average concentrations to make it easier to construct a \textit{grunnkrets}-time-specified panel dataset: given the large number of \textit{grunnkrets} and the length of pregnancy (about 40 weeks in total), the \textit{grunnkrets}-(calendar) day specified panel data is too large to process without difficulty. More importantly, my identification strategy (Section \ref{sec:identify}) relies on the variation of air pollution in a spatio-temporal unit. In a time interval as narrow as a calendar day, the variation of air pollution (and the sample size) are not sufficient to support my identification strategy.

\begin{table}[htbp]
\centering
\begin{threeparttable}
\caption{Station level weekly average ambient air pollution in Norway between 1999 and 2016 \label{tab:nilu} }
\begin{tabular}{l*{9}{c}}
\toprule
      &  \multicolumn{4}{c}{1999-2004} & &  \multicolumn{4}{c}{2005-2016}      \\  
            \cmidrule{2-5} \cmidrule{7-10}
Pollutant& mean& s.d.&  min&  max&     & mean& s.d.&  min&  max\\
\midrule
$NO$          & 58.23&       52.27&         0.00      &      369&  &    32.13&       35.05&        0.00&      629 \\
$NO_2$        &       38.39&       16.11&        2.55&      119  &   &       32.02&       19.21&        0.00&      241  \\
$NO_x$        &      126.79&       91.72&        0.00&      671  &   &       81.18&       69.96&        0.00&    1,178  \\
$PM_{10}$     &       25.81&       15.43&        6.56&      155  &   &       20.20&       11.22&        0.00&      135  \\
$PM_{2.5}$    &       13.33&        5.73&        3.72&       59  &   &        9.51&        4.88&        0.72&       88  \\
$O_3$         &       62.01&       16.18&        3.40&      119  &   &       56.23&       17.01&        0.00&      126  \\
$SO_2$        &        7.49&       10.38&        0.00&       75  &   &        8.83&       13.23&        0.00&      147  \\
\bottomrule\end{tabular}
\begin{tablenotes}
      \small
      \item Notes: (1) I separate the study period into two parts to highlight the high $NO$ concentration before 2005. (2) All pollutants are measured in $\mu g/m^3$. (3) Here $NO_x$ includes $NO$, $NO_2$ and other nitrogen oxides. (4) The raw data provided by NILU contains negative values for the concentrations. According to NILU, negative values between -5 and 0 can be treated as 0 and those below -5 (very rare) was wrongly recorded. I thereby replaced values between -5 and 0 with 0 and treat values less than -5 as omitted.
    \end{tablenotes}
\end{threeparttable}
\end{table} 
\begin{figure}[htbp]
    \centering
    \copyrightbox[b]{    \includegraphics[width=0.57\textwidth]{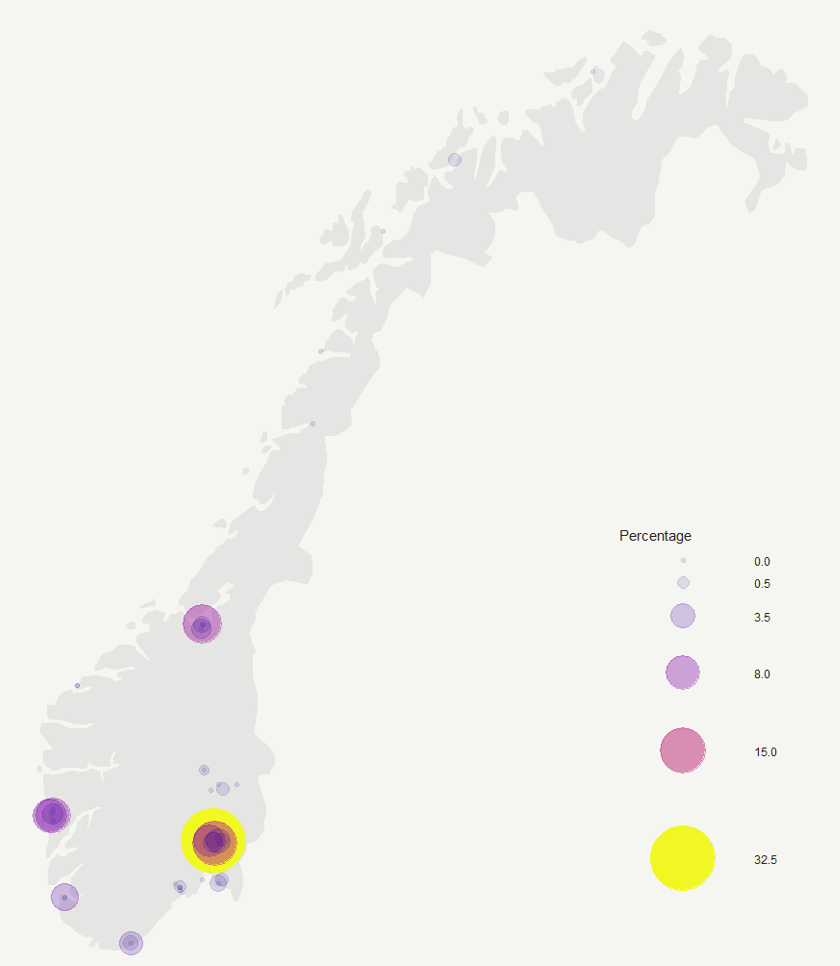}}{}
    \caption{Percentage of weeks with high ambient $NO$ concentration (> $110 \mu g / m^3$) at the station level between 1999 and 2016}
    \label{fig:high_station_bubble}
\end{figure}

Table \ref{tab:nilu} depicts the weekly average ambient air pollution concentrations in Norway from 1999 to 2016. Inspired by the graph \ref{fig:photo}, I divided the study period into two halves to emphasize the high $NO$ concentrations prior to 2005. As can be seen from the figure, the ambient air pollution levels in Norway are much lower than in many of the areas studied in the literature. According to the Table \ref{tab:standards} in the appendix, Norwegian air quality generally meets international standards. In addition, compared to air pollution levels prior to 2005, concentrations of all ambient air pollutants in Norway have decreased year by year, except $SO_2$, indicating that the Norwegian environment has been gradually improving since the new regulations came into effect in 2002.

A comparison of $NO$ and $NO_2$ concentrations before and after 2005 in Table \ref{tab:nilu} shows that the average $NO_2$ concentrations remained stable throughout the study period, while the average $NO$ concentrations before 2005 are much higher. Surprisingly, despite the significant decrease in average ambient air pollution levels over these years, the maximum weekly concentrations of $NO$ and $NO_2$ after 2005 can still reach twice the pre-2005 levels. This suggests that extreme $NO$ and $NO_2$ pollution events continued to occur after 2005. The high volatility of weekly ambient air pollution provides the conditions for determining the effects of air pollution on birth outcomes.

Figure \ref{fig:high_station_bubble} shows the percentage of weeks with high levels of $NO$ (95th percentile, or $110 \mu g / m^3$) at the monitoring station level between 1999 and 2016. Unsurprisingly, high levels of ambient $NO$ are common in urban areas. Weekly data collected at a number of monitoring stations in major cities such as Trondheim (middle bubble), Bergen, and Stavanger (two large bubbles on the west coast) show that $NO$ concentrations exceed $110 \mu g / m^3$ for about 15\% of weeks. In Oslo, the capital of Norway (yellow bubble), $NO$ concentrations exceeded $110 \mu g / m^3$ in 32\% of weeks between 1999 and 2016.

\subsection{Meteorological data}
 \label{subsec:met}

\begin{figure}[htbp]
    \centering
    \includegraphics[width=0.6\textwidth]{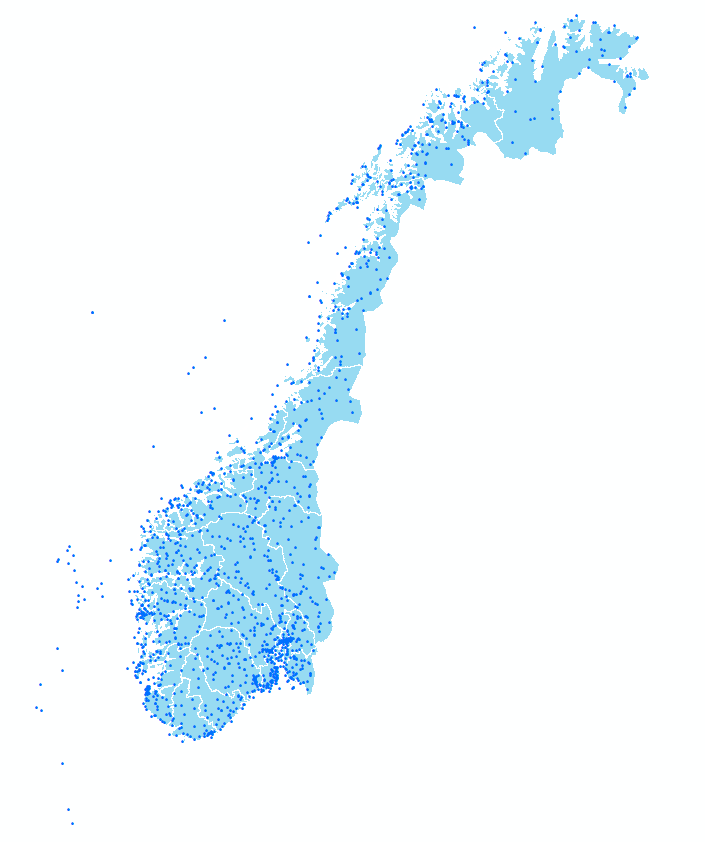}
    \caption{Meteorological detection stations in Norway}
    \label{fig:frost}
\end{figure}

My meteorological information is provided by \href{https://www.met.no/en/About-us/About-MET-Norway}{Norwegian Meteorological Institute (MET)}, the official weather forecasting institution that monitors Norway's climate and conducts research. Similar to NILU, MET owns weather detection stations across the country that record meteorological information such as temperature ($\degree{C}$), air pressure (hPa), moisture (\%), wind speed (m/s) and precipitation (mm). Once again, the high frequency meteorological data between 1999 and 2016 is averaged as weekly averages and will be interpolated at the \textit{grunnkrets} level (Subsection \ref{sec:inter_des}).

Figure \ref{fig:frost} illustrates that Norway has a total of 1,198 meteorological detection stations (not including Svalbard), which is many more than the number of air pollution monitoring stations. Moreover, most of the meteorological detection stations are established early, with some of them operating more than a century ago (although not all of them are in continuous operation). As a result, the spatial resolution of the meteorological data is considerably higher and more balanced than the data from the ambient air pollution panel data.

\section{Data interpolation and statistic description}
\label{sec:inter_des}
The above-mentioned ambient air pollution data and meteorological data are at the station level. To study the environment (\textit{grunnkrets}) where the pregnant women lived during the pregnancy, I need to interpolate the station-level data to the \textit{grunnkrets} level. This section describes the interpolation method and its performance. I use the same method to interpolate air pollution and meteorological conditions, but the challenge lies mainly in the interpolation of air pollution concentrations because there are not as many air pollution monitoring stations as there are meteorological monitoring stations. Therefore I focus on interpolation of air pollution concentrations in this section.

\subsection{Inverse Distance Weighting (IDW) interpolation}
\label{subsec:idw}

\begin{figure}[htbp]
    \centering
    \includegraphics[width=0.7\textwidth]{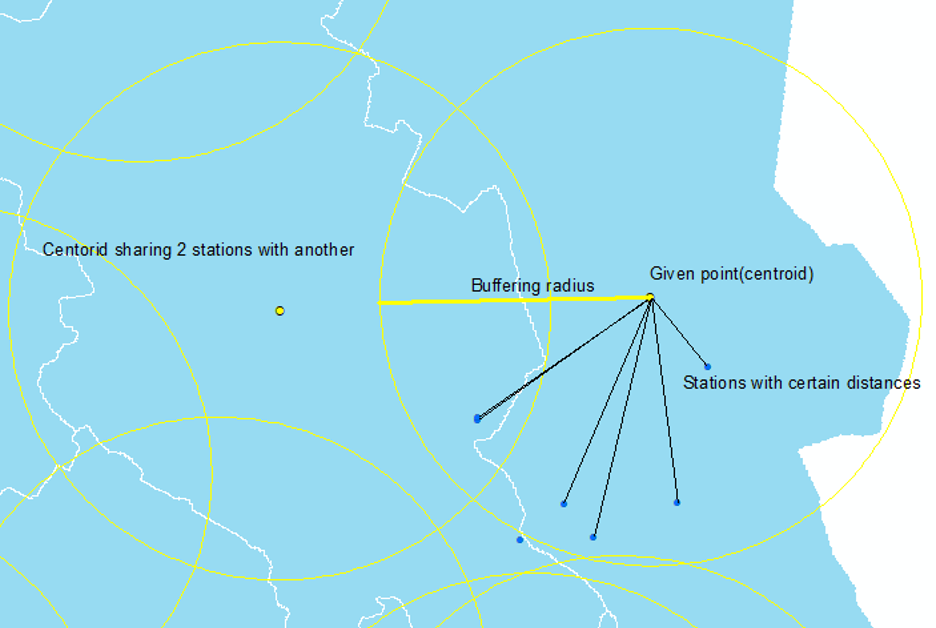}
    \caption{An example of the IDW interpolation method}
    \label{fig:idw}
\end{figure}


I use the Inverse Distance Weighting (IDW) method to interpolate the station-level pollution and meteorological data to the \textit{grunnkrets} level. As the name implies, the IDW method uses the inverse distance between \textit{grunnkrets} and the monitoring stations to weight the station-level data. Take air pollution as an example, the IDW method uses function (\ref{eq:idw}) to interpolate the ambient air pollution concentration in \textit{grunnkrets} $g$ at any time point $t$ based on the pollution concentration detected by the monitoring stations in the neighborhood of $g$ at time $t$.

\begin{equation}
Pollutio{{n}_{gt}}=\frac{\sum\limits_{i=s}^{n}{\frac{1}{d_{s}^{e}}}\times {{p}_{st}}}{\sum\limits_{i=1}^{n}{\frac{1}{d_{s}^{e}}}}
\label{eq:idw}
\end{equation}

where $n$ is the number of monitoring stations within a certain range (buffering radius) (e.g., 20 miles) around $g$. Any of these $n$ monitoring stations (station $s$) records the ambient air pollution concentration value $p_{st}$ detected at time $t$. The distance between station $s$ and \textit{grunnkrets} $g$ is $d_s$. The exponent $e$ is a power of the distance: the larger $p$, the higher the degree of weighting of the proximity monitoring station. In practice, I use the centroid of \textit{grunnkrets} to represent it.
\footnote{Note that there may be some textit{grunnkrets} that do not have monitoring stations nearby, in which case I cannot interpolate the air pollution levels for these textit{grunnkrets}. This is why my baseline sample does not cover the entire population. For \textit{grunnkrets} with only one nearby monitoring station, the interpolated concentrations are exactly the same as for the nearby (only) monitoring station.}
Figure \ref{fig:idw} visually illustrates the application of the IDW method on the map. The dark blue and bright yellow dots in Figure \ref{fig:idw} represent monitoring stations and certain \textit{grunnkrets} centroids. 
\footnote{The map in Figure \ref{fig:idw} is for illustrative purposes only. In fact, because \textit{grunnkrets} are very small and most air pollution monitoring stations are located in large cities, most \textit{grunnkrets} within these cities actually share the same monitoring stations if a 20-mile radius is used. In contrast, for many \textit{grunnkrets} in rural areas, there are no monitoring stations nearby at all.}

The Inverse Distance Weighting method is commonly applied in the literature and performs better than many other interpolation methods such as nearest neighbor, spatial averaging, and kriging method, especially when monitoring station density is relatively low \citep{jha2011evaluation,musashi2018comparison,wong2004comparison}, as this is for my ambient air pollution data.
\footnote{Note that ambient air pollution spreads in the ground atmosphere after emission, and the location of the monitoring station is not the source of the pollution. Therefore, I cannot use data from monitoring stations alone to model the dispersion of pollutants in the air. Instead, data from the monitoring stations are used to represent the exposure to pollution of residents living near the monitoring stations.}
The IDW method does not utilize the intrinsic characteristics of \textit{grunnkrets}, except for the spatio-temporal association with nearby monitoring stations. In contrast, there is also a large body of literature using land use regression (LUR) interpolation methods, which utilize data on elevation, traffic, population, and vegetation cover. The rich spatial and temporal fixed effects in the regressions of
Section \ref{sec:identify} can make the IDW approach comparable to, or even superior to, the LUR approach in the sense of a partitioned regression \citep{frisch1933partial}: If spatiotemporal fixed effects capture all features of the locations considered by the LUR method, then regressing the birth results on the LUR interpolated concentrations is equivalent to regressing the birth results on the IDW interpolated concentrations while controlling for spatiotemporal fixed effects. In simple terms, the latter is equivalent to the partitioned regression of the former.

Therefore, in my identification strategy, the coarseness of IDW interpolation compared to LUR interpolation depends mainly on the resolution of the controlled spatio-temporal fixed effects: when the fixed effects are at the \textit{grunnkrets}-(calendar) month level (i.e., the \textit{grunnkrets} and calendar-month indicators and the interaction of the two), the IDW method is not necessarily coarser than LUR interpolation because characteristics such as population and vegetation cover can be captured by fixed effects at the \textit{grunnkrets}-(calendar) month level. Another benefit of this method is that fixed effects can also capture unobservable features of a location that are ignored in the LUR method. Of course, this benefit comes at the risk of overfitting and thus requires a large sample size.

Furthermore, even though LUR provides detailed spatial resolution, it lacks temporal resolution because the information it relies on is mostly time-invariant. In contrast, the IDW interpolation method provides good temporal resolution, but the spatial resolution is limited by the number of monitors and their separation distances  \citep{marshall2008within}. It is important to note that the resolution of spatial-temporal fixed effects certainly cannot exceed the level of \textit{grunnkrets}-(calendar) weeks, as the prenatal pollution exposure in my study is also at such a level, otherwise, prenatal ambient air pollution exposure would be perfectly multicollineary with the fixed effects.

\subsection{Performance of IDW interpolation}
\label{subsec:idw_perform}

The IDW interpolation method weights the monitoring values within a certain range (radius) of the given points (centroids of $grunnkrets$).
If the radius is short, IDW interpolation relies only on detection stations close to \textit{grunnkrets}, so the interpolated values are closer to the true values; a long radius covering more stations may involve more measurement errors because some stations are too far from the \textit{grunnkrets} to precisely interpolate.  On the other hand, the radius also determines how many \textit{grunnkrets} (and thus observations) I can interpolate the air pollution concentration to, since there may not be any detection stations within a short range of a \textit{grunnkrets}. Estimates of the effect of air pollution on birth outcomes may be less precise in this case due to the lack of observations. In other words, the choice of radius thus involves a trade-off between interpolation (measurement) accuracy and identification accuracy (sample size).

I use a cross-validation strategy to test the performance of the IDW method at different radii and distance weights (exponent $e$ in Function \ref{eq:idw}). For a monitoring station $i$ that detects a pollution value $p^{real}_{it}$ in week $t$, I first interpolate the pollution value $p^{interp}_{it}$ at the same site where $i$ is located using the IDW method. The interpolation is based on other stations within a certain radius (except for station $i$ itself). I then compare the interpolated value $p^{interp}_{it}$ with the true value $p^{real}_{it}$ detected at station $i$. Intuitively, the correlation between the interpolated values and the detected true values shows how well the interpolation performs. Again, since I will use \textit{grunnkrets}-(calendar) month fixed effects in my identification strategy, cross-validation should take these fixed effects into account. Formally, I regress the actual value $p^{real}_{it}$ detected by the station on the interpolated value $p^{interp}_{it}$ (both at the weekly level), conditional on the station's location and the (calendar) month fixed effects (and the interaction of the two), as shown in Eq. \ref{eq:cv}.

\begin{equation}
p^{real}_{it} =   p^{interp}_{it} \cdot b + s_i+m_t+ s_i \cdot m_t + \epsilon_{it}
\label{eq:cv}
\end{equation}

Where $b$ is the regression coefficient of the interpolated value $p^{interp}$. $s_i$ and $m_t$ are the fixed effects of monitoring stations and (calendar) months, and $s_i \cdot m_t$ is the interaction effect. Both real and interpolated pollutant concentration data are at the calendar-week level (approximately 700 unique calendar-weeks). The $R^2$ value of the regression \ref{eq:cv} is the square of the correlation coefficient, which describes how well the interpolated and fixed effects work.

\begin{table}[htbp]
\centering

\begin{threeparttable}
\caption{Cross-validation $R^2$: how well the interpolation and fixed effects predict the real concentration (\%) }
 \label{tab:cv_p}

\begin{tabular}{c*{7}{c}}
\toprule
$e$&radius &   $NO$  & $NO_2$  &   $PM_{10}$ & $PM_{2.5}$  &   $O_3$  \\
\midrule
\multirow{5}{*}{0.1} &
10	&  86.9&        89.3&        79.8&        38.0&        - \\ 
&15	&  87.2&        90.1&        80.1&        38.3&        91.9 \\ 
&20	&  87.2&        90.1&        80.1&        38.4&        89.3 \\ 
&25	&  87.2&        90.1&        80.3&        38.4&        89.3 \\ 
&30	&  87.2&        90.1&        80.5&        38.5&        89.3 \\ 
\midrule
\multirow{5}{*}{1} &
10	&  86.7&        89.1&        79.8&        38.0&         - \\
&15	&  87.1&        89.9&        79.8&        38.3&        91.9 \\
&20	&  87.1&        90.0&        79.9&        38.3&        89.3 \\
&25	&  87.0&        90.0&        80.0&        38.3&        89.3 \\
&30	&  87.0&        90.0&        80.1&        38.4&        89.3 \\
\midrule
\multirow{5}{*}{2} &
10	&   86.3&        88.7&        79.5&        38.0&       - \\
&15	& 86.7&        89.6&        79.5&        38.2&        91.9 \\
&20	& 86.7&        89.6&        79.6&        38.2&        89.3 \\
&25	& 86.7&        89.7&        79.7&        38.2&        89.3 \\
&30	& 86.6&        89.7&        79.8&        38.3&        89.3 \\
\midrule
\multirow{5}{*}{5} &
10	&  85.3&        88.0&        78.9&        37.8&       -\\
&15	& 85.7&        89.0&        79.0&        38.0&        91.9\\
&20	& 85.7&        89.0&        79.0&        38.0&        89.3\\
&25	& 85.7&        89.1&        79.1&        38.1&        89.3\\
&30	& 85.7&        89.1&        79.3&        38.1&        89.3\\

\bottomrule
\end{tabular}
\begin{tablenotes}
      \scriptsize
      \item Notes: (1) As defined in Section \ref{subsec:idw}, $e$ is the power of distance as defined in equation; radius determines within which range the monitoring stations are used for interpolation. (2) Column 3-7 contains the $R^2$ of regression \ref{eq:cv} for different pollutants, which shows how well the real concentration is explained by the interpolation and fixed effects (3) There is no interpolation for $O_3$ with a radius of 10 miles because there is no station within such a range because $O_3$ is only monitored by a few stations. For the same reason, I didn't cross-validate $SO_2$ either.
\end{tablenotes}
\end{threeparttable}

\end{table} 

Table\ref{tab:cv_p} shows the regression $R^2$ values for different distance weights (index $e$ in Function \ref{eq:idw}) and radii for some major ambient air pollutants.
\footnote{Because the shortest distance between stations monitoring $O_3$ is greater than 10 miles, the cross-validation of $O_3$ is not applicable for a radius of 10 miles.}
With the exception of $PM_{2.5}$ in column (6), the IDW interpolation together with fixed effects predicts well the actual pollutant concentrations ($R^2\approx 90\%$). 

The power of the distance in column (1), $e$, is used as a penalty for the distance between the site and the location to be interpolated. The intuition is that although adding more sites provides more information, the interpolation may be distorted by sites that are too far away. According to Table \ref{tab:cv_p}, larger distance indices $e$ tend to slightly reduce the fit, suggesting that I should not penalize the addition of monitoring stations too much. In other words, adding more monitoring stations is relatively beneficial to improving the fit.

The radius in column (2) has little effect on the fit because most of the monitoring stations are located near large cities that are much farther apart than the radius itself. Therefore, increasing the radius from 10 miles to 40 miles may not include more monitoring stations. In addition, pollutant concentrations within a city can be highly correlated. Adding more stations within a city would not greatly improve interpolation.

Note that both the concentrations of air pollutants and spatial fixed effects are at the monitoring station level, so all spatial variation in pollutant concentrations is captured by fixed effects. The $R^2$ actually describes how well the interpolation predicts the temporal variation (within a calendar-month) of the actual pollutant concentration, or in other words, $R^2$ is the correlation index between the interpolated values and the actual concentrations after the fixed effects are excluded. My identification in the next section exploits the temporal variation in prenatal exposure to ambient air pollution in exactly the same way.

I also cross-validated the performance of the IDW method for interpolation of meteorological conditions. Because there are more meteorological monitoring stations in Norway, the interpolation is also more accurate ($R^2\ge 85\%$), as shown in Appendix Table \ref{tab:cv_m}. Based on the cross-validation results in Table \ref{tab:cv_p} and Appendix Table \ref{tab:cv_m}, I used 20 miles ($32 km$) as the base radius (which is also the same as \cite{RN296}) and $0.1$ as the default distance power, which means that the weighting is fairly uniform (i.e., close to the arithmetic mean). 
\footnote{In Section \ref{sec:robust} I also tried different radii and distance weighting indices $e$ to check the robustness of my identification strategy.}

However, the interpolation method always leads to some measurement error. Even in the cross-validation in Table \ref{tab:cv_p}, the interpolation and fixed effects do not perfectly ($R^2 = 100\%$) predict the actual concentration. 
\footnote{When it comes to identification in Section \ref{sec:identify}, since I also conditioned on the (interpolated) average meteorological conditions, which are related to the real pollutant concentrations, part of the measurement error can be taken into account.}
If the measurement error are random (classical), it would bias my estimates towards zero. However, the measurement error may not be random. Because air pollution monitoring stations are mainly located near major roads in large cities, where ambient air pollutant concentrations may be higher and more volatile, interpolation may overestimate concentration levels and volatility in areas relatively far from the monitoring stations. This is particularly true for $NO$, which is more likely to be oxidized in the ambient air after emission, as described in Section \ref{sec:review}. An overestimate of $NO$ also biases the estimate toward zero (or even have a protective effect) if people living relatively far from the main road are wealthier and have better (potential) birth outcomes.

\subsection{Data description and balance check}
\label{subsec:balance}

\begin{figure}[htbp]
    \centering
    \includegraphics[width=0.45\textwidth]{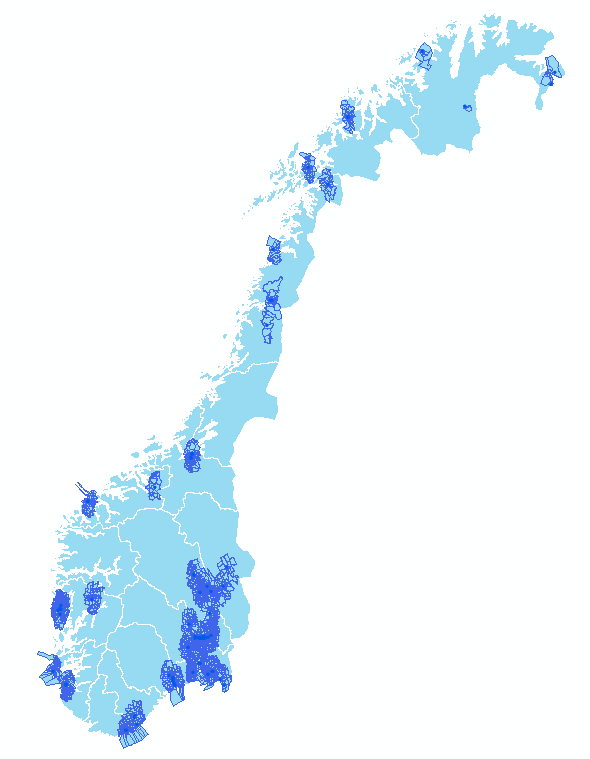}
    \caption{\textit{Grunnkrets} with at least one air pollution monitoring stations within 20 miles}
    \label{fig:cover}
\end{figure}

In this subsection, I compare my baseline sample (consisting of infants born in places where ambient air pollution can be interpolated, i.e., places with monitoring stations within 20 miles) with the rest  (54\%) of the population to assess the representativeness of my sample.

The map in Figure \ref{fig:cover} shows the distribution of $7,131$ (out of $14,016$) \textit{grunnkrets} within $20$ miles of at least one ambient air pollution monitoring station. Of these $7,131$ \textit{grunnkrets}, $91.5\%$ are actually within $15$ miles of the nearest ambient air pollution monitoring station, and $80\%$ are even within $10$ miles. The population of the $7,131$ \textit{grunnkrets} covers $67.8\%$ of all Norwegians (data from the end of 2017). As previously mentioned in Section \ref{sec:construct}, each ambient air pollution monitoring station detects only certain types of pollutants. In order to simultaneously observe (or interpolate) the main pollutants, such as $NO$, $NO_2$ and $PM$, only $5,330$ of the $7,131$ \textit{grunnkrets} could be utilized. Thus, my baseline sample represents only $46\%$ of all newborns during the study period. Weather monitoring station coverage is not an issue here because there are so many weather monitoring stations around Norway.

\begin{table}[htbp]
\centering
\caption{Balance check by interpolated $NO$ data availability \label{tab:balance_i_p}}

\begin{threeparttable}
\begin{tabular}{l*{7}{c}}
\toprule
            &  \multicolumn{3}{c}{\textbf{Baseline sample}} & &  \multicolumn{3}{c}{\textbf{Population uncovered}}      \\  
            \cmidrule{2-4} \cmidrule{6-8}
 Variable   &     mean&  s.d.&  Obs.& &     mean&  s.d.&  Obs.\\
\midrule

\multicolumn{2}{l}{\textbf{A. Infantile Info.}} \\

birth date &  2009&  4.42& 464  &&  2007&  4.97& 545 \\
gender   &         0.51&   0.50& 464   &&     0.51&   0.50& 545 \\
weight(g)   &   3,492& 591& 464   &&  3,521& 608& 544 \\
length(cm)   &    50&   2.71& 447   &&    50&   2.69& 526 \\
APGAR1&    8.7 &   1.2 &  464      &&    8.7 &  1.2 & 544 \\
APGAR5&    9.5 &  0.9  & 464     &&   9.4 &  0.9  & 544\\
\midrule

\multicolumn{2}{l}{\textbf{B. Parental Info.}} \\
parity &           1.59&   0.74& 464   &&     1.49&   0.72& 539 \\
age$_m$    &     31&      4.95& 464   &&     30&    5.22& 515 \\
edu$_m$     &     5.25&   1.55&  446  & &     4.79&     1.48&  520 \\
edu$_f$     &     5.03&   1.60&  435  & &     4.44&     1.41&  516 \\
native$_m$  &     0.88&   0.32&  464  & &     0.90&     0.30&  539 \\
native$_f$  &     0.88&   0.32&  453  & &     0.92&     0.28&  530 \\
img$_m$    &      0.70&   0.46&  464  & &     0.79&     0.41&  539 \\
img$_f$    &      0.71&   0.45&  453  & &     0.82&     0.39&  530 \\

income$_m$ &   234 &     342  &  449  & &   187&   124&  503 \\
income$_f$ &   331 &     900  &  438  & &   263&   576&  509 \\
wealth$_m$ &   484 &   3,411  &  449  & &   166& 1,353&  503 \\
wealth$_f$ &   671 &   3,793  &  438  & &   279& 1,449&  509 \\
debt$_m$   &   568 &     935  &  449  & &   308&   614&  503 \\
debt$_f$   &  1,094&   2,084  &  438  & &   785& 1,393&  509 \\
\midrule

\multicolumn{4}{l}{\textbf{C. Number of districts}} \\
municipality        & \multicolumn{3}{c}{118}    &&  \multicolumn{3}{c}{441}     \\
postcode            & \multicolumn{3}{c}{1,455}  &&  \multicolumn{3}{c}{3,054}     \\
\textit{grunnkrets} & \multicolumn{3}{c}{5,330}  &&  \multicolumn{3}{c}{13,207}      \\
\bottomrule

\end{tabular}

\begin{tablenotes}
      \small
      \item Notes: (1) Obs. is the number of observations in thousand. (2) Subscripts ``$_m$'' and ``$_f$'' denote mother and father of the newborn separately. (3) $parity$ means the number of children previously borne; Binary variable $native$ indicates Norwegian nationality; $img$ is ``immigration background'', 1 for person born in Norway to Norwegian parents, 0 for other cases; $edu$ is an ordered 0-8 categorical variable as defined by Statistics Norway: \href{https://www.ssb.no/klass/klassifikasjoner/36/}{https://www.ssb.no/klass/klassifikasjoner/36/}, e.g., edu=4 for upper secondary education. (4) (Gross) income, wealth and debt are registered 3 years before the delivery and in thousand Norwegian kroner (NOK) at current price.
    \end{tablenotes}
\end{threeparttable}
\end{table} 

Table \ref{tab:balance_i_p} compares the characteristics of the observations covered by the interpolation (baseline sample) with those of the remaining part of the population. According to Panel A of Table \ref{tab:balance_i_p}, the newborns in my baseline sample are, on average, very similar to the rest of the population, except for birth date and weight. The infants in the baseline sample were averagely born later, as the monitoring stations are established gradually over the study period. Infants in my baseline sample are also slightly lighter, probably because there are more immigrants in my sample, as Panel B shows. 

Mothers in the baseline sample have children later on average; a higher proportion of parents in the sample have higher education than the remaining 54\% of the population, and they are also wealthier and more likely to have an immigrant background or foreign nationality. Given that the interpolation covers most cities and more international areas, the parental characteristics in my sample are not particularly surprising. In other words, my baseline sample is more representative of the urban population in Norway.
\footnote{Even so, as described in Section \ref{sec:intro}, ambient air pollution levels in these urban areas are still low overall compared to air pollution levels in other countries, according to air quality guidelines.}
Therefore, the findings in my paper are not intended to be extrapolated to rural areas in Norway, but rather compared to other areas where ambient air pollution is at a comparable level.

\section{Identification strategy and model specification}
\label{sec:identify}

As mentioned earlier, prenatal ambient air pollution exposure is non-random and associated with a large number of observable or unobservable factors, such as parental characteristics, because families can decide where to live and when to have children. A simple comparison of fetuses exposed to low and high levels of pollution during the delivery period would be subject to omitted variable bias. The ideal solution would be to randomize prenatal exposure to ambient air pollutants, but this is clearly unrealistic. My identification method attempts to mimic this hypothetical experiment by using quasi-random variations in pollution exposure across time and space. Another difficulty in identification is the measurement error induced by IDW interpolation discussed in Section \ref{sec:inter_des}. 

With the National Registry data, I have sufficient power to apply rich spatio-temporal fixed effects in order to overcome both challenges to a large extent. Although I do not precisely interpolate pollution concentrations at each site, I focus only on the variation of air pollutants at a given site over a short period of time (a given month). In the case of small areas and narrow time intervals, precise self-selection of residence locations and delivery date by households is less likely to occur. Also, the abundance of temporal fixed effects improves estimation precision and compensates for the lack of accuracy of interpolation.

I use model \ref{eq:reg} to identify the effects of air pollution on birth outcomes in order to bypass the aforementioned problems of endogeneity and coarse interpolation.

\begin{equation}
\text{Outcome}_{i}=p_{i}\beta +w_i\gamma _1+X_i\gamma_2+ g_i+m_i+ g_i \times m_i+ \epsilon_i
\label{eq:reg}
\end{equation}

The dependent variables $\text{Outcome}_{i}$ in Equation \ref{eq:reg} are the birth weight, birth length, and APGAR scores of infant $i$.
The \textit{grunnkrets} where infant $i$'s mother lived in the year of delivery is known, and the variables $p_i$ and $w_i$ are the average interpolated concentrations of ambient air pollution and weather conditions in that \textit{grunnkrets} prior to the mother's delivery.
\footnote{Note that once the date of birth of baby $i$ is given, prenatal exposure to ambient air pollution is also known. Since birth outcomes are ``one-time'' rather than recurrent, i.e., there is no temporal variation in birth outcomes for individual $i$, my sample is actually pooled cross-sectional data rather than panel data. Therefore, I can omit the time subscripts in Equation \ref{eq:reg}.}
The pollutants studied in my baseline regressions include $NO$, $NO_2$, and $PM_{10}$. 
\footnote{The analysis of other pollutants such as $PM_{2.5}$, $SO_2$ and $O_3$ is included in the robustness test section (section \ref{sec:robust}) because there are fewer monitoring stations for these three pollutants and the samples are smaller.}
The controlled weather conditions are humidity, precipitation, barometric pressure, temperature, and wind level. Weather conditions are important to consider because they affect both birth outcomes and air pollution, as mentioned in Section \ref{sec:construct}.

I retraced the pregnancy based on the birth date of the newborn. Pregnancy usually lasts about $39$ weeks and is divided into three trimesters. Building on the literature, I focused on the third trimester, i.e., the $11$ weeks before delivery, which is considered critical for fetal development.
\footnote{I studied the average ambient air pollution and weather conditions for all three quarters in Appendix Table \ref{tab:3t} and found that only the last trimester has significant significant effects.}
It is also more practical to study only the last trimester because the true gestation period may not be precisely $39$ weeks. No matter how long the actual gestation period is, as long as it is longer than $11$ weeks, air pollution in the $11$ weeks before delivery is always what the mother is exposed to during the prenatal period.
\footnote{In extreme cases, pregnancy may even be shorter than $11$ weeks, and the weight of the stillbirth is also registered. I may thus wrongly specify the prenatal pollution exposure levels, but these cases are very rare.}
In addition, mothers are less likely to migrate during this time. By default, I assume that mothers live in the same place during the last trimester, as doctors do not recommend travel in the last weeks before delivery. 
\footnote{In the robustness check section, I will consider mothers moving between \textit{grunnkrets}.}

The vector $X_i$ represents the demographic and financial characteristics of the parents of newborn $i$ listed in Panel B of Table \ref{tab:balance_i_p}. Maternal age and parity are adjusted because they themselves directly affect birth outcomes, and more experienced mothers may be more aware of the effects of air pollution and thus choose lower prenatal exposures. Parents' economic status is adjusted, as wealthier parents may have better personal protection against air pollution, as well as better medical care and nutrition than other parents living in the same location, which resulted in better birth outcomes for their babies.

The terms $g_i$ and $m_i$ in the equation are the \textit{grunnkrets} and calendar-month fixed effects on birth outcomes for infant $i$ at birth, respectively, and $g_i \times m_i$ is the interaction term of these two fixed effects. Calendar-month means the month of a particular year. For example, January 2010 and January 2011 are two different calendar-months. The calendar-month fixed effect in Equation \ref{eq:reg} covers both annual and seasonal time trends. The interaction term $g_i \times m_i$ reflects the fact that certain spatial features have different effects on air pollution and birth outcomes at different times of a year.
For example, how the topography of a place affects ambient air pollutant concentrations may depend on seasonal variations in wind direction. 

In my baseline regression, there are approximately $4,000$ \textit{grunnkrets} and $200$ calendar-month indicators, but not all \textit{grunnkrets} have enough newborns in a given calendar-month to participate in regressions. Such \textit{grunnkrets}-calendar-month combinations without sufficient samples are called singletons. After excluding these singletons, there are about $10,000$ $g_i \times m_i$ combinations containing sufficient samples (about $300,000$ newborns in total). On average, in any given calendar-month, there are about $30$ births per \textit{grunnkrets}. 
\footnote{In the robustness check section, I apply more coarse spatio-temporal fixed effects, such as postal-code-(calendar) month levels, which cover more regions with smaller populations.}

Because the interpolated air pollution data is also at the \textit{grunnkrets} level, I implicitly assume that infants born in the same \textit{grunnkrets} are exposed to the same environment; after all, \textit{grunnkrets} is both small and homogeneous within it. This is particularly evident in densely populated areas, where a \textit{grunnkrets} can be so small as to encompass only a few blocks.
Once I condition on $g_i$, all spatial variations in air pollution concentrations and weather conditions are captured. Indeed, conditional on rich spatial-temporal fixed effects, the variation in prenatal exposure to ambient air pollution comes exclusively from different delivery weeks within a calendar-month. 
\footnote{The graph in Appendix Figure \ref{fig:baby} gives an example: Two infants were born in the same calendar quarter, but have different prenatal exposures to ambient $NO$ simply because they were born in different weeks of the calendar-quarter. The large amount of registry data provides me with sufficient power to use differences in prenatal exposure within a calendar-month to determine its impact on birth outcomes.}

The error term $\epsilon_i$ is allowed to correlate with infants whose mothers resided in the same \textit{grunnkrets} in the year of delivery. As a robustness check, I also allowed $\epsilon_i$ to be correlated at many different levels, including family (children of the same mother), zip-code, municipality, and the nearest monitoring station in Appendix Table \ref{tab:cluster}. In all these cases, the significance levels of the coefficients are very stable. 

My strategy relies on the conditional independence assumption (CIA), $E[p_i\perp\epsilon_i|w_i,X_i,g_i,m_i,g_i\times m_i]$, to identify the causal effect of air pollution on birth outcomes. That is, I hypothesized that after controlling for all covariates, infants would appear to be randomly exposed to different levels of ambient air pollution. Omitted factors (confounders) that affect pollution exposure $p_i$ and birth outcomes would violate the conditional independence assumption. Thanks to the rich spatio-temporal fixed effects, it is unlikely that individuals can manipulate the time and place of delivery (i.e., prenatal exposure of the baby) in such a small spatio-temporal space; nor are shocks like improvements in urban construction (new parks, hospitals, etc.) and deterioration of living conditions (new roads in the neighborhood) likely to exist briefly in such a small spatio-temporal unit without being captured.
\footnote{In robustness tests, I will show that, conditional on spatio-temporal fixed effects, prenatal air pollution exposure is effectively like a random assignment of the parental characteristics $X_i$ mentioned above to infants. That is, once spatio-temporal fixed effects are controlled for, there is no need to use covariates $X_i$ in the Equation \ref{eq:reg} to identify the effect of ambient air pollution on birth outcomes.}
Furthermore, because Norway has relatively little pollution compared to many developing countries, it is unlikely that there are other potential confounders, such as soil and water pollution, that happen to have the same variability as ambient air pollution.

However, it is important to note that if the choice of residence and timing of delivery are consequences of air pollution, then spatio-temporal fixed effects may be ``bad controls'' (i.e., covariates that are also caused by treatment) and may bias the estimated mean treatment effect. This may not be a problem because: (i) The main pollutant $NO$ in my study is colorless and not very visible to the public. (ii) The average treatment effect is a weighted average of the effects estimated in the specified units in each \textit{grunnkrets}-month. Thus, manipulations of residence and delivery time by different residents may cancel each other out. (iii) According to Appendix Table \ref{tab:time}, I find no indication of parental manipulation of delivery dates to avoid ambient air pollution. 
\footnote{I analyzed the characteristics of families who chose to give birth in different seasons and also did not find meaningful indigenous differences (results not shown). In the robustness testing section, I also discuss more about mothers moving in the year before delivery, which may be an indication of choice of residence.}

\section{Results}
\label{sec:result}

Based on regression model (\ref{eq:reg}), I estimated the effect of average ambient air pollution on birth weight in the third trimester of pregnancy. Regression results are presented in Table \ref{tab:bm_w}. Each regression in Table \ref{tab:bm_w} considers spatio-temporal fixed effects, $g,m$ and $g\times m$. Other independent variables are gradually added to the regressions to test the robustness of the model specification, and column (7) of Table \ref{tab:bm_w} is the baseline specification for the rest of this paper.

Columns (1) and (2) in Table \ref{tab:bm_w} include the mean $NO$ and $NO_2$ concentrations in the third trimester, respectively, as the only independent variables to avoid potential bad controls. In both regressions, $NO$ and $NO_2$ are negatively associated with birth weight, but only the coefficient of $NO$ is significant at the $5\%$ level of significance. Including both pollutants in column (3), the sign and significance level of the coefficient of $NO$ are unaffected; the coefficient of $NO_2$ changes sign, although it remains insignificant. It appears that prenatal exposure to $NO_2$ in the environment is not a confounder for $NO$.

\begin{table}[htbp]
\centering

\caption{The effect of ambient air pollution in the 3rd. trimester on birth weight \label{tab:bm_w} }

\begin{threeparttable}
\def\sym#1{\ifmmode^{#1}\else\(^{#1}\)\fi}

\begin{tabular}{l*{8}{c}}
\toprule
            &\multicolumn{1}{c}{(1)}&\multicolumn{1}{c}{(2)}&\multicolumn{1}{c}{(3)}&\multicolumn{1}{c}{(4)}&\multicolumn{1}{c}{(5)}&\multicolumn{1}{c}{(6)}&\multicolumn{1}{c}{(7)}\\
\midrule
$NO$        &      -0.728\sym{**} &                     &      -1.098\sym{**} &      -1.361\sym{**} &      -1.409\sym{**} &      -1.386\sym{**} &     -1.387\sym{**} \\
            &     (0.330)         &                     &     (0.515)         &     (0.582)         &     (0.585)         &     (0.590)         &    (0.611)         \\
\addlinespace
$NO_2$       &                     &      -0.823         &       1.121         &       1.080         &       0.929         &       0.100         &     -0.259         \\
            &                     &     (0.804)         &     (1.251)         &     (1.600)         &     (1.636)         &     (1.697)         &    (1.762)         \\
\addlinespace
$PM_{10}$      &                     &                     &                     &                     &       0.513         &       0.752         &      1.329         \\
            &                     &                     &                     &                     &     (1.392)         &     (1.448)         &    (1.489)         \\
\addlinespace    
parity      &                     &                     &                     &                     &                     &      71.698\sym{***}&       75.928\sym{***}\\
            &                     &                     &                     &                     &                     &     (2.304)         &    (2.428)         \\
\addlinespace    
age$_m$     &                     &                     &                     &                     &                     &       0.379         &     -0.357         \\
            &                     &                     &                     &                     &                     &     (0.370)         &    (0.392)         \\
\addlinespace    
edu$_m$        &                     &                     &                     &                     &                     &                  &         8.740\sym{***}\\
            &                     &                     &                     &                     &                     &                     &    (1.361)         \\
\addlinespace    
edu$_f$        &                     &                     &                     &                     &                     &                     &      6.678\sym{***}\\
            &                     &                     &                     &                     &                     &                     &    (1.255)         \\
\addlinespace            
weather     &     no         &      no    &               no   &               yes   &               yes   &     yes             &     yes   \\
parental    &     no         &      no    &               no   &               no    &               no    &     yes             &     yes   \\
\midrule
$r^2$       &       0.441         &       0.441         &       0.441         &       0.435         &       0.435         &       0.456         &       0.464         \\
Obs.        &     292,349         &     293,526         &     292,343         &     274,334         &     273,112         &     241,913         &     225,239         \\

\bottomrule
\end{tabular}
\begin{tablenotes}
      \small
      \item note: (1) $weather$ includes humidity, precipitation, air pressure, temperature and wind; $parental$ consists of parental economic conditions, immigration background and nationality. (2) \textit{grunnkrets} and month fixed effect (main and interaction) are controlled for in all the regressions. (3) Cluster robust standard errors at \textit{grunnkrets} level in parentheses, (4) *** $p < 0.01$, ** $p<0.05$, * $p<0.1$. (5) A radius of 20 miles and a distance power of 0.1 were used as default for pollution value interpolation. (6) Pollutants in $\mu g/m^3$, birth-weight in gram.
    \end{tablenotes}
\end{threeparttable}
\end{table} 

As discussed in Section \ref{sec:review} and Section \ref{sec:identify}, meteorological conditions affect ambient air pollutants and birth outcomes. Therefore, I control for the average meteorological conditions in the last trimester in columns (4)-(7) of Table \ref{tab:bm_w}. In column (5) I include the average concentration of another pollutant $PM_{10}$ in the last trimester before birth. The magnitude of the coefficient on $NO$ in columns (3)-(5) increases with the addition of more covariates, and remains significant at the $5\%$ level, while the other two pollutants, $NO_2$ and $PM_{10}$, have no significant effect. 
\footnote{I control for the other types of pollutants in the robustness check section and find that the inclusion of these additional controls has little effect on the coefficient of $NO$.}

In columns (6) and (7) of Table \ref{tab:bm_w}, I further include the parental characteristics introduced in Table \ref{tab:balance_i_p} in regressions. As mentioned in Section \ref{sec:construct}, to avoid endogeneity, the parents' financial status is registered three years before the year of birth. However, due to data limitations, the parents' education level may be registered after the birth and thus endogenous. Therefore, I include the parents' education separately in column (7). As expected, maternal parity and parental education level are positively associated with birth weight, but conditioning on these characteristics has no effect on the coefficient of $NO$. This supports the identification hypothesis that, given rich spatio-temporal fixed effects, prenatal air pollution exposure behaves as if it were randomly assigned to the infant. More on the manifestation of fixed effects will be discussed in the robustness checks section.

Since the coefficients on $NO$ in Tables (3)-(7) are very stable and significant at the $5\%$ level, I conclude that a $1 \mu g /m^3$ increase in mean environmental $NO$ concentration in the third trimester reduces birth weight by approximately $1.4 g$ (approximately  $1/6 \sim  1/5$ of the coefficient on parental education level).
\footnote{$edu$ is an ordered categorical variable taking values from $0$ to $8$, where $0$ indicates no education and $8$ indicates postgraduate education, as defined by SSB: \href{https://www.ssb.no/klass/klassifikasjoner/36/}{https://www.ssb.no/klass/klassifikasjoner/36/}. One additional one unit increase in $edu$ can be interpreted as an increase in education level, which is arguably more important than an additional year of education}.
For each standard deviation increase ($25.43 \mu g/m^3$) in the average ambient $NO$ concentration in the third trimester, birth weight decreases by $35g$, or $1\%$ of the average birth weight in my sample ($3500g$). The effect of $NO$ found in my study is similar in magnitude to that of other pollutants studied in the literature (Subsection \ref{subsec:birth_air}). The average concentrations of the other two pollutants $NO_2$ and $PM_{10}$ in the third trimester have no significant effect on birth weight, indicating that they are at safe concentration levels for newborns in Norway.

Based on these findings, $NO$ may pose a greater threat to newborns in Norway than other ambient air pollutants, especially in large cities such as Oslo and Bergen. In recent years, the quarterly average ambient $NO$ values in Norway have typically been $60 \mu g / m^3$ in winter. If the adverse effect of environmental $NO$ pollution on birth weight is linear, then winter $NO$ pollution may contribute to a birth weight loss of $84g$ for this group of infants on average, or $2.4\%$ of the average birth weight in Norway. In Bergen, Norway's second largest city, monthly $NO$ pollution levels can be as high as $120 \mu g /m^3$ (2019) and even reach $275 \mu g /m^3$ (2010) in some months in ``Danmarksplass'' (around the city center), which may cause even more birth weight loss.

\begin{table}[htbp]
\centering
\caption{The effect of ambient air pollution in the 3rd. trimester on birth length \label{tab:bm_l}}

\begin{threeparttable}

\def\sym#1{\ifmmode^{#1}\else\(^{#1}\)\fi}
\begin{tabular}{l*{7}{c}}
\toprule
            &\multicolumn{1}{c}{(1)}&\multicolumn{1}{c}{(2)}&\multicolumn{1}{c}{(3)}&\multicolumn{1}{c}{(4)}&\multicolumn{1}{c}{(5)}&\multicolumn{1}{c}{(6)}&\multicolumn{1}{c}{(7)}\\
\midrule
$NO$        &      -0.037\sym{**} &                     &      -0.044\sym{*}  &      -0.054\sym{*}  &      -0.053\sym{*}  &      -0.055\sym{*}  &      -0.052\sym{*}  \\
            &     (0.016)         &                     &     (0.025)         &     (0.028)         &     (0.028)         &     (0.029)         &     (0.029)         \\
\addlinespace
$NO_2$       &                     &      -0.054         &       0.022         &       0.029         &       0.033         &       0.017         &      -0.023         \\
            &                     &     (0.038)         &     (0.058)         &     (0.074)         &     (0.074)         &     (0.077)         &     (0.080)         \\
\addlinespace
$PM_{10}$      &                     &                     &                     &                     &      -0.045         &      -0.037         &      -0.004         \\
            &                     &                     &                     &                     &     (0.071)         &     (0.075)         &     (0.077)         \\
\addlinespace
parity      &                     &                     &                     &                     &                     &       1.588\sym{***}&         1.769\sym{***}\\
            &                     &                     &                     &                     &                     &     (0.113)         &     (0.119)         \\
\addlinespace
age$_m$     &                     &                     &                     &                     &                     &       0.051\sym{***}&       0.005         \\
            &                     &                     &                     &                     &                     &     (0.018)         &     (0.019)         \\
\addlinespace
edu$_m$        &                     &                     &                     &                     &                     &                     &       0.523\sym{***}\\
            &                     &                     &                     &                     &                     &                     &     (0.065)         \\
\addlinespace
edu$_f$        &                     &                     &                     &                     &                     &                     &       0.292\sym{***}\\
            &                     &                     &                     &                     &                     &                     &     (0.057)         \\
\addlinespace            
weather     &     No              &               No    &               No   &               Yes   &               Yes   &     Yes             &     Yes   \\
parental      &     No              &               No    &               No    &               No    &               No   &     Yes             &     Yes   \\

\midrule
$r^2$       &       0.445         &       0.444         &       0.445         &       0.439         &       0.438         &       0.454         &        0.461         \\
Obs.        &     276,584         &     277,752         &     276,578         &     259,813         &     258,679         &     228,890         &      212,938         \\
\bottomrule
\end{tabular}
\begin{tablenotes}
      \small
      \item Notes: (1) $weather$ includes humidity, precipitation, air pressure, temperature and wind; $parental$ consists of parental economic conditions, immigration background and nationality. (2) \textit{grunnkrets} and month fixed effect (main and interaction) are controlled for in all the regressions. (3) Cluster robust standard errors at \textit{grunnkrets} level in parentheses, (4) *** $p < 0.01$, ** $p<0.05$, * $p<0.1$. (5) A radius of 20 miles and a distance power of 0.1 were used as default for pollution value interpolation.  (6) Pollutants in $\mu g/m^3$, birth length in millimeter.
    \end{tablenotes}
\end{threeparttable}

\end{table} 

The effect of air pollution on birth length has similar patterns, as shown in Table \ref{tab:bm_l}. In columns (4)-(7) of Table \ref{tab:bm_l}, the coefficient of $NO$ is stable, hovering around $-0.052 mm$. Although the coefficients of all three pollutants are insignificant at the $5\%$ level, the coefficient of $NO$ is significant at the $10\%$ level, while the coefficients of the other two pollutants are far from significant ($t\text{-statistic}<0.5$). The coefficients on NO indicate that during the third trimester of pregnancy, every $1$ $\mu g / m^3$ increase in the ambient $NO$ concentration results in a birth length reduction of $0.052mm$ (about $1/10 \sim 1/6$ of the coefficients on parental education level). One standard deviation increase of ambient $NO$ concentration in the third trimester would reduce birth length by $1.4 mm$, which is $0.3 \%$ of the mean birth length ($500 mm$) in the baseline sample. This effect is comparable to the association between $NO_2$ and birth length found in the literature (Subsection \ref{subsec:birth_air}).

I also examined the effect of ambient air pollution during the last three trimesters on infant APGAR1 and APGAR5 scores, but did not find any significant effects (results not shown). This may be due to the small variation in APGAR scores in Norway, which is described in Subsection \ref{subsec:mfr_data}. In conclusion, I find that prenatal exposure to environmental $NO$ in the third trimester reduced birth weight and birth length, whereas prenatal exposure to ambient $NO_2$ and $PM_{10}$ are at safe levels for Norwegian newborns.

\section{Robustness check}
\label{sec:robust}

In this section, I first evaluate the sensitivity of my identification strategy to IDW interpolation, which affects both estimation and statistical inference (as it affects sample size). I then indirectly test the conditional independence assumptions underlying my identification strategy by testing for spatio-temporal fixed effects and other potential confounders. Finally, I discuss the case of mothers moving pre/post-natally, which may lead to measurement error and make spatial fixed effects a ``bad control'' \citep{angrist2009mostly}.

\subsection{Sensitivity to IDW interpolation}
\label{subsec:rbst_inter_error}

The choice of the radius for the IDW interpolation method is a trade-off between the accuracy of the interpolation itself and the precision of the estimate (because it affects sample size), as discussed in Section \ref{sec:inter_des}. The shorter the distance between the location to be interpolated and the monitoring station, the closer the interpolated value will be to the actual value detected by the station; on the other hand, the shorter the distance, the fewer the monitoring stations around the location, and therefore the fewer the locations (observations) that can be covered by the interpolation.
\begin{table}[htbp]
\centering
\caption{Benchmark regression with different IDW radius \label{tab:rbst_dis}}

\begin{threeparttable}
\def\sym#1{\ifmmode^{#1}\else\(^{#1}\)\fi}
\begin{tabular}{l*{7}{c}}
\toprule
         &\multicolumn{1}{c}{(1)}&\multicolumn{1}{c}{(2)}&\multicolumn{1}{c}{(3)}&\multicolumn{1}{c}{(4)}&\multicolumn{1}{c}{(5)}&\multicolumn{1}{c}{(6)}&\multicolumn{1}{c}{(7)}\\

radius: &\multicolumn{1}{c}{10}&\multicolumn{1}{c}{15}&\multicolumn{1}{c}{baseline}&\multicolumn{1}{c}{25}&\multicolumn{1}{c}{30}&\multicolumn{1}{c}{35}&\multicolumn{1}{c}{40}\\

\midrule
\multicolumn{7}{l}{A. Birth weight} \\

$NO$        &      -1.263\sym{*}  &      -1.348\sym{**} &      -1.387\sym{**} &      -1.358\sym{**} &      -1.036\sym{*}  &      -0.891         &      -0.687         \\
            &     (0.691)         &     (0.647)         &     (0.611)         &     (0.595)         &     (0.569)         &     (0.558)         &     (0.550)         \\
$NO_2$       &       1.914         &       0.136         &      -0.259         &      -0.288         &      -1.233         &      -1.187         &      -1.225         \\
            &     (2.002)         &     (1.889)         &     (1.762)         &     (1.692)         &     (1.591)         &     (1.530)         &     (1.515)         \\
$PM_{10}$      &       1.460         &       1.486         &       1.329         &       2.660\sym{*}  &       2.284         &       2.389\sym{*}  &       2.627\sym{*}  \\
            &     (1.721)         &     (1.581)         &     (1.489)         &     (1.452)         &     (1.413)         &     (1.383)         &     (1.361)         \\
\midrule
$r^2$       &       0.448         &       0.453         &       0.464         &       0.469         &       0.474         &       0.479         &       0.483         \\
Obs.        &     167,713         &     200,554         &     225,239         &     244,918         &     261,907         &     275,778         &     290,303         \\

\midrule
\multicolumn{7}{l}{B. Birth length} \\

$NO$        &      -0.031         &      -0.047         &      -0.052\sym{*}  &      -0.063\sym{**} &      -0.044         &      -0.035         &      -0.029         \\
            &     (0.034)         &     (0.031)         &     (0.029)         &     (0.028)         &     (0.027)         &     (0.026)         &     (0.026)         \\
$NO_2$       &       0.031         &      -0.031         &      -0.023         &      -0.020         &      -0.054         &      -0.050         &      -0.045         \\
            &     (0.092)         &     (0.086)         &     (0.080)         &     (0.076)         &     (0.072)         &     (0.069)         &     (0.068)         \\
$PM_{10}$      &      -0.006         &       0.007         &      -0.004         &       0.023         &       0.012         &       0.015         &       0.023         \\
            &     (0.090)         &     (0.083)         &     (0.077)         &     (0.074)         &     (0.072)         &     (0.070)         &     (0.069)         \\
\midrule
$r^2$       &       0.446         &       0.450         &       0.461         &       0.468         &       0.472         &       0.476         &       0.480         \\
Obs.        &     159,599         &     190,484         &     212,938         &     231,167         &     246,758         &     259,498         &     273,088         \\
\bottomrule

\end{tabular}

\begin{tablenotes}
      \small
      \item Notes: (1) IDW radius is defined in Section \ref{sec:inter_des} radius in miles. (2) All regressions are based on the benchmark model (column (7) of Table \ref{tab:bm_w} and Table \ref{tab:bm_l}). (3) Cluster robust standard errors at \textit{grunnkrets} level in parentheses. (4) *** $p < 0.01$, ** $p<0.05$, * $p<0.1$. (5) Pollutants in $\mu g/m^3$, birth-weight in gram, birth length in millimeter.
    \end{tablenotes}
\end{threeparttable}
\end{table} 

In Table \ref{tab:rbst_dis}, I try radii from $10$ to $40$ miles to check how sensitive my estimates are to the choice of radius. Panels A and B of Table \ref{tab:rbst_dis} have birth weight and birth length as explanatory variables, respectively. The regression results in column (4) of Table \ref{tab:rbst_dis} are the same as the baseline regression results (column (7) of Table \ref{tab:bm_w} and Table \ref{tab:bm_l}), i.e., both have a radius of $20$ miles. For simplicity, I report only the coefficients for the three pollutants in Table \ref{tab:rbst_dis}. We can see that the effect of $NO$ on birth weight is fairly stable and remains significant at the $5\%$ level when the radius is between $15$ and $25$ miles (columns (2)-(5)). The coefficient of the effect of $NO$ on birth length is about $-0.05 mm$ when the radius is between $15$ and $25$ miles, but is significant at the $5\%$ level only when the radius is $25$ miles. The significant effect of prenatal $NO$ exposure disappears in both Panel A and Panel B when the radius exceeds $30$ miles. This is because interpolation tends to overestimate the true pollutant fluctuations as the radius increases, as discussed in Section \ref{sec:inter_des}. In summary, identification is not sensitive to interpolation when the radius is between $15$ and $25$ miles.

\subsection{Validity of conditional independence}
\label{subsec:rbst_fixed}

My identification relies on controlling for a rich set of spatio-temporal fixed effects. As described in chapter \ref{sec:identify}, to claim that the coefficients on pollutants are their causal effects on birth outcomes, prenatal air pollution exposure in the last trimester of pregnancy should be randomly assigned conditional on fixed effects and other covariates, i.e., conditionally independent. The assumption of conditional independence is more convincing when the resolution of spatial and temporal fixed effects is higher, because in this case there is less room for household self-selection of place of residence and time of delivery. On the other hand, however, fewer babies are born in the areas and time intervals determined by the high-resolution fixed effects. Therefore, the estimates are less precise. The choice of the resolution of the spatio-temporal fixed effects involves a trade-off between the bias and precision of the estimates.

\begin{table}[htbp]
\centering
\begin{threeparttable}
\caption{Regression of $NO$ concentration in the 3rd. trimester on parental characteristics and F-test \label{tab:ftest} }

\def\sym#1{\ifmmode^{#1}\else\(^{#1}\)\fi}

\begin{tabular}{l*{7}{c}}
\toprule    
            &\multicolumn{1}{c}{(1)}&\multicolumn{1}{c}{(2)}&\multicolumn{1}{c}{(3)}&\multicolumn{1}{c}{(4)}&\multicolumn{1}{c}{(5)}&\multicolumn{1}{c}{(6)}\\
            &\multicolumn{1}{c}{no FE}&\multicolumn{1}{c}{TWFE}&\multicolumn{1}{c}{$p$-$q$}&\multicolumn{1}{c}{$p$-$m$}&\multicolumn{1}{c}{$g$-$q$}&\multicolumn{1}{c}{baseline}\\
\midrule
\multicolumn{5}{l}{A. F-test: $NO$ as dependent variable} \\

$r^2$       &       0.581         &       0.848         &      0.948           &       0.986         &       0.960         &       0.991         \\
Obs.        &     371,646         &     371,531         &    365,104          &     326,562          &     320,343         &     226,455        \\
$F(14, df_r)$ &     115.95\sym{***} &       2.78\sym{***}  &   2.52\sym{***}   &     1.33          &      1.70\sym{**}  &      1.46    \\
$df_r$      &       4,264         &       4,149         &       1,185          &      1,161         &      3,784         &       3,527         \\

\midrule
\multicolumn{5}{l}{B. Birth weight} \\

$NO$        &       0.573\sym{***}&      -0.437\sym{***}&      -0.839\sym{***}&      -0.927\sym{**} &      -0.786\sym{***}&      -1.387\sym{**} \\
            &     (0.085)         &     (0.141)         &     (0.210)         &     (0.452)         &     (0.247)         &     (0.611)         \\
$NO_2$       &      -1.616\sym{***}&      -0.133         &       0.101         &      -0.270         &      -0.118         &      -0.259         \\
            &     (0.186)         &     (0.280)         &     (0.489)         &     (1.185)         &     (0.577)         &     (1.762)         \\
$PM_{10}$      &       1.975\sym{***}&       0.462         &       1.752\sym{***}&       2.137\sym{*}  &       1.308\sym{**} &       1.329         \\
            &     (0.225)         &     (0.311)         &     (0.521)         &     (1.132)         &     (0.632)         &     (1.489)         \\

$r^2$       &       0.015         &       0.034         &       0.172         &       0.324         &       0.315         &       0.464         \\
Obs.        &     369,208         &     369,092         &     362,715         &     324,271         &     318,365         &     225,239         \\

\midrule
\multicolumn{5}{l}{C. Birth length} \\

$NO$        &       0.030\sym{***}&      -0.003         &      -0.024\sym{**} &      -0.023         &      -0.029\sym{**} &      -0.052\sym{*}  \\
            &     (0.004)         &     (0.006)         &     (0.010)         &     (0.021)         &     (0.012)         &     (0.029)         \\
$NO_2$       &      -0.000         &      -0.013         &      -0.016         &      -0.094\sym{*}  &      -0.003         &      -0.023         \\
            &     (0.009)         &     (0.012)         &     (0.022)         &     (0.053)         &     (0.026)         &     (0.080)         \\
$PM_{10}$      &       0.000         &      -0.014         &       0.040         &       0.065         &       0.030         &      -0.004         \\
            &     (0.010)         &     (0.014)         &     (0.025)         &     (0.054)         &     (0.028)         &     (0.077)         \\

$r^2$       &       0.009         &       0.032         &       0.174         &       0.324         &       0.315         &       0.461         \\
Obs.        &     355,425         &     355,311         &     348,653         &     309,678         &     304,128         &     212,938         \\

\bottomrule
\end{tabular}
\begin{tablenotes}
      \small
      \item Notes: (1) Panel A is derived from regressions of average ambient $NO$ concentration in the third trimester on parental characteristics and weather conditions with different sets of spacial-temporal fixed effects: ``no FE'' conditioned on no fixed effects; ``TWFE'' means calendar month and \textit{grunnkrets} are controlled for. In column 3-6, $q$ and $m$ represent calendar quarter and calendar month fixed effects; $g$ and $p$ mean \textit{grunnkrets} and post-zone fixed effects (both main effect and the interaction are controlled for). The F statistic is to test the joint significance of the 14 parental characteristic variables, where $df_r$ is the residual degrees of freedom. The independent variables in Panel B and Panel C are birth weight and birth length separately. (2) income and wealth in million Norwegian kroner at current price.  (3) Cluster robust standard errors at \textit{grunnkrets} level in parentheses in column 1,2,5 and 6; the  standard errors are cluster at post-zone level in column 3-4. (4) *** $p < 0.01$, ** $p<0.05$, * $p<0.1$.
    \end{tablenotes}
\end{threeparttable}
\end{table}
 

Although I cannot test the conditional independence hypothesis directly, I can infer its validity indirectly from the characteristics of the parents. In Table \ref{tab:ftest}, Panel A, I regress mean ambient $NO$ concentrations in the last trimester on the $14$ observable parental characteristics and weather conditions used in the baseline regression. Also, I control for spatio-temporal fixed effects at different resolutions in these regressions. 
The spatio-temporal fixed effects used in the first two columns  of Table \ref{tab:ftest} are: (1) ``no FE'' for no fixed effects; (2) ``TWFE'' for calendar-month and \textit{grunnkrets} fixed effects (without interaction term); In columns (3)-(6) of  Table \ref{tab:ftest} , the main effect as well as the interaction of post-zone $p$ (or \textit{grunnkrets} $g$) and calendar quarter $q$ (or calendar-month $m$) are controlled for. 
If prenatal exposure to ambient air pollution is randomly assigned conditional on these fixed effects and weather conditions, then the parental characteristics should not be jointly significant. Therefore, I tested the joint significance of the $14$ parental characteristics after regression.

According to Table \ref{tab:ftest}, once the fixed effects of postcode-(calendar)month or \textit{grunnkrets}-(calendar)month are controlled for (columns (4) and (6)), the $F$-statistic for Panel A becomes insignificant at the $10\%$ level. This means that we cannot reject the hypothesis that the $14$ parental characteristics are jointly independent of prenatal $NO$ exposure at that level of significance. In other words, in terms of parental characteristics, prenatal exposure to $NO$ appears to be randomly assigned to infants when postcode-(calendar) months or \textit{grunnkrets}-(calendar) months are given, rather than chosen by the parents themselves. In contrast, in Table \ref{tab:ftest}, columns (1)-(3) and (5), the $F$ statistic is significant at the $0.1\%$ and $5\%$ levels, implying that prenatal $NO$ exposure is associated with the characteristics of the infant's parents under the condition of lower resolution of fixed effects, when the conditional independence assumption does not seem to hold.
\footnote{The extraordinarily high $R^2$ values when the interaction term $m\times g$ and weather conditions are controlled for in column (6) are unsurprising given that with the \textit{grunnkrets}-(calendar) month fixed effects conditioned on, the only variation left in prenatal $NO$ exposure originates from the variation of birth date within a month. I thus need a large sample to provide enough power for a precise estimation.}

While the zip-code-(calendar) month and \textit{grunnkrets}-(calendar) month fixed effects perform fairly well in Table \ref{tab:ftest} Panel A, the former regression has more observations than the latter. This is because in a given calendar-month, a zip-code area is more likely to have enough observations for estimation, whereas in the smaller \textit{grunnkrets}, there may be too few births to estimate (i.e., more singletons). Such singletons that do not participate in the regression are more likely to occur due to the low population density in rural areas. Thus, when controlling for \textit{grunnkrets}-(calendar)month fixed effects, the observations represent more urban areas. In contrast, when controlling for postcode-(calendar)month fixed effects, more rural samples are able to participate in the regression due to fewer singletons. In my study, unless otherwise stated, I reserve the use of \textit{grunnkrets}-(calendar) month fixed effects in order to avoid as much as possible the selection of unobservables, especially in an area with many newborns in the zip-code area.

In panels B and C of Table \ref{tab:ftest}, I regress birth outcomes on the covariates in the baseline regression using different levels of fixed effects. By comparing columns (4) and (6) with the other columns, we can see that the effect of $NO$ is greatly underestimated when the resolution of the fixed effects is relatively low. This happens for a number of reasons, including the choice of living location and delivery date, or the random measurement error of the IDW interpolation is accentuated at coarser fixed effects, as explained in Section \ref{sec:inter_des}.
\footnote{Similar to the measurement bias of the Section \ref{sec:inter_des} difference increasing with radius, in the case of more rural samples participating in the regression, I may have underestimated the $NO$ coefficient by overestimating the pollution fluctuations in rural areas.}
It is worth noting that the effect of environmental $NO$ is stronger and more pronounced in column (6) of Table \ref{tab:ftest} than in column (4), which may be due to the random measurement error described above, but there is another possible explanation: the effect of ambient air pollution may be heterogeneous when the samples are different. As mentioned earlier, the sample in participating in the regression in column (6) is more representative of infants born in urban areas with higher levels of pollution. The marginal effect of air pollution on this group of infants may be larger.
\footnote{In Section \ref{sec:hetero}, I find that a one-unit increase in ambient $NO$ has a greater adverse effect on birth outcomes when pollution levels are higher.}

\subsection{Potential confounders}
\label{subsec:rbst_confound}

Ambient air pollutants are usually produced simultaneously by traffic and industrial production. As a result, the concentrations of these pollutants are closely correlated. In addition, the complex interactions between ambient air pollutants make it challenging to disentangle their respective effects on birth outcomes. Therefore, I further introduced three other pollutants, $PM_{2.5}$, $O_3$, and $SO_2$, in my regressions to check whether my baseline regressions are influenced by these omitted variables. 
\footnote{As suggested by Subsection \ref{subsec:concen}, it is of particular interest to examine how the correlation between $O_3$, $NO$ and $NO_2$ affects the birth results. In addition, in Appendix Table \ref{tab:ozone}, I also tried to include only these three pollutants mentioned above in the baseline regression model, and I find that only $NO$ have a significant effect on birth weight and birth length. }
In Table \ref{tab:other}, panels A and B examine the effects of environmental pollutants on birth weight and birth length, respectively. For comparison purposes, I replicate the baseline regression results (column (7) of Table \ref{tab:bm_w} and Table \ref{tab:bm_l}) in the first column of Table \ref{tab:other}. I include the average concentrations of $PM_{2.5}$, $O_3$, and $SO_2$ for the third trimester in columns (2) to (4) of Table \ref{tab:other} in the baseline regression, respectively.
In column (5), I control for $O_3$ and $SO_2$ concentrations. The concentration of $PM_{2.5}$ is added last to column (6) due to the poor performance of the interpolation shown in Table \ref{tab:cv_p}.

In Table \ref{tab:other}, columns (3)-(6), the estimation results are very imprecise because the monitoring values of $O_3$ and $SO_2$ concentrations are few, and their inclusion in the regression makes the sample size of the regression shrink dramatically. Column (3) is relatively less problematic, although the sample size is still cut in half. The magnitude and sign of the coefficients of $NO$ are similar to the results of the baseline regression. More severe sample size decreases occur in columns (4)-(6), which have only a quarter of the sample size of the baseline regression, so I lack the power to accurately identify the effect. However, the sign of the coefficient on $NO$ is still negative in these columns.


\begin{table}[htbp]
\centering

\begin{threeparttable}
\caption{The effects of parental diabetes history and other ambient air pollutants in the 3rd. trimester on birth outcomes \label{tab:other} }
\def\sym#1{\ifmmode^{#1}\else\(^{#1}\)\fi}

\begin{tabular}{l*{7}{c}}
\toprule
            &\multicolumn{1}{c}{(1)}&\multicolumn{1}{c}{(2)}&\multicolumn{1}{c}{(3)}&\multicolumn{1}{c}{(4)}&\multicolumn{1}{c}{(5)} &\multicolumn{1}{c}{(6)}&\multicolumn{1}{c}{(7)} \\
            &\multicolumn{1}{c}{baseline}&\multicolumn{1}{c}{$PM_{2.5}$}&\multicolumn{1}{c}{$O_3$}&\multicolumn{1}{c}{$SO_2$}&\multicolumn{1}{c}{$SO_3 \& O_2$} &\multicolumn{1}{c}{all}&\multicolumn{1}{c}{diabetes} \\

\midrule
\multicolumn{6}{l}{A. Birth weight} \\

$NO$        &     -1.387\sym{**}  &  -1.199\sym{*}  &      -1.289         &      -2.211         &      -1.488         &      -0.958       &     -1.541\sym{**}      \\
            &    (0.611)          & (0.671)         &     (1.135)         &     (1.910)         &     (2.058)         &     (2.247)       &    (0.612)              \\

$NO_2$       &     -0.259         &    0.520         &       0.917         &       5.324         &       4.857         &       4.881      &      -0.220              \\
            &    (1.762)          & (1.906)         &     (2.808)         &     (4.982)         &     (5.285)         &     (5.400)       &    (1.772)              \\

$PM_{10}$    &        1.329       &      2.622         &       5.110\sym{*}  &       4.670         &       6.174         &       8.079    &         0.942              \\
            &    (1.489)          & (1.666)         &     (2.880)         &     (4.313)         &     (4.625)         &     (5.221)       &    (1.506)              \\

$PM_{2.5}$  &                      &        0.589         &                     &                     &                     &      -7.799    &                            \\
            &                     &    (2.117)         &                     &                     &                     &    (12.976)       &                        \\

$O_3$       &                      &                    &       0.017         &                     &       1.184         &       1.809      &                      \\
            &                     &                    &     (1.530)         &                     &     (2.780)         &     (3.039)       &                     \\

$SO_2$      &                      &                    &                     &     -12.456         &     -18.437         &     -20.247      &                      \\
            &                     &                    &                     &    (12.271)         &    (24.294)         &    (38.584)       &                     \\

\midrule
$r^2$       &         0.464         &      0.461         &       0.451         &       0.435         &       0.431         &       0.431       &      0.458           \\
Obs.        &       225,239         &    203,845         &     115,434         &      55,829         &      52,194         &      51,985       &    239,561           \\

\midrule
\multicolumn{6}{l}{B.Birth length} \\

$NO$        &     -0.052\sym{*}    &   -0.041         &      -0.023         &      -0.089         &      -0.031         &       0.012          &    -0.053\sym{*}      \\
            &    (0.029)           &  (0.032)         &     (0.052)         &     (0.089)         &     (0.096)         &     (0.106)          &   (0.029)             \\

$NO_2$       &     -0.023          &     0.027         &       0.027         &       0.284         &       0.197         &       0.217         &     -0.032             \\
            &    (0.080)           &  (0.086)         &     (0.129)         &     (0.218)         &     (0.236)         &     (0.242)          &   (0.081)             \\

$PM_{10}$      &     -0.004        &       0.093         &       0.119         &       0.095         &       0.211         &       0.367       &       -0.008             \\
            &    (0.077)           &  (0.087)         &     (0.143)         &     (0.209)         &     (0.229)         &     (0.255)          &   (0.078)             \\

$PM_{2.5}$   &                         &       -0.106         &                     &                     &                     &      -0.669       &                           \\
             &                      &      (0.095)         &                     &                     &                     &     (0.619)          &                        \\

$O_3$        &                       &                      &       0.047         &                     &       0.219         &       0.276\sym{*}  &                         \\
             &                      &                      &     (0.073)         &                     &     (0.138)         &     (0.148)          &                        \\

$SO_2$       &                       &                      &                     &      -0.145         &      -0.648         &      -0.856         &                         \\
             &                      &                      &                     &     (0.591)         &     (1.054)         &     (1.780)          &                        \\

\midrule
$r^2$       &      0.461           &      0.459         &       0.452         &       0.440         &       0.437         &       0.437        &      0.462         \\
Obs.        &    212,938           &    192,552         &     108,828         &      51,966         &      48,580         &      48,395        &    211,382         \\
\bottomrule
\end{tabular}

\begin{tablenotes}
      \small
      \item Notes: (1) Regressions are based on the benchmark model and include other types of pollutants in column 1-6 and parental diabetes history in column 7. (2) The independent variables in Panel A and Panel B are birth weight and birth length separately. (3) Cluster robust standard errors at \textit{grunnkrets} level in parentheses. (4) *** $p < 0.01$, ** $p<0.05$, * $p<0.1$. (5) All pollutants are in $\mu g/m^3$, birth-weight in gram, birth length in millimeter.
    \end{tablenotes}
\end{threeparttable}

\end{table}


The coefficient of $NO_2$ in column (4) of Table \ref{tab:other} Panel B is even significantly positive, probably because there is a negative correlation between $NO$ and $NO_2$ in the process of reaching the photochemical equilibrium state in the samples of this group. Because $NO_2$ is toxic anyway, this negative coefficient is not justified. And when the sample size is large enough, as in other columns, this negative correlation disappears, and the effect of $NO_2$ is not significant. Notably, according to Table \ref{tab:other}, prenatal exposure to environmental $SO_2$ appears to have a strong (but not significant due to small sample size) adverse effect on birth weight and length; after all, environmental $SO_2$ is clearly a health issue in the few areas where it is monitored.

In addition to other pollutants, genetic diversity may be one of the potential confounding factors. According to the literature, air pollution has been associated with diabetes. Also, parental diabetes has been shown to have an effect on birth weight. Thus, parental diabetes is also a potential missed confounder. This phenomenon is known as genetic pleiotropy. If the only reason for the association of air pollution with birth outcomes is this genetic pleiotropy, $NO$ would have an impact on birth outcomes once the parental history of diabetes is taken into account. Thanks to the detailed registration data, I can additionally control for parental diabetes history in column (7) of Table \ref{tab:other}. Here, parental diabetes history is a binary variable, indicating parental type I and type II diabetes. As we see in column (7), controlling for parental diabetes history does not affect the coefficients of $NO$ in panels A and B. Therefore, I conclude that parental diabetes is not a channel through which $NO$ affects birth weight and birth length.

In summary, based on the regression results in Table \ref{tab:other}, I do not find other pollutants such as $PM_{2.5}$, $O_3$, or $SO_2$ as confounders of $NO$. Furthermore, in addition to $NO$, prenatal exposure to $SO_2$ seems to have a negative effect on birth outcomes, whereas in the Norwegian environment, $PM$ and $O_3$ are at safe levels for newborns. 
I also find that parental diabetes is not a mechanism by which $NO$ concentration affects birth weight and birth length.
It is worth mentioning that designating $NO$ and $NO_2$ as the same pollutant, i.e., $NO_x$, may underestimate the negative health effects of $NO$, since it is $NO$ and not $NO_2$ that has adverse effects on the fetus, and the two pollutants are positively correlated, as mentioned in the literature review section.
\footnote{I tried to identify the effect of $NO_x$ on birth outcomes. The coefficients for $NO_x$ are$-0.771g$ for birth weight and $-0.03 mm$ for birth length (both significant at the $5\%$ level, regression results not shown in the table), both of which are approximately half the coefficients in the baseline regression.}

\subsection{Mothers relocating prior to childbirth}
\label{subsec:move}

Mothers moving between \textit{grunnkrets} may cause serious measurement errors. If mothers move to a new area after childbirth, but within the same year, the address of mothers registered at the end of the year (\textit{grunnkrets}) is not where they lived in the last trimester, that is, I incorrectly located where mothers lived before childbirth and also incorrectly measured the level of prenatal ambient air pollution exposure. On the other hand, moving before pregnancy may make the spatial fixation effect a ``bad control'' because pregnant women may move to avoid ambient air pollution.

In my data, I can observe in which \textit{grunnkrets} (annual data) the mother lived and in which municipality the baby was born between 2000 and 2016. In the baseline analysis, I assume that the mother stays in the same place for the last trimester (and that the reason for choosing this place to live is not due to ambient air pollution), but in reality, it is common for mothers to move between \textit{grunnkrets}. To investigate the above, I further divided the baseline sample into four groups based on the location of the mother and the place of birth of the infant, as shown in Table \ref{tab:mlocation}:

\begin{table}[htbp]
\centering

\caption{Sub-groups by maternal location in the year of childbirth  \label{tab:mlocation}}
\begin{threeparttable}

\begin{tabular}{lc|c|c|}

                              &  \multicolumn{1}{c}{}   & \multicolumn{2}{c}{mothers delivered in another municipality}     \\
                              & \multicolumn{1}{c}{}    & \multicolumn{1}{c}{yes}        &       \multicolumn{1}{c}{no}   \\ \cline{3-4} 
                              &  \multirow{2}{*}{yes}   & Group (1) move after delivery        &    Group  (2) move before delivery    \\       
mothers lived in another  &                         &          4.2\%               &            10.3\%                \\      \cline{3-4}
\textit{grunnkrets} last year &  \multirow{2}{*}{no}    & Group (3) no move, far hospital &   Group  (4) no move, local hospital  \\     
                              &                         &          24.9\%              &            60.6\%                \\      \cline{3-4} 

\end{tabular}

\begin{tablenotes}
      \small
      \item Notes: (1) Infants born between 2001-2016 are separated into 4 sub-groups (1)-(4) in the table. Those who were born in 2000 are not included because maternal location in 1999 are not traceable in my data. (2) The percentage following is the proportion each group accounts for. (3) Mothers who moved before delivery but chose to give birth in a different city also belongs to group (1), since I cannot distinguish them based on the data. (4) Because the mother's location is updated annually, in groups (3) and (4), there may be mothers who move several times during the year and return to their original \textit{grunnkrets} at the end of the year, although this is unlikely to happen in the last trimester before delivery.
    \end{tablenotes}
\end{threeparttable}
\end{table}
 

\begin{itemize}
    \item Group (1) moved immediately after delivery: The baby was born in a municipality other than the one where the mother lived in the year of delivery and the mother lived in two different \textit{grunnkrets} in the year of delivery and the year before;
    \item Group (2) moved before delivery: The baby was born in the same municipality where the mother lived in the year of delivery, but the mother lived in two different places in the year of delivery and the year before; 
    \item Group (3) delivered in hospitals far away: The mother did not move in the year of delivery, but the infant is born in another municipality; 
    \item Group (4) did not move: The mother did not move in the year of delivery, and the infant is also born in the same municipality where the mother lived.
\end{itemize}

Prenatal air pollution exposure for infants in group (1) may not have been measured correctly because the mothers likely moved after delivery (in the year of delivery). If so, the mothers' registered address (\textit{grunnkrets}) is not where they lived during the last trimester of pregnancy. Group 2 does not challenge my measurements as long as most mothers in the group moved during the first two trimesters rather than the last one. Mothers in group (3) gave birth in a different municipality than where they lived, but they did not move at all. This could have been self-selected or due to a lack of capacity at the local hospital. If the hospital is self-selected, there would be endogeneity problems. Mothers in group (4) lived in the same place in the year of delivery and before delivery and gave birth in the municipality where they lived, which is closer to the description of those who live stably in Norway.


\begin{table}[htbp]
\centering
\caption{Robustness check: mothers moving between \textit{grunnkrets}  \label{tab:nomove}}

\begin{threeparttable}
\def\sym#1{\ifmmode^{#1}\else\(^{#1}\)\fi}

\begin{tabular}{l*{5}{c}}
\toprule
            &\multicolumn{1}{c}{(1)}&\multicolumn{1}{c}{(2)}&\multicolumn{1}{c}{(3)}&\multicolumn{1}{c}{(4)}&\multicolumn{1}{c}{(5)} \\
           &\multicolumn{1}{c}{Group 1}&\multicolumn{1}{c}{Group 2}&\multicolumn{1}{c}{Group 3}&\multicolumn{1}{c}{Group 4} &\multicolumn{1}{c}{Corrected} \\            
\midrule

\multicolumn{5}{l}{A. Birth weight} \\

$NO$        &       1.011         &      -3.354         &      -2.083         &      -0.157  &   -1.267\sym{*}     \\
            &     (3.949)         &     (2.600)         &     (1.787)         &     (1.000)  &  (0.747)            \\
$NO_2$      &       2.025         &      -0.205         &      -2.776         &      -1.160  &   -0.240            \\
            &    (10.078)         &     (7.466)         &     (4.430)         &     (2.901)  &  (2.230)            \\
$PM_{10}$   &      17.791         &      -7.858         &       9.236\sym{**} &       3.119  &    2.364            \\
            &    (12.918)         &     (6.091)         &     (4.708)         &     (2.529)  &  (1.824)            \\

\midrule                          
$r^2$       &       0.695         &       0.620         &       0.561         &       0.509  &    0.469            \\
Obs.        &       6,373         &      15,140         &      37,542         &      90,416  &  143,675            \\

\midrule
\multicolumn{5}{l}{B.Birth length} \\

$NO$        &     -0.122         &      -0.040         &      -0.124         &      -0.002   &  -0.039            \\
            &    (0.176)         &     (0.164)         &     (0.076)         &     (0.045)   & (0.036)            \\
$NO_2$      &      0.566         &      -0.611         &       0.109         &      -0.142   &  -0.097            \\
            &    (0.429)         &     (0.557)         &     (0.215)         &     (0.127)   & (0.104)            \\
$PM_{10}$   &      0.885         &      -0.890         &       0.061         &       0.158   &   0.047            \\
            &    (0.598)         &     (0.705)         &     (0.193)         &     (0.125)   & (0.100)            \\
                         
\midrule                         
$r^2$       &      0.705         &       0.599         &       0.562         &       0.502   &   0.461            \\
Obs.        &      5,833         &      14,142         &      34,792         &      85,704   & 136,600            \\
\midrule
\multicolumn{5}{l}{C. Scenario} \\

$move_m$   &       yes            &       yes           &      no             &      no      &   -                \\
$move_b$   &       yes            &        no           &     yes             &     no       &   no               \\

\bottomrule
\end{tabular}

\begin{tablenotes}
      \small
      \item Notes: (1) Regressions are based on the benchmark model for sub-samples labeled in panel C, where $move_m$ indicates if the mothers moved between \textit{grunnkrets} in the year and the year before delivery; $move_b$ indicates if the babies were born in the same municipality as where the mothers live in the year of delivery. (2) The independent variables in Panel A and Panel B are birth weight and birth length separately. (3) Cluster robust standard errors at \textit{grunnkrets} level in parentheses. (4) *** $p < 0.01$, ** $p<0.05$, * $p<0.1$. (5) All pollutants are in $\mu g/m^3$, birth-weight in gram, birth length in millimeter.
    \end{tablenotes}
\end{threeparttable}

\end{table}
 

Table \ref{tab:nomove} reports the results of regressions by groups as discussed above. As usual, panels A and B of the table are for birth weight and length, respectively. Panel C indicates the subgroups of the sample involved in the regression. For example, in column (3) of Table \ref{tab:nomove}, the indicator $move_m = yes$ means that the mother lived in two different \textit{grunnkrets} in the year of delivery and the year before; the indicator $move_b = no$ means that the infant was born in the same municipality where the mother lived in the year of delivery. Thus, column (2) corresponds to group (2) in Table \ref{tab:mlocation}, i.e., mothers who moved prior to delivery.

Column (1) of Table \ref{tab:nomove} reports the regression results for those whose prenatal exposure is wrongly measured (sample size of $1/40$ of the baseline sample). The coefficients of pollutants makes no sense.
Column (2) of Table \ref{tab:nomove} indicates that infants of mothers who moved before delivery are more sensitive (although not significantly so) to ambient air pollution in the last trimester. This is explainable if the mothers in group (2) are those who are vulnerable to ambient air pollution. 
The mothers in column (3) of Table \ref{tab:nomove} gave birth in a different municipality than where they lived, even though they themselves did not move at all in the year of delivery. Based on the coefficients in column (3), the babies of these mothers appear to be more vulnerable to ambient air pollution. This coincides with the self-selection of the maternity hospital. Perhaps vulnerable mothers and infants are the reason for choosing a distant hospital. In column (4), infants whose mothers did not move and gave birth in the same municipality are less affected by prenatal ambient air pollution exposure. This may be due to certain unobservable traits of these mothers, such as the fact that they may be long-time local residents of Norway. 
\footnote{Heterogeneity in the effect of prenatal ambient air pollution exposure on birth outcomes is discussed further in the next section.}

The last column of Table \ref{tab:nomove} excludes group (1) from the baseline sample. The prenatal exposure should thus be measured correctly. The results in this column are almost identical to the baseline regression results. In summary, neither the measurement error due to the mother's move nor the self-selection of hospitals appeared to affect the baseline estimates after taking into account the birthplace of the infant.

\section{Heterogeneity}
\label{sec:hetero}

This section examines the heterogeneous effects of prenatal exposure to ambient air pollution on birth outcomes across subgroups categorized by demographics and ambient air pollution levels. It is important to note that splitting the sample into subgroups reduces the number of observations within a \textit{grunnkrets}-month. As a result, more singletons are excluded from the regression, and the precision of the estimates is expected to be reduced.

\subsection{Heterogeneity by demographics}
\label{subsec:hetero_group}

Table \ref{tab:hetero} report regressions of subgroups with different demographic characteristics. The first two columns in Table \ref{tab:hetero} are regressions by gender grouping of infants. Consistent with the literature, the $NO$ effects in column (1) are greater and more significant than those in column (1), implying that male newborns appear to be more susceptible to prenatal ambient air pollution than female newborns.

\begin{table}[htbp]
\centering
\begin{threeparttable}
\caption{Heterogeneous effect of maternal exposure to ambient air pollution in the 3rd. trimester on birth outcomes     \label{tab:hetero} }
\def\sym#1{\ifmmode^{#1}\else\(^{#1}\)\fi}
\begin{tabular}{l*{8}{c}}
\toprule
            &\multicolumn{1}{c}{(1)}&\multicolumn{1}{c}{(2)}&\multicolumn{1}{c}{(3)}&\multicolumn{1}{c}{(4)}&\multicolumn{1}{c}{(5)}&\multicolumn{1}{c}{(6)}&\multicolumn{1}{c}{(7)}&\multicolumn{1}{c}{(8)}\\
             &\multicolumn{1}{c}{boy}&\multicolumn{1}{c}{girl}&\multicolumn{1}{c}{non-ntv}&\multicolumn{1}{c}{native}&\multicolumn{1}{c}{$<icm_m^{50}$}&\multicolumn{1}{c}{$>icm_m^{50}$}&\multicolumn{1}{c}{$<icm_m^{25}$} &\multicolumn{1}{c}{$>icm_m^{75}$}\\

\midrule
\multicolumn{8}{l}{A. Birth weight} \\

$NO$        &      -1.857\sym{*}  &      -1.672         &      -2.291         &      -0.893         &      -1.787\sym{*}  &      -1.376         &      -3.526         &      -0.856         \\
            &     (1.084)         &     (1.107)         &     (1.643)         &     (0.852)         &     (1.022)         &     (1.226)         &     (2.189)         &     (2.333)         \\
$NO_2$       &      -0.326         &       2.038         &       2.871         &      -1.628         &      -3.334         &      -0.476         &      -1.208         &       0.564         \\
            &     (3.068)         &     (3.101)         &     (4.277)         &     (2.324)         &     (2.924)         &     (3.261)         &     (5.792)         &     (6.457)         \\
$PM_{10}$      &       0.183         &       2.556         &      -3.310         &       2.229         &       2.326         &       5.847\sym{*}  &      -0.398         &       1.650         \\
            &     (2.742)         &     (2.864)         &     (4.149)         &     (1.965)         &     (2.479)         &     (3.026)         &     (4.991)         &     (5.479)         \\
\midrule
$r^2$       &       0.520         &       0.521         &       0.559         &       0.517         &       0.534         &       0.534         &       0.610         &       0.591         \\
Obs.        &      80,311         &      73,638         &      46,703         &     121,495         &      77,258         &      89,165         &      22,531         &      36,030         \\

\midrule
\multicolumn{8}{l}{B. Birth length} \\

$NO$        &      -0.079         &      -0.046         &      -0.198\sym{**} &      -0.039         &      -0.093\sym{*}  &      -0.062         &      -0.249\sym{**} &      -0.061         \\
            &     (0.051)         &     (0.050)         &     (0.082)         &     (0.039)         &     (0.050)         &     (0.058)         &     (0.115)         &     (0.111)         \\
$NO_2$       &       0.064         &      -0.022         &       0.397\sym{**} &      -0.128         &      -0.129         &       0.087         &       0.087         &       0.154         \\
            &     (0.140)         &     (0.143)         &     (0.196)         &     (0.106)         &     (0.148)         &     (0.145)         &     (0.351)         &     (0.298)         \\
$PM_{10}$      &       0.039         &      -0.097         &      -0.078         &      -0.005         &       0.046         &       0.134         &      -0.343         &      -0.075         \\
            &     (0.124)         &     (0.161)         &     (0.195)         &     (0.093)         &     (0.170)         &     (0.139)         &     (0.347)         &     (0.276)         \\
\midrule
$r^2$       &       0.463         &       0.466         &       0.462         &       0.466         &       0.464         &       0.467         &       0.465         &       0.470         \\
Obs.        &     194,486         &     157,291         &     195,830         &     148,232         &     195,498         &     155,014         &     180,669         &     121,101         \\
\bottomrule

\end{tabular}

\begin{tablenotes}
      \small
      \item Notes: (1) Regression is based on the benchmark model. (2) ``non-ntv'' means at least one of the parents have immigration background or non-Norwegian nationality; ``$icm_m^p$'' is the maternal income at $p$th percentile. (3) The independent variables in Panel A and Panel B are birth weight and birth length separately. (4) Cluster robust standard errors at \textit{grunnkrets} level in parentheses, (5) *** $p < 0.01$, ** $p<0.05$, * $p<0.1$. (6) All pollutants are in $\mu g/m^3$, birth-weight in gram, birth length in millimeter.
    \end{tablenotes}
\end{threeparttable}
\end{table}
 

In columns (3) and (4) of Table \ref{tab:hetero}, I split the sample into two groups based on immigrant background and nationality, where ``non-ntv'' (non-native) is defined as having at least one parent with an immigrant background or with non-Norwegian nationality, while ``native'' means that both parents are Norwegian and have no history of immigration. Comparing the results in columns (3) and (4) with the baseline regression, we can see that environmental $NO$ concentrations have a greater marginal adverse effect ($2-4$ times) on birth outcomes for non-native infants. In contrast, for native infants, the marginal effect of maternal $NO$ exposure is smaller and less pronounced than in the baseline regression. One possible explanation is that immigrants are more exposed to ambient air pollution than natives due to their occupation  and the effect of air pollutants on birth outcomes is non-linear (marginal increment). 
\footnote{For example, $12.6\%$ of immigrants in Norway are in primary occupations, compared to $2.7\%$ of native Norwegians. Data from Statistics Norway: \href{https://www.ssb.no/en/arbeid-og-lonn/sysselsetting/statistikk/sysselsetting-blant-innvandrere-registerbasert}{https://www.ssb.no/en/arbeid-og-lonn/sysselsetting/statistikk/sysselsetting-blant-innvandrere-registerbasert}}

I further examine the heterogeneity of the effect of air pollution in terms of mothers' income in columns (5)-(8) of Table \ref{tab:hetero}.
\footnote{ I also classify the sample according to the father's income, both parents' income and wealth, all with relatively similar results (not shown). The reason for using mother's income by default is that more observations register information about the mother.} 
The annual after-tax income of mothers in columns (5) and (6) is below and above the mean ($icm^{50}_m$), respectively. It can be seen that infants whose mothers have below-average income appear to be more vulnerable to prenatal $NO$ exposure than those who are financially well off. To highlight this heterogeneity, I further compared newborns whose mothers' income is below the first quartile ($<icm^{25}_m$, the worse off) and above the third quartile ($>icm^{75}_m$, the better off). Not surprisingly, the marginal adverse effect of $NO$ on birth weight is much larger and more significant in poorer conditioned infants. This result is consistent with the findings for immigrants in columns (3)-(4).

In summary, I find that prenatal exposure to ambient $NO$ is more detrimental to male than female infants. Also, the marginal adverse effects of ambient $NO$ on fetuses are larger and more significant for families with immigrant background/nationality and/or lower incomes.

\subsection{Heterogeneity by prenatal exposure level}
\label{subsec:hetero_extent}

The effects of ambient air pollutants on birth outcomes may be nonlinear. For example, below certain safe levels, even long-term prenatal exposure may not affect birth outcomes. On the contrary, at high pollution levels, short-term exposures may also cause serious harm. In this subsection, I first examine the response of birth outcomes to air pollutant concentrations in infants exposed to different average levels of air pollution during the last trimester. Note that there are $11$ weeks in the last trimester, the high average trimester prenatal exposure levels may come from very high air pollution in just a few weeks (while the other weeks have very low pollution levels. Appendix Figure \ref{fig:extent} illustrates this scenario). Therefore, I studied further to see if the adverse health effects on birth outcomes were driven by these occasional high levels of air pollution events.
\footnote{I similarly examine the effects of occasional high level air pollution events of $NO_2$ and $PM_{10}$ on birth outcomes in the Appendix Table \ref{tab:extent} and find that these events do not affect birth outcomes.} 


\begin{table}[htbp]
\centering

\begin{threeparttable}
\caption{Regression by maternal exposure extent to ambient $NO$ in the 3rd. trimester \label{tab:extent_tmavg} }

\def\sym#1{\ifmmode^{#1}\else\(^{#1}\)\fi}

\begin{tabular}{l*{5}{c}}
\toprule
$NO$ conc.  &\multicolumn{1}{c}{<115}&\multicolumn{1}{c}{<90}&\multicolumn{1}{c}{<78}&\multicolumn{1}{c}{<56} \\
\midrule

\multicolumn{5}{l}{A. Birth weight} \\

$NO$        &      -1.282\sym{*}  &      -1.464\sym{*}  &      -1.073         &       0.566         \\
            &     (0.669)         &     (0.781)         &     (0.867)         &     (1.119)         \\
$NO_2$       &       0.175         &       0.843         &       0.258         &      -0.437         \\
            &     (1.805)         &     (1.866)         &     (1.968)         &     (2.110)         \\
$PM_{10}$      &       1.433         &       1.250         &       1.601         &       1.032         \\
            &     (1.497)         &     (1.601)         &     (1.661)         &     (1.793)         \\
\midrule
$r^2$       &       0.465         &       0.467         &       0.468         &       0.471         \\
Obs.        &     222,602         &     212,126         &     199,251         &     163,691         \\

\midrule
\multicolumn{5}{l}{B.Birth length} \\

$NO$        &      -0.044         &      -0.038         &      -0.020         &       0.085         \\
            &     (0.031)         &     (0.036)         &     (0.041)         &     (0.054)         \\
$NO_2$       &      -0.015         &       0.009         &      -0.011         &      -0.093         \\
            &     (0.082)         &     (0.086)         &     (0.090)         &     (0.100)         \\
$PM_{10}$      &       0.001         &      -0.021         &      -0.035         &      -0.038         \\
            &     (0.077)         &     (0.082)         &     (0.086)         &     (0.093)         \\
\midrule
$r^2$       &       0.462         &       0.464         &       0.465         &       0.469         \\
Obs.        &     210,501         &     200,909         &     189,049         &     155,822         \\

\bottomrule
\end{tabular}

\begin{tablenotes}
      \small
      \item Notes: (1) Regressions are based on the benchmark model in sub-samples with different average maternal exposure extent to ambient $NO$ in the last trimester. (2) The independent variables in Panel A and Panel B are birth weight and birth length separately. (3) Cluster robust standard errors at \textit{grunnkrets} level in parentheses. (4) *** $p < 0.01$, ** $p<0.05$, * $p<0.1$. (5) All pollutants are in $\mu g/m^3$, birth-weight in gram, birth length in millimeter.
    \end{tablenotes}
\end{threeparttable}

\end{table}
 

Columns (1)-(4) of Table \ref{tab:extent_tmavg} are regressions for sub-samples with mean prenatal $NO$ exposure levels below the $99th$, $90th$, $75th$, and $50th$ percentiles in the last trimester. It appears that the exclusion of observations with the highest prenatal $NO$ exposure does not change the coefficient of $NO$ much, especially for birth weight. When prenatal exposure is below average (column (4)), the $NO$ coefficients in panels A and B are no longer negative, indicating that the last trimester of below-average prenatal $NO$ exposure is safe. The other two pollutants, $NO_2$ and $PM_{10}$, have no significant effect on birth outcomes. Also, according to Table \ref{tab:extent_tmavg}, the marginal effect of last-trimester ambient $NO$ pollution on birth outcomes appears to be greater when the average $NO$ concentration level is higher.

For observations with above-average prenatal last-trimester $NO$ exposure ($>56\mu g/m^3$), I further divided them into groups based on the number of ``high-concentration environmental $NO$ events'' they experienced in the last trimester. Here a ``high ambient $NO$ event'' is defined as a week in which the ambient $NO$ concentration is above certain percentiles, such as the $99th$ ($170 \mu g/m^3$) and $95th$ ($110 \mu g/m^3$) percentiles. The regression results, as well as the mean prenatal $NO$ exposure in the last trimester corresponding to each subgroup, are shown in Table \ref{tab:extent_triNwH}. Taking columns (1) and (4) of the table as examples, the newborns in both columns are exposed to above-average levels of ambient $NO$ in the last trimester prenatally, but only the infants in column (4) experienced ``high-level ambient $NO$ events'' (3 weeks), whereas the infants in column (1) are not exposed to any high levels of ambient $NO$ pollution events.

\begin{table}[htbp]
\centering
\begin{threeparttable}
\caption{How high-level ambient $NO$ pollution events in the third trimester affect birth outcomes given last-trimester-averaged $NO>56 \mu g/m^3$ \label{tab:extent_triNwH} }
\def\sym#1{\ifmmode^{#1}\else\(^{#1}\)\fi}
\begin{tabular}{l*{10}{c}}
\toprule
events:  &\multicolumn{4}{c}{$99th^+$ percentile events} &&\multicolumn{4}{c}{$95th^+$ percentile events} \\
            \cmidrule{2-5} \cmidrule{7-10}   
weeks $\le$:    &\multicolumn{1}{c}{0}&\multicolumn{1}{c}{1}&\multicolumn{1}{c}{2}&\multicolumn{1}{c}{3}&&\multicolumn{1}{c}{0}&\multicolumn{1}{c}{1}&\multicolumn{1}{c}{2} &\multicolumn{1}{c}{3}\\
\midrule
\multicolumn{5}{l}{A. Average ambient $NO$ level in trimester 3} \\
            &    69.9         &     73.5         &     76.1         &     76.8        & &      63.3         &    65.4         &     69 & 71.8         \\

\midrule
\multicolumn{5}{l}{B. Birth weight} \\

$NO$        &      -5.346\sym{***}&      -2.723\sym{**} &      -2.558\sym{**} &      -1.817\sym{*}  & &      -3.882         &      -6.317\sym{*}  &      -4.644\sym{**} &      -1.691         \\
            &     (1.800)         &     (1.387)         &     (1.191)         &     (1.073)         & &     (5.780)         &     (3.324)         &     (2.076)         &     (1.504)         \\
$NO_2$      &       2.514         &      -0.411         &      -1.229         &      -2.777         & &      12.134         &      13.994\sym{*}  &       6.678         &      -1.200         \\
            &     (5.311)         &     (4.404)         &     (3.941)         &     (3.767)         & &    (13.898)         &     (7.764)         &     (5.529)         &     (4.680)         \\
$PM_{10}$   &       0.795         &       0.459         &       0.042         &      -0.594         & &      19.110         &      -2.425         &       2.417         &      -3.929         \\
            &     (4.904)         &     (4.049)         &     (3.723)         &     (3.695)         & &    (13.309)         &     (7.878)         &     (5.551)         &     (4.521)         \\
\midrule
$r^2$       &       0.477         &       0.471         &       0.468         &       0.467         & &       0.486         &       0.486         &       0.481         &       0.477         \\
Obs.        &      36,588         &      49,596         &      54,373         &      55,182         & &       8,834         &      20,235         &      35,268         &      44,776         \\

\midrule
\multicolumn{5}{l}{C. Birth length} \\

$NO$        &      -0.253\sym{***}&      -0.142\sym{**} &      -0.140\sym{***}&      -0.117\sym{**} & &      0.090         &      -0.180         &      -0.151         &      -0.104         \\
            &     (0.085)         &     (0.064)         &     (0.054)         &     (0.051)         & &    (0.260)         &     (0.156)         &     (0.098)         &     (0.071)         \\
$NO_2$      &       0.308         &       0.130         &       0.085         &       0.012         & &      0.198         &       0.735\sym{**} &       0.302         &       0.180         \\
            &     (0.237)         &     (0.194)         &     (0.173)         &     (0.168)         & &    (0.651)         &     (0.366)         &     (0.252)         &     (0.205)         \\
$PM_{10}$   &       0.001         &       0.049         &       0.092         &       0.063         & &      0.207         &      -0.376         &       0.132         &      -0.100         \\
            &     (0.232)         &     (0.194)         &     (0.172)         &     (0.172)         & &    (0.585)         &     (0.359)         &     (0.251)         &     (0.214)         \\
\midrule
$r^2$       &       0.464         &       0.462         &       0.459         &       0.459         & &      0.481         &       0.481         &       0.471         &       0.468         \\
Obs.        &      33,968         &      45,907         &      50,340         &      51,054         & &      8,201         &      18,770         &      32,685         &      41,450         \\
\bottomrule

\end{tabular}

\begin{tablenotes}
      \small
      \item Notes: (1) Regression is based on the benchmark model in observations whose average maternal $NO$ exposure in the last trimester is greater than the average ($56 \mu g/m^3$). These observations are further classified in to sub-samples according to the number of ``high-level $NO$ pollution events'', which is defined as weeks with average ambient $NO$ concentration higher than $99th/95th$ percentile (170 $\mu g/m^3$ and $110  \mu g/m^3$ separately) of the weekly $NO$ concentration in the last trimester. (2) The average ambient $NO$ level in the last trimester for each sub-group is in Panel A. The independent variables in Panel B and Panel C are birth weight and birth length separately. (3) Cluster robust standard errors at \textit{grunnkrets} level in parentheses, (4) *** $p < 0.01$, ** $p<0.05$, * $p<0.1$. (5) All pollutants are in $\mu g/m^3$, birth-weight in gram, birth length in millimeter.
    \end{tablenotes}
\end{threeparttable}
\end{table}
 

Interestingly, columns (1)-(4) of Table \ref{tab:extent_triNwH} suggest that the more ``ambient $NO$ concentrations above the $99th$ percentile event'' in the sample with above-average prenatal $NO$ exposure in the last trimester, the smaller the marginal effect of $NO$, although the average $NO$ in each subgroup concentration increases from column (1) to column (4). When there is no such ``high ambient $NO$ event'' in the last trimester (column (1)), the marginal effect of $NO$ is four times greater than in the baseline regression. This suggests that chronic exposure to relatively high levels of ambient $NO$ pollution (column (1)) is more detrimental than occasional high levels of ambient $NO$ pollution events for fetuses whose mothers are exposed to above-average levels of ambient $NO$ pollution in the last trimester. The same pattern is shown in columns (6)-(8) of Table \ref{tab:extent_triNwH}. In these columns, ``high ambient $NO$ events'' are defined as weeks when the weekly average ambient $NO$ concentration is above the $95th$ percentile. The estimate in column (5) is very imprecise because there are too few observations.

Similarly, for observations with below $78\mu g/m^3$ (75th percentile) prenatal $NO$ exposure, I grouped them into four groups based only on the number of $NO$ pollution events above the $95th$ percentile concentration in the last trimester.
\footnote{I also tried to use observations with below-average prenatal $NO$ exposure ($<56 \mu g/m^3$) in the last trimester, as in Table \ref{tab:extent_triNwL}, but ``high ambient $NO$ events'' are too few in these observations and the estimated noise is too large.}
\footnote{Events above the $99th$ percentile are not studied here because they are extremely rare and the sample size is not sufficient for regression.}
The regression results for each subgroup for the last three months of prenatal $NO$ exposure are presented in Table \ref{tab:extent_triNwL}. As seen in Panel A, the mean ambient $NO$ concentrations are low for all four subgroups. Compared to Table \ref{tab:extent_triNwH}, the sign of the coefficient on $NO$ is insignificant and can even be positive when there are fewer than three high-level $NO$ pollution events (columns (1)-(3)), whereas when the number of high-level $NO$ pollution events increases to three, the magnitude of the adverse effect becomes much larger than in the baseline regression and is significant at the $5\%$ level. It appears that for those with very low prenatal $NO$ exposure, occasional environmental $NO$ concentration events ``above the $95th$ percentile'' in the last trimester also adversely affect their birth weight and length.

\begin{table}[htbp]
\centering
\begin{threeparttable}
\caption{How $NO$ $95th^+$ percentile events in the third trimester affect birth outcomes given last-trimester-averaged $NO<78 \mu g/m^3$ \label{tab:extent_triNwL} }
\def\sym#1{\ifmmode^{#1}\else\(^{#1}\)\fi}
\begin{tabular}{l*{5}{c}}
\toprule
weeks  $\le$:     &\multicolumn{1}{c}{0}&\multicolumn{1}{c}{1}&\multicolumn{1}{c}{2} &\multicolumn{1}{c}{3} \\
\midrule
\multicolumn{5}{l}{A. Average ambient $NO$ level in trimester 3} \\
            &  30.1       &   33    &   34.6  &  35.1   \\
\midrule
\multicolumn{5}{l}{B. Birth weight} \\

$NO$        &      -0.869         &       5.724         &       0.894         &      -4.566\sym{**} \\
            &     (0.906)         &    (13.204)         &     (1.227)         &     (2.213)         \\
$NO_2$       &       0.293         &       9.550         &      -0.450         &       3.297         \\
            &     (1.976)         &    (38.983)         &     (2.120)         &     (7.254)         \\
$PM_{10}$      &       1.412         &     -24.531         &       0.577         &       9.481         \\
            &     (1.668)         &    (38.419)         &     (1.766)         &     (6.676)         \\
\midrule
$r^2$       &       0.468         &       0.552         &       0.468         &       0.498         \\
Obs.        &     195,790         &       2,673         &     165,227         &      29,473         \\

\midrule
\multicolumn{5}{l}{C. Birth length} \\

$NO$        &      -0.009         &       0.509         &       0.105\sym{*}  &      -0.276\sym{***}\\
            &     (0.043)         &     (0.649)         &     (0.058)         &     (0.105)         \\
$NO_2$       &      -0.004         &      -0.488         &      -0.056         &       0.331         \\
            &     (0.091)         &     (1.886)         &     (0.101)         &     (0.319)         \\
$PM_{10}$      &      -0.042         &      -1.580         &      -0.089         &       0.185         \\
            &     (0.086)         &     (1.753)         &     (0.094)         &     (0.301)         \\
\midrule
$r^2$       &       0.464         &       0.597         &       0.466         &       0.492         \\
Obs.        &     185,866         &       2,445         &     156,990         &      27,671         \\
\bottomrule

\end{tabular}

\begin{tablenotes}
      \small
      \item Notes: (1) Regression is based on the benchmark model in observations whose average maternal $NO$ exposure in the last trimester is less than the $78 \mu g/m^3$. These observations are further classified in to sub-samples according to the number of ``high-level $NO$ pollution events'', which is defined as weeks with average ambient $NO$ concentration higher than $99th/95th$ percentile ($170 \mu g/m^3$ and $110 \mu g/m^3$ separately) of the weekly $NO$ concentration in the last trimester. (2) The average ambient $NO$ level in the last trimester for each sub-group is in Panel A. The independent variables in Panel B and Panel C are birth weight and birth length separately. (3) Cluster robust standard errors at \textit{grunnkrets} level in parentheses, (4) *** $p < 0.01$, ** $p<0.05$, * $p<0.1$. (5) All pollutants are in $\mu g/m^3$, birth-weight in gram, birth length in millimeter.
    \end{tablenotes}
\end{threeparttable}
\end{table}
 

By combining Table \ref{tab:extent_triNwH} and Table \ref{tab:extent_triNwL}, I conclude that for the sample with above-average prenatal environmental $NO$ exposure levels in the last trimester, long-term exposure to relatively high ambient $NO$ levels caused more harm than occasional high ambient $NO$ events, whereas for observations with relatively low average last trimester prenatal environmental $NO$ exposure levels, occasional high ambient $NO$ events, if present, are also harmful to birth outcomes.

\clearpage
\section{Conclusion}
\label{conclude}

In this paper, by using the variance in prenatal ambient air pollution exposure levels among infants born within a specific calendar-month in the same sub-zip-code area, I find that exposure to ambient nitric oxide ($NO$) in the last trimester of pregnancy can significantly reduce birth weight and length in Norwegian children born between 2000 and 2016. On average, each standard deviation increase ($25.4 \mu g /m^3$) in prenatal exposure to $NO$ resulted in a $1\%$ decrease in birth weight and a $0.3\%$ decrease in birth length, which is similar to the effects of other studies on the effects of ambient air pollutants. Pollution levels of other types of ambient air pollutants, such as $NO_2$, $PM_{10}$ and $O_3$, appear to be safe for Norwegian fetuses. Prenatal exposure to $SO_2$ in the environment appears to have a negative effect on birth weight and length, but there are not enough observations to make a precise estimate. I do not find an effect of prenatal ambient air pollution exposure on APGAR scores.

A limitation of the study methodology in this paper is that the addresses of pregnant women are updated annually, and I may not have been able to accurately determine where the mothers resided during pregnancy. This may be the reason why I find no significant effect of prenatal exposure to ambient air pollution in the first two months, although the literature suggests that most abnormal fetal development occurs in the last trimester. Using the mother's workplace address in the year of birth, which is regularly recorded by social welfare and tax agencies, may be a solution for the future.

The affinity of $NO$ for hemoglobin may be a contributor to this adverse effect. The diffusion of inhaled $NO$ into the blood of the pregnant woman through the alveoli and capillaries oxidizes the Fe(II) of red blood cell hemoglobin (Hb) to the Fe(III) state, forming methemoglobin (MetHb) and impairing oxygen transport. As a result, the fetus is exposed to methemoglobin through the placental barrier. Although ambient air pollution is associated with diabetes, which in turn affects birth weight, I have not found any evidence that ambient $NO$ has such a mechanism. It would be interesting to further confirm the mechanisms by which environmental $NO$ pollution affects the fetus. Though the literature has found a link between birth outcomes and long-term health outcomes, it is not clear whether reduced birth weight and length due to ambient air pollution can also affect long-term health outcomes. Understanding the mechanisms by which pollutants affect birth outcomes can help assess their long-term effects.

In addition, I find that both average ambient $NO$ in the last trimester and occasional high ambient $NO$ pollution events can be harmful to the fetus. Although ambient air quality in Norway is generally high and has been improving in recent years, there are weeks with high ambient $NO$ concentrations that are harmful to the last trimester fetus. As found in the literature, reductions in birth weight and length may have a negative impact on the long-term health status of children. This poses a challenge to environmental pollution management and policy development: not only to focus on average pollutant levels, but also to pay attention to the containment of short-term high pollution events.

Prenatal exposure to ambient $NO$ also has heterogeneous effects on different groups. Consistent with the literature, I find that male infants are more susceptible to environmental $NO$ pollution than female infants. Infants from economically and/or ethnically disadvantaged families are more affected than children from better-off families. This is not surprising because most immigrants live in large cities like Oslo, where ambient $NO$ concentrations are occasionally quite high during some weeks, and as mentioned earlier, the marginal effect of ambient $NO$ on birth outcomes is greater when air pollution levels are high. Another possible explanation is that less privileged mothers are physically more vulnerable to the effects of ambient air pollution. Due to the nature of their work, they may also engage in more outdoor activities. Future studies may examine why newborns of poorly conditioned parents are more vulnerable to ambient air pollution and how to protect them. If infants' long-term health is made worse by prenatal exposure, and thus disadvantaged in the labor market in the future, they may be more likely to be exposed to the same harmful environment - parental and offspring air pollution exposure reinforcing each other and create a poor-health (and also poverty) trap.

\clearpage

\bibliographystyle{aer}
\bibliography{air}

@article{bowling1986climatology,
  title={Climatology of high-latitude air pollution as illustrated by Fairbanks and Anchorage, Alaska},
  author={Bowling, Sue Ann},
  journal={Journal of climate and applied meteorology},
  pages={22--34},
  year={1986},
  publisher={JSTOR}
}

@article{schjoldager1979observations,
  title={Observations of high ozone concentrations in Oslo, Norway, during the summer of 1977},
  author={Schjoldager, J{\o}rgen},
  journal={Atmospheric Environment (1967)},
  volume={13},
  number={12},
  pages={1689--1696},
  year={1979},
  publisher={Elsevier}
}

@Article{RN329,
  author  = {Almond, Douglas and Chay, Kenneth Y and Lee, David S},
  journal = {The Quarterly Journal of Economics},
  title   = {The costs of low birth weight},
  year    = {2005},
  issn    = {1531-4650},
  number  = {3},
  pages   = {1031-1083},
  volume  = {120},
  doi     = {10.1162/003355305774268228},
  type    = {Journal Article},
}

@Article{almond2018childhood,
  author  = {Almond, Douglas and Currie, Janet and Duque, Valentina},
  journal = {Journal of Economic Literature},
  title   = {Childhood circumstances and adult outcomes: Act II},
  year    = {2018},
  number  = {4},
  pages   = {1360--1446},
  volume  = {56},
  doi     = {10.1257/jel.20171164},
}

@Article{RN303,
  author  = {Altshuller, Aubrey P},
  journal = {Journal of the air pollution control association},
  title   = {Thermodynamic considerations in the interactions of nitrogen oxides and oxy-acids in the atmosphere},
  year    = {1956},
  issn    = {0002-2470},
  number  = {2},
  pages   = {97-100},
  volume  = {6},
  doi     = {10.1080/00966665.1956.10467740},
  type    = {Journal Article},
}

@Article{RN331,
  author  = {Andersen, Zorana J and Raaschou-Nielsen, Ole and Ketzel, Matthias and Jensen, Steen S and Hvidberg, Martin and Loft, Steffen and TjÃ¸nneland, Anne and Overvad, Kim and SÃ¸rensen, Mette},
  journal = {Diabetes care},
  title   = {Diabetes incidence and long-term exposure to air pollution: a cohort study},
  year    = {2012},
  issn    = {0149-5992},
  number  = {1},
  pages   = {92-98},
  volume  = {35},
  doi     = {https://doi.org/10.2337/dc11-1155},
  type    = {Journal Article},
}

@Article{RN309,
  author  = {Avery, Alexander Austin},
  journal = {Environmental health perspectives},
  title   = {Infantile methemoglobinemia: reexamining the role of drinking water nitrates},
  year    = {1999},
  issn    = {0091-6765},
  number  = {7},
  pages   = {583-586},
  volume  = {107},
  doi     = {10.1289/ehp.99107583},
  type    = {Journal Article},
}

@Article{RN319,
  author  = {Bell, Michelle L and Ebisu, Keita and Belanger, Kathleen},
  journal = {Environmental health perspectives},
  title   = {Ambient air pollution and low birth weight in Connecticut and Massachusetts},
  year    = {2007},
  issn    = {0091-6765},
  number  = {7},
  pages   = {1118-1124},
  volume  = {115},
  doi     = {10.1289/ehp.9759},
  type    = {Journal Article},
  url     = {https://www.ncbi.nlm.nih.gov/pmc/articles/PMC1913584/pdf/ehp0115-001118.pdf},
}

@Article{beltran2014associations,
  author    = {Beltran, Alyssa J and Wu, Jun and Laurent, Olivier},
  journal   = {International journal of environmental research and public health},
  title     = {Associations of meteorology with adverse pregnancy outcomes: a systematic review of preeclampsia, preterm birth and birth weight},
  year      = {2014},
  number    = {1},
  pages     = {91--172},
  volume    = {11},
  doi       = {https://doi.org/10.3390/ijerph110100091},
  publisher = {Multidisciplinary Digital Publishing Institute},
}

@Article{RN288,
  author  = {Bobak, Martin},
  journal = {Environmental health perspectives},
  title   = {Outdoor air pollution, low birth weight, and prematurity},
  year    = {2000},
  issn    = {0091-6765},
  number  = {2},
  pages   = {173-176},
  volume  = {108},
  doi     = {10.1289/ehp.00108173},
  type    = {Journal Article},
  url     = {https://www.ncbi.nlm.nih.gov/pmc/articles/PMC1637893/pdf/envhper00303-0121.pdf},
}

@Article{RN330,
  author  = {Brook, Robert D and Jerrett, Michael and Brook, Jeffrey R and Bard, Robert L and Finkelstein, Murray M},
  journal = {Journal of occupational and environmental medicine},
  title   = {The relationship between diabetes mellitus and traffic-related air pollution},
  year    = {2008},
  issn    = {1076-2752},
  number  = {1},
  pages   = {32-38},
  volume  = {50},
  doi     = {10.1097/jom.0b013e31815dba70},
  type    = {Journal Article},
}

@Article{RN320,
  author  = {Chen, Lei and Yang, Wei and Jennison, Brian L and Goodrich, Andy and Omaye, Stanley T},
  journal = {Inhalation toxicology},
  title   = {Air pollution and birth weight in northern Nevada, 1991-1999},
  year    = {2002},
  issn    = {0895-8378},
  number  = {2},
  pages   = {141-157},
  volume  = {14},
  doi     = {10.1080/089583701753403962},
  type    = {Journal Article},
  url     = {https://www.tandfonline.com/doi/pdf/10.1080/089583701753403962?needAccess=true},
}

@Article{RN341,
  author  = {Clark, Reese H and Kueser, Thomas J and Walker, Marshall W and Southgate, W Michael and Huckaby, Jeryl L and Perez, Jose A and Roy, Beverly J and Keszler, Martin and Kinsella, John P},
  journal = {New England Journal of Medicine},
  title   = {Low-dose nitric oxide therapy for persistent pulmonary hypertension of the newborn},
  year    = {2000},
  issn    = {0028-4793},
  number  = {7},
  pages   = {469-474},
  volume  = {342},
  doi     = {10.1056/nejm200002173420704},
  type    = {Journal Article},
}

@Article{RN296,
  author  = {Currie, Janet and Neidell, Matthew},
  journal = {The Quarterly Journal of Economics},
  title   = {Air pollution and infant health: what can we learn from California's recent experience?},
  year    = {2005},
  issn    = {1531-4650},
  number  = {3},
  pages   = {1003-1030},
  volume  = {120},
  doi     = {10.1162/003355305774268219},
  type    = {Journal Article},
}

@Article{cyrys2012variation,
  author    = {Cyrys, Josef and Eeftens, Marloes and Heinrich, Joachim and Ampe, Christophe and Armengaud, Alexandre and Beelen, Rob and Bellander, Tom and Beregszaszi, Timea and Birk, Matthias and Cesaroni, Giulia and others},
  journal   = {Atmospheric Environment},
  title     = {Variation of NO2 and NOx concentrations between and within 36 European study areas: results from the ESCAPE study},
  year      = {2012},
  pages     = {374--390},
  volume    = {62},
  doi       = {10.1016/j.atmosenv.2012.07.080},
  publisher = {Elsevier},
}

@Article{estarlich2011residential,
  author    = {Estarlich, Marisa and Ballester, Ferran and Aguilera, Inmaculada and Fern{\'a}ndez-Somoano, Ana and Lertxundi, Aitana and Llop, Sabrina and Freire, Carmen and Tard{\'o}n, Adonina and Basterrechea, Mikel and Sunyer, Jordi and others},
  journal   = {Environmental health perspectives},
  title     = {Residential exposure to outdoor air pollution during pregnancy and anthropometric measures at birth in a multicenter cohort in Spain},
  year      = {2011},
  number    = {9},
  pages     = {1333--1338},
  volume    = {119},
  doi       = {10.1289/ehp.1002918},
  publisher = {National Institute of Environmental Health Sciences},
}

@article{RN332,
  title={Association between ambient air pollution and diabetes mellitus in Europe and North America: systematic review and meta-analysis},
  author={Eze, Ikenna C and Hemkens, Lars G and Bucher, Heiner C and Hoffmann, Barbara and Schindler, Christian and K{\"u}nzli, Nino and Schikowski, Tamara and Probst-Hensch, Nicole M},
  journal={Environmental health perspectives},
  volume={123},
  number={5},
  pages={381--389},
  year={2015},
  doi     = {10.1289/ehp.1307823},
  publisher={NLM-Export}
}

@Article{frisch1933partial,
  author    = {Frisch, Ragnar and Waugh, Frederick V},
  journal   = {Econometrica: Journal of the Econometric Society},
  title     = {Partial time regressions as compared with individual trends},
  year      = {1933},
  pages     = {387--401},
  doi       = {10.2307/1907330},
  publisher = {JSTOR},
}

@Article{ghosh2013prenatal,
  author    = {Ghosh, Jo Kay C and Heck, Julia E and Cockburn, Myles and Su, Jason and Jerrett, Michael and Ritz, Beate},
  journal   = {American journal of epidemiology},
  title     = {Prenatal exposure to traffic-related air pollution and risk of early childhood cancers},
  year      = {2013},
  number    = {8},
  pages     = {1233--1239},
  volume    = {178},
  doi       = {10.1093/aje/kwt129},
  publisher = {Oxford University Press},
}

@Article{ghosh2012assessing,
  author    = {Ghosh, Jo Kay C and Wilhelm, Michelle and Su, Jason and Goldberg, Daniel and Cockburn, Myles and Jerrett, Michael and Ritz, Beate},
  journal   = {American journal of epidemiology},
  title     = {Assessing the influence of traffic-related air pollution on risk of term low birth weight on the basis of land-use-based regression models and measures of air toxics},
  year      = {2012},
  number    = {12},
  pages     = {1262--1274},
  volume    = {175},
  doi       = {10.1093/aje/kwr469},
  publisher = {Oxford University Press},
}

@Article{RN337,
  author  = {Gibson, QH and Roughton, FJW},
  journal = {The Journal of Physiology},
  title   = {The kinetics and equilibria of the reactions of nitric oxide with sheep haemoglobin},
  year    = {1957},
  number  = {3},
  pages   = {507},
  volume  = {136},
  doi     = {10.1113/jphysiol.1957.sp005777},
  type    = {Journal Article},
}

@Article{RN325,
  author  = {Godfrey, Keith M and Barker, David JP},
  journal = {The American journal of clinical nutrition},
  title   = {Fetal nutrition and adult disease},
  year    = {2000},
  issn    = {0002-9165},
  number  = {5},
  pages   = {1344S-1352S},
  volume  = {71},
  doi     = {10.1093/ajcn/71.5.1344s},
  type    = {Journal Article},
}

@Article{RN340,
  author  = {Griffiths, Mark JD and Evans, Timothy W},
  journal = {New England Journal of Medicine},
  title   = {Inhaled nitric oxide therapy in adults},
  year    = {2005},
  issn    = {0028-4793},
  number  = {25},
  pages   = {2683-2695},
  volume  = {353},
  doi     = {10.1056/nejmra051884},
  type    = {Journal Article},
}

@Article{guidotti1978higher,
  author    = {Guidotti, Tee Lamont},
  journal   = {Environmental research},
  title     = {The higher oxides of nitrogen: inhalation toxicology},
  year      = {1978},
  number    = {3},
  pages     = {443--472},
  volume    = {15},
  doi       = {10.1016/0013-9351(78)90125-1},
  publisher = {Elsevier},
}

@Article{RN302,
  author  = {Hack, Maureen and Klein, Nancy K and Taylor, H Gerry},
  journal = {The future of children},
  title   = {Long-term developmental outcomes of low birth weight infants},
  year    = {1995},
  issn    = {1054-8289},
  pages   = {176-196},
  doi     = {10.2307/1602514},
  type    = {Journal Article},
}

@Article{RN327,
  author  = {Hattersley, Andrew T and Tooke, John E},
  journal = {The Lancet},
  title   = {The fetal insulin hypothesis: an alternative explanation of the association of low bir thweight with diabetes and vascular disease},
  year    = {1999},
  issn    = {0140-6736},
  number  = {9166},
  pages   = {1789-1792},
  volume  = {353},
  doi     = {10.1016/s0140-6736(98)07546-1},
  type    = {Journal Article},
}

@Article{henschel2015trends,
  author    = {Henschel, Susann and Le Tertre, Alain and Atkinson, Richard W and Querol, Xavier and Pandolfi, Marco and Zeka, Ariana and Haluza, Daniela and Analitis, Antonis and Katsouyanni, Klea and Bouland, Catherine and others},
  journal   = {Atmospheric Environment},
  title     = {Trends of nitrogen oxides in ambient air in nine European cities between 1999 and 2010},
  year      = {2015},
  pages     = {234--241},
  volume    = {117},
  doi       = {10.1016/j.atmosenv.2015.07.013},
  publisher = {Elsevier},
}

@book{angrist2009mostly,
  title={Mostly harmless econometrics: An empiricist's companion},
  author={Angrist, Joshua D and Pischke, J{\"o}rn-Steffen},
  year={2009},
  publisher={Princeton university press}
}

@Article{RN289,
  author  = {van den Hooven, Edith H and Pierik, Frank H and de Kluizenaar, Yvonne and Willemsen, Sten P and Hofman, Albert and van Ratingen, Sjoerd W and Zandveld, Peter YJ and Mackenbach, Johan P and Steegers, Eric AP and Miedema, Henk ME},
  journal = {Environmental health perspectives},
  title   = {Air pollution exposure during pregnancy, ultrasound measures of fetal growth, and adverse birth outcomes: a prospective cohort study},
  year    = {2012},
  issn    = {0091-6765},
  number  = {1},
  pages   = {150-156},
  volume  = {120},
  doi     = {10.1289/ehp.1003316},
  type    = {Journal Article},
  url     = {https://www.ncbi.nlm.nih.gov/pmc/articles/PMC3261932/pdf/ehp.1003316.pdf},
}

@Article{jedrychowski2009gender,
  author    = {Jedrychowski, Wieslaw and Perera, Frederica and Mrozek-Budzyn, Dorota and Mroz, Elzbieta and Flak, Elzbieta and Spengler, Jack D and Edwards, Susan and Jacek, Ryszard and Kaim, Irena and Skolicki, Zbigniew},
  journal   = {Environmental research},
  title     = {Gender differences in fetal growth of newborns exposed prenatally to airborne fine particulate matter},
  year      = {2009},
  number    = {4},
  pages     = {447--456},
  volume    = {109},
  doi       = {10.1016/j.envres.2009.01.009},
  publisher = {Elsevier},
}

@ARTICLE{jha2011evaluation,
  author = {Jha, Dilip Kumar and Sabesan, M and Das, Anup and Vinithkumar, NV
	and Kirubagaran, R},
  title = {Evaluation of Interpolation Technique for Air Quality Parameters
	in Port Blair, India.},
  journal = {Universal journal of environmental research \& technology},
  year = {2011},
  volume = {1},
  number = {3}
}

@Article{kim2004traffic,
  author    = {Kim, Janice J and Smorodinsky, Svetlana and Lipsett, Michael and Singer, Brett C and Hodgson, Alfred T and Ostro, Bart},
  journal   = {American journal of respiratory and critical care medicine},
  title     = {Traffic-related air pollution near busy roads: the East Bay Children's Respiratory Health Study},
  year      = {2004},
  number    = {5},
  pages     = {520--526},
  volume    = {170},
  doi       = {10.1164/rccm.200403-281oc},
  publisher = {American Thoracic Society},
}

@Article{kimbrough2017no,
  author    = {Kimbrough, Sue and Owen, R Chris and Snyder, Michelle and Richmond-Bryant, Jennifer},
  journal   = {Atmospheric Environment},
  title     = {NO to NO2 conversion rate analysis and implications for dispersion model chemistry methods using Las Vegas, Nevada near-road field measurements},
  year      = {2017},
  pages     = {23--34},
  volume    = {165},
  doi       = {10.1016/j.atmosenv.2017.06.027},
  publisher = {Elsevier},
}

@Article{koehler2010aspects,
  author    = {Koehler, C and Ginzkey, C and Friehs, G and Hackenberg, S and Froelich, K and Scherzed, A and Burghartz, M and Kessler, M and Kleinsasser, N},
  journal   = {Toxicology and applied pharmacology},
  title     = {Aspects of nitrogen dioxide toxicity in environmental urban concentrations in human nasal epithelium},
  year      = {2010},
  number    = {2},
  pages     = {219--225},
  volume    = {245},
  doi       = {10.1016/j.taap.2010.03.003},
  publisher = {Elsevier},
}

@Article{RN328,
  author  = {Lindsay, Robert S and Dabelea, Dana and Roumain, Janine and Hanson, Robert L and Bennett, Peter H and Knowler, William C},
  journal = {Diabetes},
  title   = {Type 2 diabetes and low birth weight: the role of paternal inheritance in the association of low birth weight and diabetes},
  year    = {2000},
  issn    = {0012-1797},
  number  = {3},
  pages   = {445-449},
  volume  = {49},
  doi     = {10.2337/diabetes.49.3.445},
  type    = {Journal Article},
}

@Article{lowenstein1994nitric,
  author    = {Lowenstein, Charles J and Dinerman, Jay L and Snyder, Solomon H},
  journal   = {Annals of internal medicine},
  title     = {Nitric oxide: a physiologic messenger},
  year      = {1994},
  number    = {3},
  pages     = {227--237},
  volume    = {120},
  doi       = {10.7326/0003-4819-120-3-199402010-00009},
  publisher = {American College of Physicians},
}

@article{RN315,
  title={Ambient air pollution exposure, residential mobility and term birth weight in Oslo, Norway},
  author={Madsen, Christian and Gehring, Ulrike and Walker, Sam Erik and Brunekreef, Bert and Stigum, Hein and N{\ae}ss, {\O}yvind and Nafstad, Per},
  journal={Environmental research},
  volume={110},
  number={4},
  pages={363--371},
  year={2010},
  publisher={Elsevier}
}

@Article{marshall2008within,
  author    = {Marshall, Julian D and Nethery, Elizabeth and Brauer, Michael},
  journal   = {Atmospheric Environment},
  title     = {Within-urban variability in ambient air pollution: comparison of estimation methods},
  year      = {2008},
  number    = {6},
  pages     = {1359--1369},
  volume    = {42},
  doi       = {10.1016/j.atmosenv.2007.08.012},
  publisher = {Elsevier},
}

@Article{RN300,
  author  = {McCormick, Marie C},
  journal = {New England journal of medicine},
  title   = {The contribution of low birth weight to infant mortality and childhood morbidity},
  year    = {1985},
  issn    = {0028-4793},
  number  = {2},
  pages   = {82-90},
  volume  = {312},
  doi     = {10.1056/nejm198501103120204},
  type    = {Journal Article},
}

@Article{RN306,
  author  = {Mohorovic, Lucijan},
  journal = {Environmental health perspectives},
  title   = {The level of maternal methemoglobin during pregnancy in an air-polluted environment},
  year    = {2003},
  issn    = {0091-6765},
  number  = {16},
  pages   = {1902-1905},
  volume  = {111},
  doi     = {10.1289/ehp.6055},
  type    = {Journal Article},
}

@Article{RN305,
  author  = {Mohorovic, Lucijan and Materljan, Eris and Brumini, Gordana},
  journal = {The Journal of Maternal-Fetal \& Neonatal Medicine},
  title   = {Consequences of methemoglobinemia in pregnancy in newborns, children, and adults: issues raised by new findings on methemoglobin catabolism},
  year    = {2010},
  issn    = {1476-7058},
  number  = {9},
  pages   = {956-959},
  volume  = {23},
  type    = {Journal Article},
}

@Article{musashi2018comparison,
  author  = {Musashi, Jaka Pratama and Pramoedyo, Henny and Fitriani, Rahma},
  journal = {Cauchy},
  title   = {Comparison of inverse distance weighted and natural neighbor interpolation method at air temperature data in Malang Region},
  year    = {2018},
  number  = {2},
  pages   = {48--54},
  volume  = {5},
  doi     = {10.18860/ca.v5i2.4722},
}

@article{RN326,
  title={The association of indicators of fetal growth with visual acuity and hearing among conscripts},
  author={Olsen, J{\o}rn and S{\o}rensen, Henrik Toft and Steffensen, Flemming Hald and Sabroe, Svend and Gillman, Matthew W and Fischer, Peer and Rothman, Kenneth J},
  journal={Epidemiology},
  volume={12},
  number={2},
  pages={235--238},
  year={2001},  
  doi     = {10.1097/00001648-200103000-00017},
  publisher={LWW}
}

@Article{roberts2012seasonal,
  author    = {Roberts--Semple, Dawn and Song, Fei and Gao, Yuan},
  journal   = {Atmospheric Pollution Research},
  title     = {Seasonal characteristics of ambient nitrogen oxides and ground--level ozone in metropolitan northeastern New Jersey},
  year      = {2012},
  number    = {2},
  pages     = {247--257},
  volume    = {3},
  doi       = {10.5094/apr.2012.027},
  publisher = {Elsevier},
}

@Article{salam2005birth,
  author    = {Salam, Muhammad T and Millstein, Joshua and Li, Yu-Fen and Lurmann, Frederick W and Margolis, Helene G and Gilliland, Frank D},
  journal   = {Environmental health perspectives},
  title     = {Birth outcomes and prenatal exposure to ozone, carbon monoxide, and particulate matter: results from the Childrenâ€™s Health Study},
  year      = {2005},
  number    = {11},
  pages     = {1638--1644},
  volume    = {113},
  doi       = {10.1289/ehp.8111},
  publisher = {National Institute of Environmental Health Sciences},
}

@Article{shah2011air,
  author    = {Shah, Prakesh S and Balkhair, Taiba and Knowledge Synthesis Group on Determinants of Preterm/LBW births and others},
  journal   = {Environment international},
  title     = {Air pollution and birth outcomes: a systematic review},
  year      = {2011},
  number    = {2},
  pages     = {498--516},
  volume    = {37},
  doi       = {10.1016/j.envint.2010.10.009},
  publisher = {Elsevier},
}

@Article{RN308,
  author  = {Speakman, Eric D and Boyd, James C and Bruns, David E},
  journal = {Clinical chemistry},
  title   = {Measurement of methemoglobin in neonatal samples containing fetal hemoglobin},
  year    = {1995},
  issn    = {0009-9147},
  number  = {3},
  pages   = {458-461},
  volume  = {41},
  doi     = {10.1093/clinchem/41.3.458},
  type    = {Journal Article},
}

@Article{spicer1977photochemical,
  author    = {Spicer, Chester W},
  journal   = {Atmospheric Environment (1967)},
  title     = {Photochemical atmospheric pollutants derived from nitrogen oxides},
  year      = {1977},
  number    = {11},
  pages     = {1089--1095},
  volume    = {11},
  doi       = {10.1016/0004-6981(77)90239-6},
  publisher = {Elsevier},
}

@Article{stieb2012ambient,
  author    = {Stieb, David M and Chen, Li and Eshoul, Maysoon and Judek, Stan},
  journal   = {Environmental research},
  title     = {Ambient air pollution, birth weight and preterm birth: a systematic review and meta-analysis},
  year      = {2012},
  pages     = {100--111},
  volume    = {117},
  doi       = {10.1016/j.envres.2012.05.007},
  publisher = {Elsevier},
}

@Article{tyrrell13,
  author    = {Tyrrell, Jessica S and Yaghootkar, Hanieh and Freathy, Rachel M and Hattersley, Andrew T and Frayling, Timothy M},
  journal   = {International journal of epidemiology},
  title     = {Parental diabetes and birthweight in 236 030 individuals in the UK Biobank Study},
  year      = {2013},
  number    = {6},
  pages     = {1714--1723},
  volume    = {42},
  doi       = {10.1093/ije/dyt220},
  publisher = {Oxford University Press},
}

@Article{RN263,
  author  = {Wang, Xiaobin and Ding, Hui and Ryan, Louise and Xu, Xiping},
  journal = {Environmental health perspectives},
  title   = {Association between air pollution and low birth weight: a community-based study},
  year    = {1997},
  issn    = {0091-6765},
  number  = {5},
  pages   = {514-520},
  volume  = {105},
  doi     = {10.1289/ehp.97105514},
  type    = {Journal Article},
  url     = {https://www.ncbi.nlm.nih.gov/pmc/articles/PMC1469882/pdf/envhper00318-0060.pdf},
}

@Article{weinberger2001toxicology,
  author    = {Weinberger, Barry and Laskin, Debra L and Heck, Diane E and Laskin, Jeffrey D},
  journal   = {Toxicological Sciences},
  title     = {The toxicology of inhaled nitric oxide},
  year      = {2001},
  number    = {1},
  pages     = {5--16},
  volume    = {59},
  doi       = {https://doi.org/10.1093/toxsci/59.1.5},
  publisher = {Oxford University Press},
}

@BOOK{world2010guidelines,
  title = {WHO guidelines for indoor air quality: selected pollutants},
  publisher = {World Health Organization. Regional Office for Europe},
  year = {2010},
  author = {WHO}
}

@TechReport{RN324,
  author      = {WHO},
  institution = {World Health Organization},
  title       = {WHO Air quality guidelines for particulate matter, ozone, nitrogen dioxide and sulfur dioxide: global update 2005: summary of risk assessment},
  year        = {2006},
  type        = {Report},
}

@book{who2021global,
  title={WHO global air quality guidelines: particulate matter (PM2. 5 and PM10), ozone, nitrogen dioxide, sulfur dioxide and carbon monoxide},
  author={WHO},
  year={2021},
  publisher={World Health Organization}
}

@TechReport{who2000,
  author = {WHO},
  title  = {Air quality guidelines for Europe},
  year   = {2000},
  type   = {Report},
}

@Article{RN301,
  author  = {Wilcox, Allen J},
  journal = {International journal of epidemiology},
  title   = {On the importance and the unimportance of birthweight},
  year    = {2001},
  issn    = {1464-3685},
  number  = {6},
  pages   = {1233-1241},
  volume  = {30},
  doi     = {10.1093/ije/30.6.1233},
  type    = {Journal Article},
}

@Article{wong2004comparison,
  author    = {Wong, David W and Yuan, Lester and Perlin, Susan A},
  journal   = {Journal of Exposure Science \& Environmental Epidemiology},
  title     = {Comparison of spatial interpolation methods for the estimation of air quality data},
  year      = {2004},
  number    = {5},
  pages     = {404--415},
  volume    = {14},
  doi       = {10.1038/sj.jea.7500338},
  publisher = {Nature Publishing Group},
}

@Article{xie2004intercomparison,
  author    = {Xie, Pinhua and Liu, Wenqing and Fu, Qiang and Wang, Ruibin and Liu, Jianguo and Wei, Qingnong},
  journal   = {Advances in atmospheric sciences},
  title     = {Intercomparison of NO x, SO 2, O 3, and aromatic hydrocarbons measured by a commercial DOAS system and traditional point monitoring techniques},
  year      = {2004},
  number    = {2},
  pages     = {211},
  volume    = {21},
  doi       = {10.1007/bf02915707},
  publisher = {Springer},
}

\appendix \clearpage
\section*{Appendix}
\setcounter{section}{0}
\setcounter{subsection}{0}
\setcounter{figure}{0}
\setcounter{table}{0}
\setcounter{equation}{0}
\renewcommand{\thetable}{A\arabic{table}}
\renewcommand{\thefigure}{A\arabic{figure}}
\renewcommand*{\theHsection}{\thesection}
\renewcommand*{\theHsubsection}{\thesubsection}
\renewcommand*{\theHtable}{\thetable}
\renewcommand*{\theHfigure}{\thefigure}

\begin{figure}[htbp]
    \centering
    \includegraphics[width=0.9\textwidth]{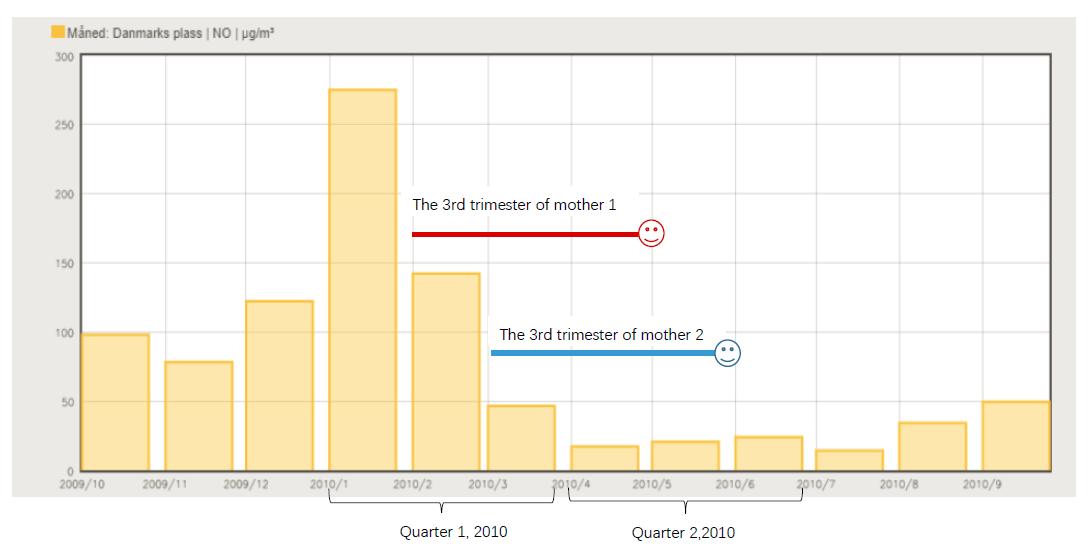}
    \caption{Two babies both born in the 2nd quarter of 2010, but have different prenatal exposure to $NO$}
    \label{fig:baby}
\end{figure}

\newpage

\begin{figure}[htbp]
    \centering
    \includegraphics[width=0.9\textwidth]{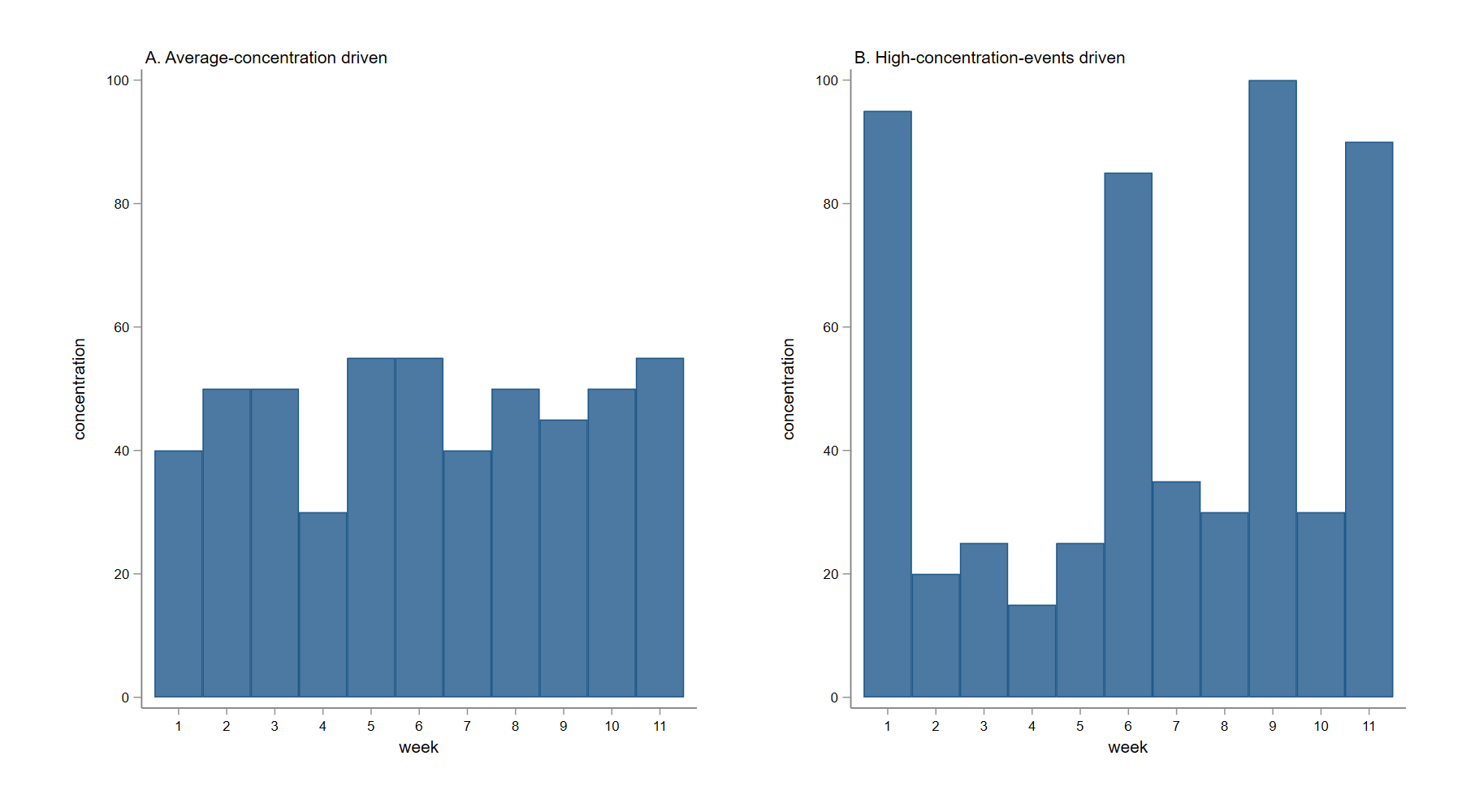}
    \caption{Given the same trimester-average concentration: relatively high concentrations in all weeks vs. occasionally very high concentration events}
    \label{fig:extent}
\end{figure}
\clearpage

\begin{table}[htbp]
\centering
\caption{Air quality standards in different countries \label{tab:standards}}

\begin{threeparttable}
\begin{tabular}{cccccccccc}
\toprule
Pollutant&\multicolumn{4}{c}{averaging period}& &EU &USA &China &WHO    \\ 
\midrule
\multirow{3}{*}{$NO_2$}          & && \multicolumn{2}{l}{annual}   && 40    & 100 & 40   &  40    \\
     							 & && \multicolumn{2}{l}{daily}    &&  -     & -   & 80   & -       \\
          						 & && \multicolumn{2}{l}{hourly}   && 200   & 188 & 200  &  200    \\
\hline        
\multirow{2}{*}{$PM_{10}$}	     & && \multicolumn{2}{l}{annual}    &&  40   &  -  &  70  &  20    \\
         						 & && \multicolumn{2}{l}{daily}    &&  50   & 150 & 150  &  50   \\
\hline
\multirow{2}{*}{$PM_{2.5}$}      & && \multicolumn{2}{l}{annual}    &&  25   &  15 &  35  &  20    \\
           						 & && \multicolumn{2}{l}{daily}    &&   -    & 35  &  75  &  25   \\

\hline
\multirow{2}{*}{$SO_2$}     	 & && \multicolumn{2}{l}{daily}    &&   -  &    125  & 150  &  20   \\ 
		  						 & && \multicolumn{2}{l}{hourly} &&  350  &  196&  500 &   -    \\
\hline
\multirow{2}{*}{$CO$}     	     & && \multicolumn{2}{l}{daily}    &&   -    & -  & 4  &  -   \\ 
		  						 & && \multicolumn{2}{l}{8-hour} &&  10   &  10 &  -   &  -     \\
\hline

\multirow{1}{*}{$O_3$}       						 & && \multicolumn{2}{l}{8-hour} &&  120  &  137&  160 &  100   \\

\bottomrule\end{tabular}
\begin{tablenotes}
      \small
      \item Notes: (1) $CO$ in $mg/m^3$; all pollutants the other pollutants are measured in $\mu g/m^3$. (2) There is no clear guideline for ambient $NO$ in all these standards.  (3) There are different safety standards in the U.S. and the one in the is meant to “provide public health protection, including protecting the health of ``sensitive" populations such as asthmatics, children, and the elderly”. (4) There are 2 classes of air quality standard in China and I adopt the one for residential zones (``2nd class”). (5) Data source: \href{https://ec.europa.eu/environment/air/quality/standards.htm}{European Commission}, \href{https://www.eea.europa.eu/themes/data-and-maps/figures/air-quality-standards-under-the}{European Environment Agency},  \href{https://www.epa.gov/criteria-air-pollutants/naaqs-table\#2.}{U.S. Environmental Protection Agency}, \href{http://www.mee.gov.cn/ywgz/fgbz/bz/bzwb/dqhjbh/dqhjzlbz/201203/t20120302_224165.shtml}{Ministry of Ecology and Environment People's Republic of China}, \href{https://www.who.int/news-room/fact-sheets/detail/ambient-(outdoor)-air-quality-and-health#:~:text=The\%202005\%20WHO\%20Air\%20quality,related\%20deaths\%20by\%20around\%2015\%25.}{World Health Organization},
    \end{tablenotes}
\end{threeparttable}
\end{table} 

\clearpage

\begin{table}[htbp]
\centering

\begin{threeparttable}

\caption{Cross-validation $R^2$: how well the interpolation and fixed effects predict the real weather condition (\%) \label{tab:cv_m} }

\def\sym#1{\ifmmode^{#1}\else\(^{#1}\)\fi}
\begin{tabular}{l*{7}{c}}
\toprule
p   &   miles  &   $humi$  &   $precip$    &   $press$ &   $temp$  &   $wind$\\
\midrule
\multirow{5}{*}{0.1} &
10  &  91.4&        88.7&       100.0&        99.3&        89.2 \\
&15 &  90.9&        85.8&       100.0&        99.2&        88.6 \\
&20 &  90.5&        86.1&       100.0&        99.2&        88.8 \\
&25 &  88.6&        85.1&        99.9&        99.1&        89.1 \\
&30 &  87.6&        80.4&        99.9&        99.1&        90.9 \\
\midrule
\multirow{5}{*}{1} &
                                                                            
10  &    91.5&        88.7&       100.0&        99.2&        89.2 \\
&15 &    90.8&        85.8&       100.0&        99.2&        88.6 \\
&20 &    90.4&        86.2&       100.0&        99.1&        88.7 \\
&25 &    88.7&        85.3&        99.9&        99.1&        89.0 \\
&30 &    87.7&        80.6&        99.9&        99.1&        90.8 \\ 
\midrule
\multirow{5}{*}{2} &
                                                                            
10  &   91.5&        88.5&       100.0&        99.2&        89.2 \\
&15 &   90.8&        85.7&       100.0&        99.2&        88.6 \\
&20 &   90.3&        86.1&       100.0&        99.1&        88.6 \\
&25 &   88.6&        85.2&        99.9&        99.1&        88.9 \\
&30 &   87.6&        80.6&        99.9&        99.1&        90.6 \\
\midrule
\multirow{5}{*}{5} &
                                                                            
10  &   91.5&        88.2&       100.0&        99.2&        89.2 \\
&15 &   90.7&        85.4&       100.0&        99.1&        88.5 \\
&20 &   90.2&        85.6&       100.0&        99.1&        88.4 \\
&25 &   88.4&        84.6&        99.9&        99.0&        88.6 \\
&30 &   87.3&        80.1&        99.9&        99.0&        90.3 \\

\bottomrule
\end{tabular}
\begin{tablenotes}
      \small
      \item Notes: (1) As defined in Section \ref{subsec:idw}, $e$ is the power of distance as defined in equation; radius determines within which range the monitoring stations are used for interpolation. (2) Column 3-7 contains the $R^2$ of in regression \ref{eq:cv} for different meteorological conditions, which shows how well the real conditions is explained by the interpolation and fixed effects.
    \end{tablenotes}
\end{threeparttable}

\end{table} 

\clearpage

\begin{table}[htbp]
\centering
\caption{Month of childbirth in the sample and the rest part of the population \label{tab:time}}

\begin{tabular}{c*{5}{c}}
\toprule
            &  \multicolumn{2}{c}{Baseline sample} & &  \multicolumn{2}{c}{Population uncovered}      \\  
            \cmidrule{2-3} \cmidrule{5-6}
Month    &    Freq.&     Percent&   &    Freq.&     Percent\\
\midrule
1        &    38,134&  8.21&&    45,213&  8.30\\
2        &    35,973&  7.75&&    42,481&  7.80\\
3        &    39,380&  8.48&&    46,224&  8.48\\
4        &    39,699&  8.55&&    46,769&  8.58\\
5        &    42,001&  9.05&&    47,633&  8.74\\
6        &    41,450&  8.93&&    46,587&  8.55\\
7        &    43,027&  9.27&&    49,247&  9.04\\
8        &    40,780&  8.78&&    48,254&  8.86\\
9        &    37,867&  8.16&&    46,993&  8.63\\
10       &    37,745&  8.13&&    44,271&  8.13\\
11       &    34,500&  7.43&&    40,722&  7.47\\
12       &    33,748&  7.27&&    40,418&  7.42\\
\midrule
Total    &  \multicolumn{2}{c}{464,304} &&  \multicolumn{2}{c}{544,812}\\
\bottomrule
\end{tabular}
\end{table} 

Table \ref{tab:time} shows the distribution of birth months in my sample. The month of delivery is evenly distributed, with slightly more babies being born in the summer months. Some parents may prefer to have their babies in early summer (May-July), either to avoid the summer heat or to take parental leave between summer vacations (so they can enjoy a "double summer vacation"). Information such as the fact that babies born in the summer get more vitamin D and are therefore healthier (which is essential in high latitude countries such as Norway) may also be at play here. I also compared the demographics of parents with ``summer babies'' and those without ``summer babies'' and found no significant differences.

\clearpage

\begin{table}[htbp]
\centering

\caption{Benchmark regression with standard errors clustered at different levels \label{tab:cluster} }

\begin{threeparttable}
\def\sym#1{\ifmmode^{#1}\else\(^{#1}\)\fi}

\begin{tabular}{l*{9}{c}}
\toprule
            &\multicolumn{1}{c}{(1)}&\multicolumn{1}{c}{(2)}&\multicolumn{1}{c}{(3)}&\multicolumn{1}{c}{(4)}&\multicolumn{1}{c}{(5)}&\multicolumn{1}{c}{(6)}&\multicolumn{1}{c}{(7)}&\multicolumn{1}{c}{(8)}&\multicolumn{1}{c}{(9)}\\
cluster:  &\multicolumn{1}{c}{p}&\multicolumn{1}{c}{muni}&\multicolumn{1}{c}{family}&\multicolumn{1}{c}{q-g}&\multicolumn{1}{c}{m-p}&\multicolumn{1}{c}{q-p}&\multicolumn{1}{c}{s}&\multicolumn{1}{c}{m-s}&\multicolumn{1}{c}{q-s}\\
\midrule
\multicolumn{7}{l}{A.Birth weight} \\
$NO$        &      -1.387\sym{*}  &      -1.387\sym{***}&      -1.387\sym{**} &      -1.387\sym{**} &      -1.387\sym{**} &      -1.387\sym{**} &      -1.387\sym{**} &      -1.387\sym{**} &      -1.387\sym{**} \\
            &     (0.068)         &     (0.008)         &     (0.035)         &     (0.027)         &     (0.026)         &     (0.033)         &     (0.011)         &     (0.011)         &     (0.017)         \\
\addlinespace
$NO_2$       &      -0.259         &      -0.259         &      -0.259         &      -0.259         &      -0.259         &      -0.259         &      -0.259         &      -0.259         &      -0.259         \\
            &     (0.902)         &     (0.841)         &     (0.881)         &     (0.891)         &     (0.881)         &     (0.890)         &     (0.876)         &     (0.876)         &     (0.887)         \\
\addlinespace
$PM_{10}$      &       1.329         &       1.329         &       1.329         &       1.329         &       1.329         &       1.329         &       1.329         &       1.329         &       1.329         \\
            &     (0.455)         &     (0.244)         &     (0.412)         &     (0.318)         &     (0.413)         &     (0.315)         &     (0.364)         &     (0.421)         &     (0.331)         \\
\midrule
\multicolumn{7}{l}{B.Birth length} \\

$NO$        &      -0.052         &      -0.052\sym{***}&      -0.052\sym{*}  &      -0.052\sym{*}  &      -0.052\sym{*}  &      -0.052\sym{*}  &      -0.052\sym{**} &      -0.052\sym{**} &      -0.052\sym{**} \\
            &     (0.141)         &     (0.000)         &     (0.087)         &     (0.065)         &     (0.089)         &     (0.069)         &     (0.035)         &     (0.041)         &     (0.023)         \\
\addlinespace
$NO_2$       &      -0.023         &      -0.023         &      -0.023         &      -0.023         &      -0.023         &      -0.023         &      -0.023         &      -0.023         &      -0.023         \\
            &     (0.816)         &     (0.710)         &     (0.784)         &     (0.770)         &     (0.779)         &     (0.769)         &     (0.804)         &     (0.800)         &     (0.792)         \\
\addlinespace
$PM_{10}$      &      -0.004         &      -0.004         &      -0.004         &      -0.004         &      -0.004         &      -0.004         &      -0.004         &      -0.004         &      -0.004         \\
            &     (0.967)         &     (0.944)         &     (0.962)         &     (0.956)         &     (0.961)         &     (0.956)         &     (0.952)         &     (0.954)         &     (0.943)         \\
\bottomrule
\end{tabular}
\begin{tablenotes}
      \small
      \item Notes: (1) The standard errors are clustered at level: p(postcode), muni(municipality), family(children of the same mother), q-g(calendar quarter and \textit{grunnkrets}), m-p(calendar month and postcode), q-p(calendar quarter and postcode), s(nearest monitoring station), m-s(calendar month and nearest monitoring station) and q-s(calendar quarter and nearest monitoring station). (2) The independent variables in Panel A and Panel B are birth weight and birth length separately. (3) *** $p < 0.01$, ** $p<0.05$, * $p<0.1$. (4) All pollutants are in $\mu g/m^3$, birth-weight in gram, birth length in millimeter.
    \end{tablenotes}
\end{threeparttable}
\end{table}




\clearpage
\begin{table}[htbp]
\centering
\caption{Air pollution exposure during the whole pregnancy \label{tab:3t}}

\begin{threeparttable}
\def\sym#1{\ifmmode^{#1}\else\(^{#1}\)\fi}

\begin{tabular}{l*{9}{c}}
\toprule
   &&\multicolumn{3}{c}{birth weight}& &\multicolumn{3}{c}{birth length} \\
            \cmidrule{3-5} \cmidrule{7-9}
\multicolumn{2}{l}{trimester} &\multicolumn{1}{c}{all}&\multicolumn{1}{c}{same}&\multicolumn{1}{c}{stay}  &&\multicolumn{1}{c}{all}&\multicolumn{1}{c}{same}&\multicolumn{1}{c}{stay} \\

\midrule

\multirow{6}{*}{$1st$}   & $NO$          &       0.958         &       1.181         &       0.884     &  &        0.022         &       0.017         &       0.049       \\
                         &               &     (0.773)         &     (0.970)         &     (1.093)     &  &      (0.036)         &     (0.046)         &     (0.052)       \\
                         & $NO_2$        &      -1.568         &      -2.520         &      -2.226     &  &       -0.065         &      -0.072         &      -0.133       \\
                         &               &     (2.058)         &     (2.582)         &     (2.765)     &  &      (0.097)         &     (0.127)         &     (0.125)       \\
                         & $PM_{10}$     &       1.151         &       1.826         &      -0.396     &  &        0.105         &       0.088         &       0.108       \\
                         &               &     (1.882)         &     (2.192)         &     (2.600)     &  &      (0.091)         &     (0.103)         &     (0.120)       \\
\addlinespace
\multirow{6}{*}{$2nd$}   & $NO$           &      0.699         &       0.803         &       0.063     &  &       -0.046         &      -0.091         &      -0.001            \\
                         &                &    (1.268)         &     (1.585)         &     (1.676)     &  &      (0.060)         &     (0.073)         &     (0.078)            \\
                         & $NO_2$         &      2.587         &       1.747         &       0.331     &  &        0.238\sym{*}  &       0.203         &       0.158            \\
                         &                &    (2.866)         &     (3.571)         &     (3.673)     &  &      (0.138)         &     (0.166)         &     (0.169)            \\
                         & $PM_{10}$      &      1.225         &       4.367         &       2.223     &  &       -0.048         &       0.154         &      -0.083            \\
                         &                &    (3.182)         &     (3.731)         &     (4.267)     &  &      (0.146)         &     (0.171)         &     (0.189)            \\
\addlinespace
\multirow{6}{*}{$3rd$}   & $NO$           &     -1.010         &      -0.777         &      -0.692     &  &       -0.063         &      -0.066         &      -0.048           \\
                         &                &    (0.858)         &     (1.052)         &     (1.102)     &  &      (0.040)         &     (0.049)         &     (0.052)           \\
                         & $NO_2$         &       0.786         &      -0.003         &      -0.566    &  &         0.071         &      -0.024         &       0.023          \\
                         &                &    (2.168)         &     (2.718)         &     (2.801)     &  &      (0.102)         &     (0.133)         &     (0.127)           \\
                         & $PM_{10}$      &      1.610         &       3.475         &       3.984     &  &       -0.019         &       0.101         &       0.077             \\
                         &                &    (1.952)         &     (2.402)         &     (2.657)     &  &      (0.104)         &     (0.136)         &     (0.125)            \\
\midrule
\multicolumn{2}{c}{ $r^2$ }               &      0.464         &       0.469         &       0.502      &  &       0.462         &       0.461         &       0.501            \\
\multicolumn{2}{c}{  Obs. }               &    211,935         &     135,535         &     135,714      &  &     200,292         &     128,855         &     127,816           \\
             
\bottomrule
\end{tabular}
\begin{tablenotes}
      \small
      \item Notes: (1)The three trimesters are regressed in one model, not separately. (2) $all$ means all observations in the working data is used; $same$ means mothers lived in the same municipalities in the delivery year as where the children were born, $stay$ keeps only those whose mothers resided in the same \textit{grunnkrets} in the delivery year and the year before delivery. (3) Cluster robust standard errors at the \textit{grunnkrets} level in parentheses. (4) *** $p < 0.001$, ** $p<0.01$, * $p<0.05$. (5) All pollutants are in $\mu g/m^3$, birth-weight in gram, birth length in millimeter.

    \end{tablenotes}
\end{threeparttable}
\end{table} 

In Table \ref{tab:3t}, I include average ambient air pollution levels and weather conditions throughout the third trimester of pregnancy in the baseline regression model. The samples in columns (3) and (5) are the baseline sample (``all''), while the infants in columns (4) and (6) (``same'') are born in the same city as their mother's residence in the year of delivery, as a robustness check (as discussed in Section \ref{sec:identify}.).

When air pollution and weather conditions in the first two quarters are included, the $NO$ coefficients in the last three months remain very similar to the baseline regressions of birth weight and birth length. Although most of the coefficients in Table \ref{tab:3t} are not significant at the $10\%$ level, the coefficients on $NO_2$ and $PM_{10}$ are less significant, i.e., the magnitude of the effect is smaller and the standard error is larger. Based on the findings in Table \ref{tab:3t}, I included only the last three months of air pollution levels and weather conditions in the baseline analysis.

\clearpage

\begin{table}[htbp]
\centering

\begin{threeparttable}

\caption{The effects of ambient $O_3$ in the 3rd. trimester on birth outcomes \label{tab:ozone} }

\def\sym#1{\ifmmode^{#1}\else\(^{#1}\)\fi}

\begin{tabular}{l*{5}{c}}
\toprule
            &\multicolumn{1}{c}{(1)}&\multicolumn{1}{c}{(2)}&\multicolumn{1}{c}{(3)}&\multicolumn{1}{c}{(4)} \\
\midrule
\multicolumn{4}{l}{A. Birth weight} \\

$O_3$        &       0.911         &      -0.141         &      -0.121         &      -0.335         \\
            &     (0.955)         &     (1.143)         &     (1.370)         &     (1.383)         \\
$NO$        &                     &      -0.985\sym{*}  &      -1.006         &      -1.796\sym{*}  \\
            &                     &     (0.529)         &     (0.769)         &     (1.038)         \\
$NO_2$       &                     &                     &                     &       2.808         \\
            &                     &                     &                     &     (2.528)         \\
\midrule
$r^2$       &       0.439         &       0.423         &       0.417         &       0.417         \\
Obs.        &     172,352         &     150,629         &     144,298         &     144,298         \\

\midrule
\multicolumn{4}{l}{B.Birth length} \\

$O_3$        &       0.052         &       0.010         &       0.032         &       0.024         \\
            &     (0.048)         &     (0.056)         &     (0.067)         &     (0.068)         \\
$NO$        &                     &      -0.038         &      -0.021         &      -0.050         \\
            &                     &     (0.024)         &     (0.034)         &     (0.047)         \\
$NO_2$       &                     &                     &                     &       0.104         \\
            &                     &                     &                     &     (0.115)         \\
\midrule
$r^2$       &       0.445         &       0.432         &       0.424         &       0.424         \\
Obs.        &     162,243         &     142,259         &     136,416         &     136,416         \\
\bottomrule
\end{tabular}

\begin{tablenotes}
      \small
      \item Notes: (1) Regressions are based on the benchmark model but include only $O_3$, $NO$ and $NO_2$ in order to examine the correlation between the pollutants (2) The independent variables in Panel A and Panel B are birth weight and birth length separately. (3) Cluster robust standard errors at \textit{grunnkrets} level in parentheses. (4) *** $p < 0.01$, ** $p<0.05$, * $p<0.1$. (5) All pollutants are in $\mu g/m^3$, birth-weight in gram, birth length in millimeter.
    \end{tablenotes}
\end{threeparttable}

\end{table}

\clearpage
\begin{table}[htbp]
\centering
\caption{Excluding mothers exposed to high-level ambient air pollution in 3rd. trimester \label{tab:extent}}

\begin{threeparttable}

\def\sym#1{\ifmmode^{#1}\else\(^{#1}\)\fi}
\begin{tabular}{l*{12}{c}}
\toprule
            &\multicolumn{1}{c}{(1)}&\multicolumn{1}{c}{(2)}&&\multicolumn{1}{c}{(3)}&\multicolumn{1}{c}{(4)}&&\multicolumn{1}{c}{(5)}&\multicolumn{1}{c}{(6)}&&\multicolumn{1}{c}{(7)}&\multicolumn{1}{c}{(8)} \\
            &\multicolumn{2}{c}{$NO$} &&\multicolumn{2}{c}{$NO_2$} &&\multicolumn{2}{c}{$PM_{10}$} &&\multicolumn{2}{c}{$NO_2$ \& $PM_{10}$}\\
            \cmidrule{2-3} \cmidrule{5-6} \cmidrule{8-9} \cmidrule{11-12}
            &\multicolumn{1}{c}{$<99p.$}&\multicolumn{1}{c}{$<95p.$}&&\multicolumn{1}{c}{$<99p.$}&\multicolumn{1}{c}{$<95p.$} &&\multicolumn{1}{c}{$<99p.$}&\multicolumn{1}{c}{$<95p.$} &&\multicolumn{1}{c}{$<99p.$}&\multicolumn{1}{c}{$<95p.$}\\

\midrule
\multicolumn{8}{l}{A.Birth weight} \\
$NO$        &      -1.896\sym{**} &       0.756         & &      -2.342\sym{***}&      -2.717\sym{***}& &      -1.240\sym{*}  &      -1.448         & &      -2.298\sym{***}&      -2.012         \\
            &     (0.784)         &     (1.151)         & &     (0.693)         &     (0.954)         & &     (0.707)         &     (0.920)         & &     (0.785)         &     (1.224)         \\
$NO_2$      &       0.765         &      -0.445         & &       1.750         &       0.693         & &       0.472         &       1.626         & &       2.933         &       1.139         \\
            &     (1.886)         &     (2.118)         & &     (1.906)         &     (2.551)         & &     (2.017)         &     (2.545)         & &     (2.207)         &     (3.336)         \\
$PM_{10}$   &       1.140         &       0.537         & &       1.716         &       0.745         & &       0.871         &      -0.287         & &       0.592         &       1.429         \\
            &     (1.620)         &     (1.763)         & &     (2.079)         &     (2.995)         & &     (1.620)         &     (1.863)         & &     (2.274)         &     (3.583)         \\
\midrule
$r^2$       &       0.467         &       0.468         & &       0.464         &       0.468         & &       0.467         &       0.470         & &       0.466         &       0.470         \\
Obs.        &     205,170         &     165,555         & &     206,770         &     156,122         & &     206,417         &     163,289         & &     190,339         &     127,154         \\

\midrule
\multicolumn{8}{l}{B.Birth length} \\

$NO$        &      -0.053         &       0.082         & &      -0.085\sym{**} &      -0.099\sym{**} & &      -0.053         &      -0.012         & &      -0.082\sym{**} &      -0.055         \\
            &     (0.038)         &     (0.055)         & &     (0.033)         &     (0.045)         & &     (0.034)         &     (0.043)         & &     (0.037)         &     (0.054)         \\
$NO_2$      &       0.007         &      -0.056         & &       0.068         &       0.008         & &       0.014         &       0.020         & &       0.124         &       0.006         \\
            &     (0.087)         &     (0.101)         & &     (0.084)         &     (0.111)         & &     (0.092)         &     (0.117)         & &     (0.097)         &     (0.145)         \\
$PM_{10}$   &      -0.054         &      -0.096         & &      -0.046         &      -0.117         & &      -0.017         &      -0.128         & &      -0.078         &      -0.086         \\
            &     (0.083)         &     (0.094)         & &     (0.099)         &     (0.137)         & &     (0.083)         &     (0.097)         & &     (0.107)         &     (0.163)         \\
\midrule
$r^2$       &       0.463         &       0.466         & &       0.462         &       0.466         & &       0.464         &       0.467         & &       0.465         &       0.470         \\
Obs.        &     194,486         &     157,291         & &     195,830         &     148,232         & &     195,498         &     155,014         & &     180,669         &     121,101         \\
\bottomrule

\end{tabular}
\begin{tablenotes}
      \small
      \item Notes: (1) Regressions are based on the benchmark regression model in babies whose mother were not exposed to high-level ambient air pollution. Here $99p.$ and $95p.$ means that the weekly maternal exposure to certain pollutants in the third trimester is all below 99th and 95th percentile. (2) Cluster robust standard errors at \textit{grunnkrets} level in parentheses, (3) *** $p < 0.01$, ** $p<0.05$, * $p<0.1$. (5) All pollutants in $\mu g/m^3$, birth-weight in gram, birth length in millimeter.

    \end{tablenotes}
\end{threeparttable}

\end{table} 
The regressions in Table \ref{tab:extent} are for certain environmental contaminants in the subsample where mothers are not exposed to any high levels of environmental contamination events (weekly mean concentrations >$99th/95th$ percentile) in the last trimester. For example, weekly prenatal environmental $NO$ exposures in the last trimester ($11$ weeks total) in column (1) are below the $99th$ percentile. Columns (1) and (2) are copied from columns (1) and (5) in Table \ref{tab:extent_tmavg} for comparison purposes. The coefficient on $NO$ is no longer negative when all $11$ weeks of prenatal exposure to $NO$ in the last trimester are below the $95th$ percentile, and the exclusion of high levels of environmental $NO_2$ and $PM_{10}$ pollution events does not affect the sign of the coefficient on $NO$. This means that it is the high levels of $NO$, not $NO_2$ or $PM_{10}$, that reduces the birth outcome.

\end{document}